\documentclass[a4paper]{article}

\usepackage[table]{xcolor} 
\usepackage{setspace} 
\usepackage{amsfonts}
\usepackage{verbatim}
\usepackage{graphicx}
\usepackage{lscape} 
\usepackage{subfigure}
\usepackage{multirow}

\usepackage{amsfonts, amsmath, hyperref, parskip, times}
\usepackage[centering]{geometry} 
\usepackage{epstopdf} 
\usepackage{booktabs}

\newtheorem{theorem} {Theorem} [section]
\newtheorem{proposition}{Proposition}[section]

\newtheorem{definition}{Definition}[section]
\newtheorem{example}{Example}[section]

\hypersetup{
  colorlinks,
  linkcolor=blue,
  urlcolor=blue
}

\renewcommand{\title}[1]{\begin{center}{\bf \LARGE #1}\end{center}}

\newcommand{\keywords}{\paragraph{Keywords:}}

\begin{document}
\pagestyle{plain}  

\begin{doublespace}
\title{\textit{Distribution and Symmetric Distribution} \\[8pt] Regression Model for Interval-Valued Variables}
\end{doublespace}

\begin{center}
  {\bf S\'{o}nia Dias$^{1,^\star}$, Paula Brito$^{2}$}
\end{center}

\begin{affiliations}
\noindent
1. Escola Superior de Tecnologia e Gest\~{a}o do Instituto Polit\'{e}cnico Viana do Castelo \&  LIAAD-INESC TEC, Universidade do Porto, Portugal \\[-2pt]
2. Faculdade de Economia \& LIAAD-INESC TEC, Universidade do Porto, Portugal \\[-2pt]
$^\star$Contact author: \href{mailto:foo@bar.com}{sdias@estg.ipvc.pt}
\end{affiliations}

\begin{abstract}

Symbolic Data Analysis works with variables for which each unit or class of units takes a finite set of values/categories, an interval or a distribution (an  histogram, for instance). When to each observation corresponds an empirical distribution, we have a histogram-valued variable;  it reduces to the case of an interval-valued variable if each unit takes values on only one interval with probability equal to one. \textit{Distribution and Symmetric Distribution} is a linear regression model proposed for histogram-valued variables that may be particularized to interval-valued variables. This model is defined for $n$ explicative variables and is based on the distributions considered within the intervals. In this paper we study the special case where the Uniform distribution is assumed in each observed interval. As in the classical case, a goodness-of-fit measure is deduced from the model. Some illustrative examples are presented. A simulation study allows discussing interpretations of the behavior of the model for this variable type.\medskip
\end{abstract}

\keywords data with variability; linear regression; Symbolic Data Analysis; quantile functions.

\section{Introduction}\label{s1}

About 30 years ago Schweizer advocated that ``distributions are the numbers of the future". Following in his footsteps, Diday generalized the classical concept of variables in Multivariate Data Analysis and introduced \textit{ Symbolic Data Analysis} \cite{diday88}. The extensive and complex data that emerged in the last decades made it necessary to extend and generalize the classical concept of data sets. Data tables where the cells contain a single quantitative or categorical value were no longer sufficient. More complex data tables were needed, with cells that include more accurate and complete information. Each cell should express the variability of the records of each observed unit. These tables are called \textit{symbolic data tables} \cite{diday88} and their cells may contain finite sets of values/categories, intervals or distributions.The corresponding variables are named \textit{symbolic variables}. In this case, the objects may be one unit (first-level units) or classes of units (higher-level units). \textit{Symbolic variables} can be classified as multi-valued quantitative/qualitative variables when each unit or class of units takes a finite set of values/categories; interval-valued variables when the values that the variable takes are intervals; modal multi-valued variable when to each (first-level or higher-level) unit corresponds a probability/frequency/weight distribution.
Histogram-valued variables constitute a particular case of this latter kind of symbolic variables where to each entity under analysis corresponds an empirical distribution. However if, for all observations, each unit takes values on only one interval with probability/frequency one, the histogram-valued variable is then reduced to the particular case of an interval-valued variable.
Table \ref{table1} is an example of a \textit{symbolic data table} where the entities under analysis are healthcare centers (higher-level units).
This table results from the aggregation (contemporary aggregation, \cite{tesear08}) of records in a classical data table where the observed units are the patients (first-level units) of each healthcare center.

\renewcommand{\arraystretch}{1.5}
\begin{table}[h!]
  \centering
\begin{tabular}{ccccc} \toprule
Healthcare  & Gender & Age & Number of & Waiting time for   \\
centers & & & emergency consults & consult (in minutes) \\ \hline
A & $\{F,\frac{1}{2}; M,\frac{1}{2}\}$ & $\left[25,53\right]$  & \{0,1,2\} & $\{\left[15,30\right[, 0.25;  $ \\
&&&& $\left[30,45\right[, 0.5; \ge 60, 0.25 \}$  \\ \hline
B &  $\{F,\frac{2}{3}; M,\frac{1}{3}\}$ & $\left[33,68\right]$ & \{0,1,4,5,10\}  & $\{\left[0,15\right[,0.25; \left[15,30\right[, 0.25; $  \\
&&&& $\left[30,45\right[, 0.25; \left[45,60\right[,0.25 \} $  \\ \hline
C &  $\{F,\frac{2}{5}; M,\frac{3}{5}\}$ & $\left[20,75\right]$  & \{0,1,7,14\}  &  $\{ \left[0,15\right[,0.33; $  \\
&&&& $\left[30,45\right[, 0.33; \ge 60, 0.33 \} $  \\  \toprule
  \end{tabular}
 \caption{Symbolic data table with information corresponding to three healthcare centers.}\label{table1}
 \end{table}

The symbolic variables in Table \ref{table1} are classified as follows: age is a interval-valued variable; number of emergency consults is a multi-valued quantitative variable; gender and waiting time for consult are modal-valued variables. The waiting time for consult is more precisely a histogram-valued variable. Alternatively, we could compute and record only the mean, median, maximum or mode of the observed values in each healthcare center, but in this case the variability of the data would be lost.

In other situations we may have multiple records associated to each unit that may be the result of several observations performed in one day/month/year. If we want to study this variable, and as an alternative to summarizing all values in just one value - and thereby losing the information of the variability - and if the observed order is not pertinent, we may aggregate the information referring to one specific period of time (temporal aggregation, \cite{tesear08}). Thereby each unit (first-level unit) may be associated to an interval of values (interval-valued variable) or to a distribution (histogram-valued variable).

In recent years, statistical concepts and methods for analyzing such symbolic data have been developed \cite{noirbri11,dinoir08, bidi07, bidi03, bodi00}. Interval-valued variables are the most studied among symbolic variable types. Even though distributions are the ``numbers of the future", it does not appear simple to work with these elements. Typically,  concepts and methods for interval-valued variables are  defined first, and only then an attempt to generalize them to histogram-valued variables is made. This approach is also used because  histogram-valued variables are considered to be a generalization of interval-valued variables. In this study the approach is different. We will consider interval-valued variables as a particular case of histogram-valued variables, and we will particularize the linear regression model proposed for histogram-valued variables, the \textit{Distribution and Symmetric Distribution} Regression Model \cite{dibr12}. In this paper, we will consider that the ``values" associated to each observation of the explicative and response interval-valued variables are uniformly distributed across each interval; however, other distributions may be considered.

In the framework of symbolic data analysis,  the linear regression models for interval-valued variables previously proposed are very different from the one presented here. The most noteworthy of the proposed models are the \textit{Center Method} \cite{bidi00}; the \textit{MinMax Method} \cite{bidi02}; the \textit{Center and Range Method} \cite{netocar08} and the \textit{Constrained Center and Range Method} \cite{netocar10}. In all these methods it is possible to predict a response variable from $n$ explicative variables. The referred models do not treat the intervals as such, they require the adjustment of classical linear regression models for the lower and upper bounds or for the center and half range. In other words, these models are based on the difference between real values and do not quantify the closeness between intervals. Therefore, the elements estimated by the models may fail to build an interval; to solve this problem the most recent model imposes non-negative constraints in the linear regression between the half ranges of the intervals \cite{netocar10}. Recently, a Particular Swarm Optimization (PSO) algorithm has been applied to estimate the parameters of the linear regression models mentioned above and this new method provides satisfactory results \cite{yang11}.
In 2011, Giordani proposed a new approach to linear regression for interval-valued variables based on the Lasso technique, named \textit{Lasso-IR method} \cite{giordani11}. As in the \textit{Center and Range Method} and \textit{Constrained Center and Range Method} in this new approach the linear relationship between interval-valued variables also considers two regression models, one for the centers and another for the half ranges. However, in this case, the parameters of the models are related and although the model imposes constraints on the linear regression between the half ranges, it does not impose a direct linear relationship between them.
Another limitation of all linear regression models referred to above, is that no goodness-of-fit measure is deduced from the models.
The limitations described above and the complexity inherent to  working with histograms may prevent a generalization of the models to histogram-valued variables.

Most linear regression models proposed to interval-valued variables and histogram-valued variables in the context of \textit{Symbolic Data Analysis} are descriptive. The development of non-descriptive methods is still an open research topic for almost all kinds of symbolic variables. However, some papers recently published propose probabilistic models for interval-valued variables and inference studies were presented (see \cite{ahn12, netocar11, brsi11, tesexu10}). Of these works the research of Lima Neto \textit{et al.} should be emphasized \cite{netocar11}. In this study the authors represent an interval-valued variable $Y$ as a bivariate vector $(Y_{1},Y_{2})$ where $Y_{1}$ and $Y_{2}$ are one-dimensional random variables. The \textit{Bivariate symbolic regression Models} proposed in this work, are a generalization of the theory regression models. In this case, the authors assume that the response interval-valued variable belongs to the bivariate exponential family of distributions. The models proposed by Lima Neto \textit{et al.} \cite{netocar11} do not have some of the problems associated to the descriptive models previously proposed. They guarantee that the upper bound of the estimated interval is always greater than or equal to the lower bound, a goodness-of-fit measure was deduced; a definition of residuals for intervals is performed and inference techniques were also proposed (residual analysis and diagnostic measures).

Other studies also investigate linear regression models for other data where the observations also take the form of intervals, i.e., imprecise data.
It is however important to underline that these data are different from symbolic data. Although the type of observations are the same, i.e.,  intervals, their meaning and the way they are built is different. Imprecise data occur when each interval associated to each unit under analysis represents the uncertain value associated to the record. For example, they may result from the measure of distances or longitudes with imprecise instruments. In this context, ``the intervals are a imprecise perception of real values non observable" \cite{moo66}. All linear regression models defined for imprecise data predict one interval from other intervals using interval arithmetic \cite{moo66}.
The use of this arithmetic is probably one of the reasons that makes the generalization of the models to $n$ explicative variables difficult. The first linear regression models proposed for this kind of elements were simple linear regression models defined in a descriptive context \cite{diamond90, gil02}.  More recently, developments for the analysis of  imprecise data have been made in an inferencial framework. Random intervals or interval-valued random sets variables are defined as a generalization of random variables (real-valued random variables) when the outcomes that result from a random experience are described by a compact set instead of a real number. Some linear regression models between random imprecise elements have been proposed, we may cite the populacional \textit{Model MRLS} \cite{rodriguez07} and the more flexible \textit{Model M} \cite{fernandez11}. However, these models only allow predicting one response interval-valued variable from one explicative valued variable and always induce direct linear relationships between the half ranges of the intervals as the \textit{Constrained Center and Range Method}.

The remainder of the paper is organized as follows. Section 2 introduces the representation of intervals by quantile functions and presents the new approach for a linear regression model with interval-valued variables. Section 3 reports two simulation studies and discusses their results. In Section 4, some illustrative applications are presented. Finally, Section 5 concludes the paper, pointing out directions for future research.
\section{\textit{DSD Model} for interval-valued variables} \label{s2}

The \textit{Distribution and Symmetric Distribution (DSD)} Regression Model is a linear regression model for histogram-valued variables proposed in \cite{dibr12}. Since interval-valued variables are a particular case of histogram-valued variables we may apply the \textit{DSD Model} to interval-valued variables. The innovations of the \textit{DSD regression model} for interval-valued variables that we propose in this paper are as follows. First and foremost, the model works with intervals and considers the distribution within the intervals; in this paper the Uniform distribution is assumed, but other distributions may also be considered. Then, the intervals are represented by quantile functions. Also, the model allows predicting a response variable from $n$ explicative variables and the predicted range of values always constitutes an interval; the linear relationships between the centers and half ranges induced by the model between the intervals are different although related. Furthermore, it is possible to deduce a goodness-of-fit measure from the model.
The fact that we shall be using a representation of the intervals by quantile functions makes it important to make  a short introduction to these functions.

\subsection{Quantile functions}\label{s2.1}

When we have a interval-valued variable $Y$, to each unit $j$ corresponds one ``symbolic value" (range of the values) that may be represented by an interval $I_{Y(j)}$ or by the respective quantile function $\Psi^{-1}_{Y(j)}.$

\begin{definition}\label{def2.1}
 $Y$ is a interval-valued variable when to each unit $j \in \left\{1,\ldots,m\right\}$ corresponds an interval $Y(j)$ of real numbers. $Y(j)$ may be represented by the interval \cite{bidi03}:
$$I_{Y(j)}=\left[\underline{I}_{Y(j)},\overline{I}_{Y(j)}\right] $$
where $\underline{I}_{Y(j)}$ and  $\overline{I}_{Y(j)}$  are the lower and upper bounds of the interval $I_{Y(j)},$ respectively.
\end{definition}

It is also possible to represent the interval $Y(j)$ by its center $c_{{Y}(j)}=\frac{\overline{I}_{Y(j)}+\underline{I}_{Y(j)}}{2}$  and half-range $r_{{Y}(j)}=\frac{\overline{I}_{Y(j)}-\underline{I}_{Y(j)}}{2}.$ In this case,
\[I_{Y(j)}=\left[c_{{Y}(j)}-r_{{Y}(j)}; c_{{Y}(j)}+r_{{Y}(j)}\right]\]

Alternatively, considering the distribution within the intervals, they may be represented by quantile functions.
Assuming an Uniform distribution in all intervals $Y(j),$ we may represent each interval $Y(j)$ by a linear function with domain $\left[0,1\right]$ as follows:

\[
\Psi_{Y(j)}^{-1}(t)=\underline{I}_{Y(j)}+\left(\overline{I}_{Y(j)}-\underline{I}_{Y(j)}\right)t, \qquad 0\leq t \leq 1
\]

or using the center $c_{Y(j)}$ and half-range $r_{Y(j)}$ of the interval as

\[
\Psi_{Y(j)}^{-1}(t)=c_{Y(j)}+r_{Y(j)}(2t-1), \qquad 0\leq t \leq 1.
\]

The representation of the intervals by linear functions was presented by Bertoluzza \textit{et al.} \cite{beall95}, that termed it parametrization of the interval. More recently, and particularizing this representation from the piece-wise function that represents histograms, Irpino and Verde \cite{irve06} named the linear function as a quantile function, the inverse cumulative distribution function.

Since in all intervals, the lower bound is always less than or equal than the upper bound, $\underline{I}_{Y(j)} \leq \overline{I}_{Y(j)},$ the quantile function that represents an interval is always a non-decreasing function \cite{dibr12}.
This behavior is illustrated in \textit{Example \ref{ex1}}.

\begin{example} \label{ex1}
Consider again the interval-valued variable ``Age", $Y_{2}$ in \textit{Table \ref{table1}}, and the respective intervals corresponding to each of the three healthcare centers. The observed value of this interval-valued variable $Y_{2}$ for Healthcare center A, may be represented by:

\[Y_{2}(A)=\left[25;53\right]\] \quad {\rm or} \quad \[\Psi^{-1}_{Y_{2}(A)}=25+28t \quad {\rm with} \quad 0\leq t \leq 1.\]

The quantile functions that represent the intervals of the ages associated to each healthcare center are represented in \textit{Figure \ref{fig1}}.

\begin{figure}[h!]
\begin{center}
    \includegraphics[width=0.75\textwidth]{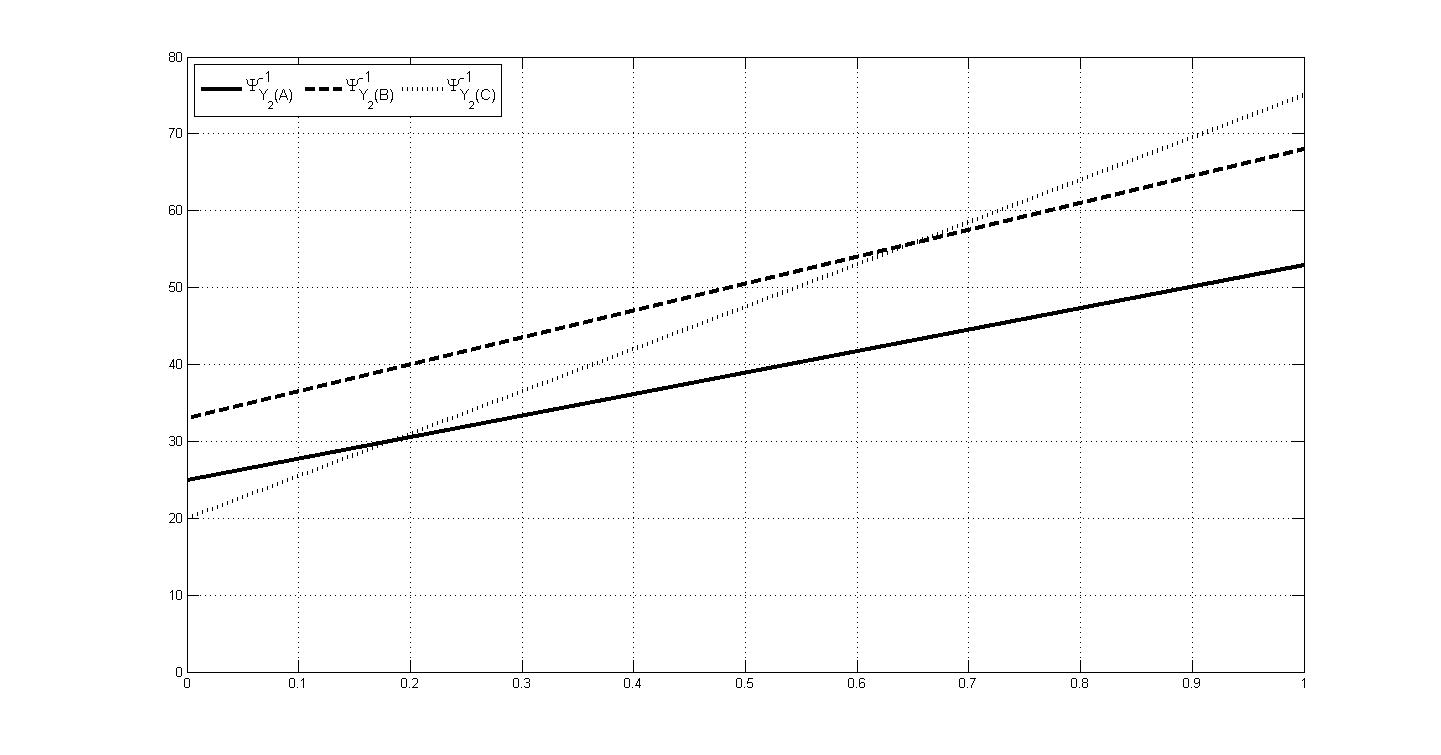}
    \caption{Representation of the quantile functions $\Psi^{-1}_{Y_{2}(A)};$ $\Psi^{-1}_{Y_{2}(B)};$ $\Psi^{-1}_{Y_{2}(C)}$ in \textit{Table \ref{table1}}.}
   \label{fig1}
\end{center}
\end{figure}

\end{example}

\begin{figure}[h!]
\begin{center}
    \includegraphics[width=0.75\textwidth]{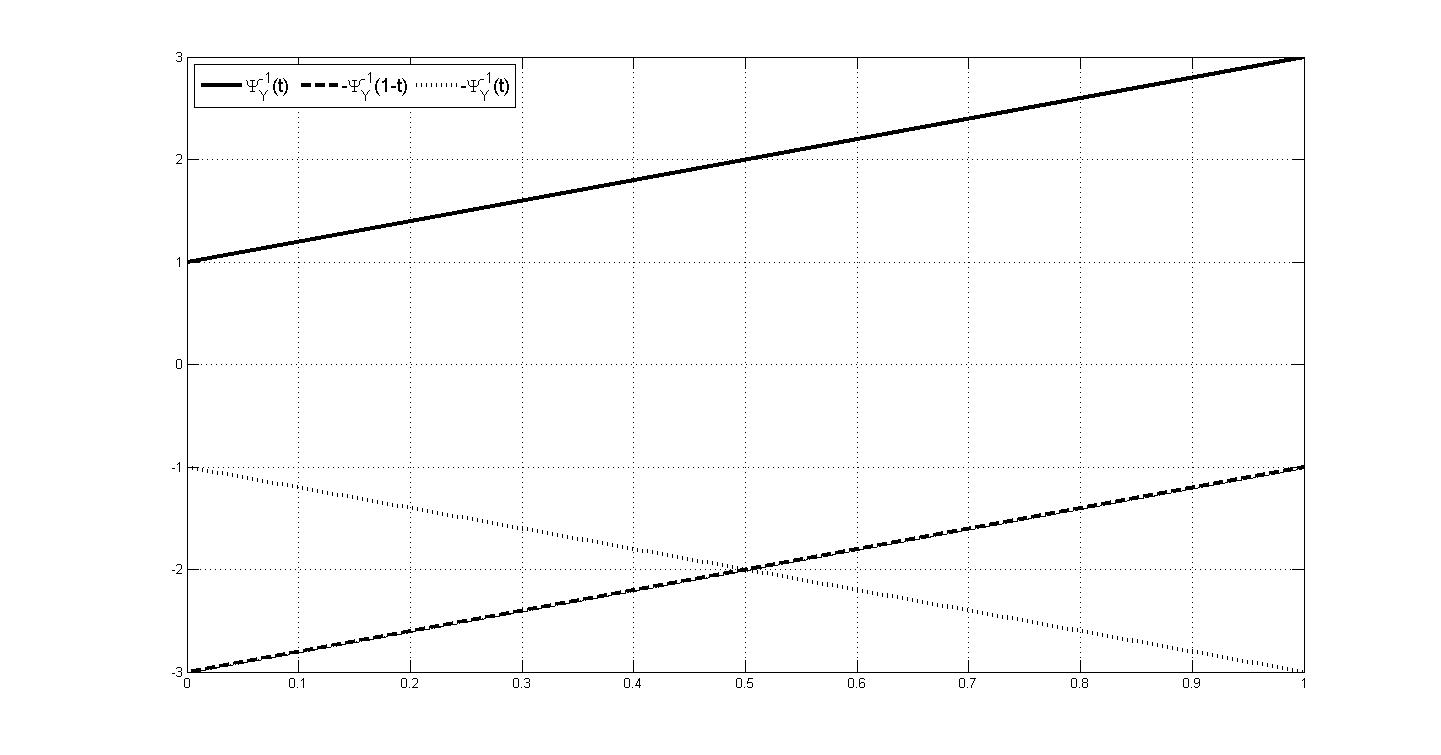}
      \caption{Representation of the functions $\Psi^{-1}(t);$ $-\Psi^{-1}(1-t)$ and $-\Psi^{-1}(t)$.}
\label{fig2}
\end{center}
\end{figure}

\begin{figure}[h!]
\begin{center}
    \includegraphics[width=0.75\textwidth]{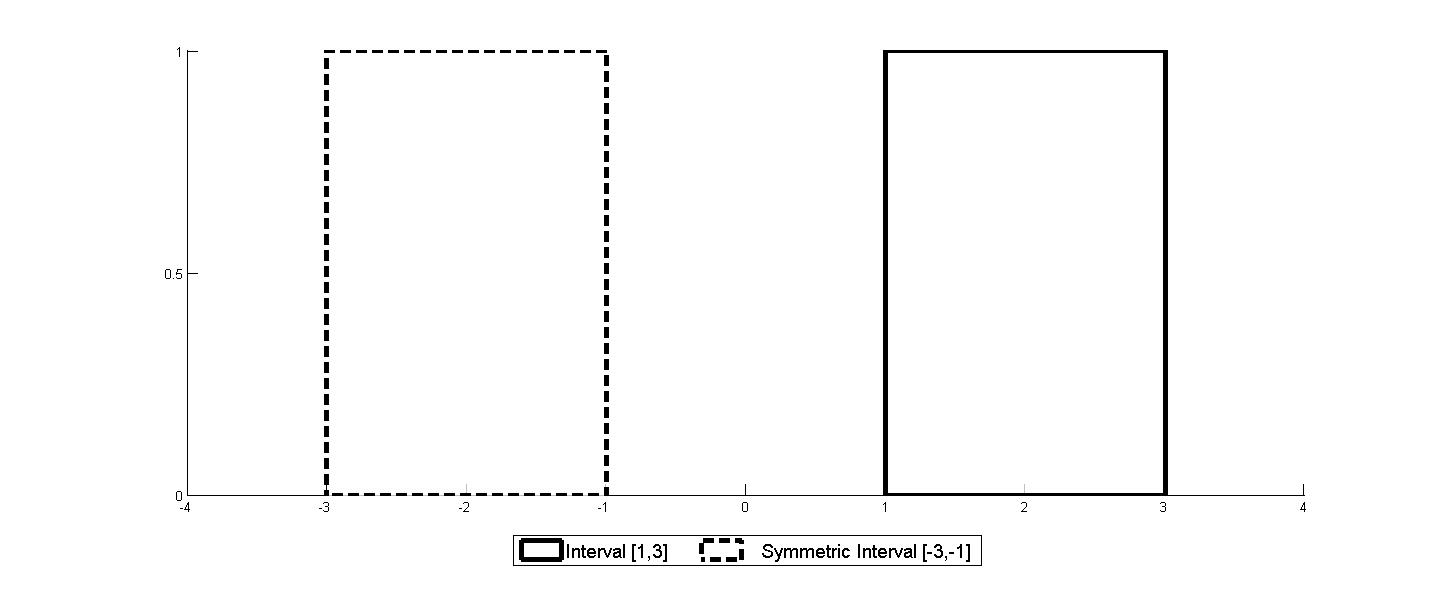}
      \caption{Representation of the intervals $I=[1,3]$ and $-I=[-3,-1]$.}
\label{fig3}
\end{center}
\end{figure}

In the \textit{DSD Model} proposed in this paper we will work with quantile functions.  As these functions are linear functions with domain $\left[0,1\right]$ we shall use the usual function arithmetic. However, when we use functions' arithmetic to operate with quantile functions, problems may arise. Quantile functions are non-decreasing functions, the addition of quantile functions is a non-decreasing function, but when we multiply a quantile function by a negative number, we obtain a function that is not non-decreasing \cite{dibr12} (See \textit{Figure \ref{fig2}}).
So, the problem arises of how to obtain the symmetric of the quantile function associated with a given interval. Consider the interval $I$ and let $-I$ the respective symmetric. If $\Psi^{-1}(t)$ with $t \in [0,1]$ is the quantile function that represents $I$,
the quantile function that represents  $-I$ is $-\Psi^{-1}(1-t)$ with $t \in [0,1].$
\textit{Figures \ref{fig2} and \ref{fig3}} illustrate this situation.

It is important to underline that some properties met by the usual symmetric elements are not met when these elements are ranges of values. The addition of  interval $I$ with $-I$ or the respective quantile functions, is not the null interval; because of this, the difference between ranges of values does not provide information on how dissimilar the intervals are. The difference between two equal intervals is an interval with symbolic mean zero \cite{bidi07}, that is, an interval with center zero and symmetric bounds.

The reasons given above show why the difference between two ranges of values is not a good solution to measure the dissimilarity between intervals as in classical statistics. In classical linear regression, to quantify the error between the observed values $y_{j}$ and the predicted values $\widehat{y}_{j},$ the difference between two real numbers, $e_{j}=y_{j}-\widehat{y}_{j},$ is used. In this case, the model that estimates the values $\widehat{y}_{j}$ minimizes $\displaystyle\sum\limits_{j=1}^m (y_{j}-\widehat{y}_{j})^2$. For intervals as well as for histograms, rather than using the difference between these ``symbolic values" we will evaluate the dissimilarity using a distance. As for the case of histogram-valued variables, the Mallows distance is used \cite{dibr12}.

\begin{definition}\label{def2.2}
Given two quantile functions $\Psi_{X(j)}^{-1}$ and $\Psi_{Y(j)}^{-1}$ that represent the range of values that the interval-valued variables $X$ and $Y$ take at observation $j$, the square of the Mallows distance is defined as follows \cite{mall72}:

\[\label{eq2.5A}
D^{2}_M(\Psi_{X(j)}^{-1},\Psi_{Y(j)}^{-1})=\int_{0}^{1}(\Psi_{X(j)}^{-1}(t)-\Psi_{Y(j)}^{-1}(t))^2dt \\
\]

\end{definition}

Considering a Uniform distribution across the intervals, Irpino and Verde \cite{irve06} rewrite the square of the Mallows distance as follows:

\begin{proposition}\label{p2.0}
Given two quantile functions $\Psi_{X(j)}^{-1}$ and $\Psi_{Y(j)}^{-1}$ that represent the interval-valued variables $X$ and $Y$ to each observation $j$, the square of the Mallows distance may be expressed by \cite{irve06}:

\[\label{eq2.5B}
D^{2}_M(\Psi_{X(j)}^{-1},\Psi_{Y(j)}^{-1})= (c_{X(j)}-c_{Y(j)})^2+\frac{1}{3}(r_{X(j)}-r_{Y(j)})^2
\]

\noindent where $c_{Y(j)},c_{X(j)}$ are the centers and $r_{Y(j)},r_{X(j)}$ are the half-ranges of the intervals $X(j)$ and $Y(j),$ respectively, with $j \in \left\{1,2,\ldots,m\right\}$ \medskip

\end{proposition}

Is important to underline that using this distance to measure the similarity between intervals is not new. It is a particular case of the Bertoluzza distance, used in literature to measure the distance between two intervals \cite{beall95}; in the linear regression models proposed for interval-valued random sets, a generalization of this distance is also used \cite{fernandez11, rodriguez07}.

\subsection{The \textit{DSD Model}}\label{s2.2}

The linear regression model proposed by Dias and Brito
\cite{dibr12} for histogram-valued variables uses quantile functions
to represent the distributions that the histogram-valued variables take.
For each unit it is possible to predict response  quantile functions from other quantile functions. However, the parameters of the model would have
to be non-negative to ensure that the predicted functions are non-decreasing functions,
in which case, the linear regression would always be direct. This does not happen
because the model not only includes the quantile functions that represent the distributions
that the explicative histogram-valued variables $X_{k}(j)$ take for each unit $j$, $\Psi^{-1}_{X_{k}(j)},$ but also the quantile functions
that represent its symmetric histogram. The presence of these two quantile functions associated to the same unit $j$ allows obtaining a direct or inverse linear relation between histogram/interval-valued variables even though the coefficients in the model are all positive.

The \textit{DSD linear regression model} for histogram-valued variables proposed by Dias and Brito \cite{dibr12}
may be particularized to interval-valued variables, as follows.

\begin{definition}\label{def2.3}

Consider the interval-valued variables $X_{1}; X_{2}; \ldots; X_{p}$. The quantile functions that represent the range of values that these variables take for each unit $j$ are denoted $\Psi_{X_1(j)}^{-1}(t),$ $\Psi_{X_2(j)}^{-1}(t),\ldots,$ $\Psi_{X_p(j)}^{-1}(t)$ and the quantile functions that represent the respective symmetric interval associated to each unit of the referred variables are denoted $-\Psi_{X_1(j)}^{-1}(1-t),-\Psi_{X_2(j)}^{-1}(1-t),\ldots,$ $-\Psi_{X_p(j)}^{-1}(1-t),$ with $t \in [0,1].$
Each quantile function $\Psi_{Y(j)}^{-1},$ may be expressed as follows:

\[
\Psi_{Y(j)}^{-1}(t)=\Psi_{\widehat{Y}(j)}^{-1}(t)+\varepsilon_{j}(t).
\]

\noindent where $\Psi_{\widehat{Y}(j)}^{-1}(t)$ is the predicted quantile function for the unit $j,$ obtained from

\[
\Psi_{\widehat{Y}(j)}^{-1}(t)=\gamma+\sum_{k=1}^{p}\alpha_{k}\Psi_{X_{k}(j)}^{-1}(t)-\sum_{k=1}^{p}\beta_{k}\Psi_{X_{k}(j)}^{-1}(1-t)
\]

\noindent with $t \in \left[0,1\right];$ $\alpha_{k},\beta_{k} \geq 0,$  $k \in \left\{1,2,\ldots,p \right\}$ and $\gamma \in \mathbb{R}.$ \medskip

This linear regression model is named \textbf{Distribution and Symmetric Distribution (DSD) Regression Model}.
\end{definition}

Particularizing \textit{Definition \ref{def2.3}} to the situation studied in this paper, where we assume uniformity within the intervals, the predicted quantile function $\Psi_{\widehat{Y}(j)}^{-1}$ is defined as follows:

\begin{equation} \label{eq2.7}
\Psi_{\widehat{Y}(j)}^{-1}(t)=\sum_{k=1}^{p}\left(\alpha_{k} -\beta_{k}\right)c_{X_{k}(j)}+
\gamma+\sum_{k=1}^{p}\left(\alpha_{k}+\beta_{k}\right) r_{X_{k}(j)} \left(2t-1\right)
\end{equation}

\noindent with $t \in \left[0,1\right];$ $\alpha_{k},\beta_{k} \geq 0,$  $k \in \left\{1,2,\ldots,p \right\}$ and $\gamma \in \mathbb{R}.$ \medskip

For each unit $j,$ the predicted interval $I_{\widehat{Y}(j)}$ may be obtained from

\begin{equation}\label{eq2.8}
I_{\widehat{Y}(j)}=\left[\displaystyle\sum_{k=1}^{p}\left(\alpha_{k} \underline{I}_{X_{k}(j)}-\beta_{k} \overline{I}_{X_{k}(j)}\right)+\gamma,\displaystyle\sum_{k=1}^{p}\left(\alpha_{k} \overline{I}_{X_{k}(j)}-\beta_{k} \underline{I}_{X_{k}(j)}\right)+\gamma \right]
\end{equation}

The error, for each unit $j$, is a function, but not necessarily a quantile function, given by $\varepsilon_{j}(t)=\Psi_{Y(j)}^{-1}(t)-\Psi_{\widehat{Y}(j)}^{-1}(t).$

By including in the model both the distribution of the explicative interval-valued variables and the respective symmetric, the linear relationship between the intervals is not necessarily direct, even though positivity restrictions are imposed on the parameters.
According to the \textit{DSD Model}, the center $c_{\widehat{Y}}(j)$ and half-range $r_{\widehat{Y}}(j)$  (or the bounds) of the predicted interval-valued variable may be described by a classical linear regression for the centers $c_{X_{k}}(j)$  and half-ranges $r_{X_{k}}(j)$ (or the bounds) of the explicative interval-valued variables. These linear regressions are the follows:

\begin{equation}\label{eq2.9}
c_{\widehat{Y}}(j)=\sum_{k=1}^{p}\left(\alpha_{k}-\beta_{k}\right)c_{X_{k}(j)}+\gamma
\end{equation}

\begin{equation}\label{eq2.10}
r_{\widehat{Y}}(j)=\sum_{k=1}^{p}\left(\alpha_{k}+\beta_{k}\right) r_{X_{k}(j)}.
\end{equation}

\noindent with  $\alpha_{k},\beta_{k} \geq 0,$  and $\gamma \in \mathbb{R}.$\medskip

From \textit{Equations} (\ref{eq2.9}) and (\ref{eq2.10}) we may observe that the parameters that define the linear regressions between the centres and half ranges of the intervals are not the same but are related. In spite of the fact that this model is defined between intervals and the relationship between the intervals may be direct or inverse, it always induces a direct linear relationship between the half ranges of the intervals. The direct or inverse relationship between the interval-valued variables is always in accordance with the linear relationship between the centers. The interval-valued variables $X_{k}$ are in direct linear relationship with $\widehat{Y}$ when $\alpha_{k}>\beta_{k}$ and the linear relation is inverse if $\alpha_{k}<\beta_{k}.$

The non-negative parameters of the \textit{DSD model}, in \textit{Definition \ref{def2.3}},
are determined solving a quadratic optimization problem, subject to non-negativity constraints on the unknowns.
The distance used to quantify the dissimilarity between
the predicted and the observed quantile function is the Mallows Distance \cite{mall72}.

Consider the centers $c_{Y_{j}}$ and half-ranges $r_{Y_{j}}$ of the observed intervals $I_{Y_{j}}$
and the predicted intervals $I_{\widehat{Y}_{j}}$ defined in Equation (\ref{eq2.8}).
The quadratic optimization problem that is necessary to solve to obtain the parameters of the model is then:

\begin{equation}
\begin{array}{lll}
\mathbf{\min} &  & \displaystyle \sum_{j=1}^{m} \left[\left( c_{Y(j)}-\sum_{k=1}^{p} \left(\alpha_{k}-\beta_{k}\right) c_{X_{k}(j)}-\gamma \right)^2+\frac{1}{3}\left( r_{Y(j)}-\sum_{k=1}^{p} \left(\alpha_{k}+\beta_{k}\right) r_{X_{k}(j)} \right)^2\right]\\
& & \\
 \mathbf{ s. t.} & & \alpha_{k},\beta_{k} \geq 0,   k\in\{1,2,\ldots,p\}\\
  & & \gamma \in \mathbb{R}
\end{array}
\label{eq2.11}
 \end{equation}\medskip

The optimization problem in (\ref{eq2.11}) may also rewritten in matricial form as a classical constraint quadratic optimization problem (see \cite{dibr12}). However this problem may also be defined as a constraint least square problem.

Consider the vectors of length $m,$ of the observed centers and half ranges of the response variable $Y:$
\[\mathbf{y^c}=\left(c_{Y(1)}, \ldots ,  c_{Y(m)} \right)^{T} \quad {\rm and} \quad \mathbf{y^r}=\left(r_{Y(1)} ,  \ldots ,  r_{Y(m)} \right)^{T};\]

\noindent the vector of length $2p+1,$ of the parameters of the model:

\[\mathbf{b}=\left(\alpha_{1} , \beta_{1} , \ldots , \alpha_{p} , \beta_{p} , \gamma\right)^{T}.\]

From vectors $\mathbf{x^{c}_{j}}$ and $\mathbf{x^{r}_{j}}$ defined by
\[\mathbf{x^{c}_{j}}=\left(c_{X_{1}(j)}, -c_{X_{1}(j)}, \ldots, c_{X_{p}(j)} , -c_{X_{p}(j)}, 1 \right)^{T}  \quad {\rm and} \quad
\mathbf{x^{r}_{j}}=\left(r_{X_{1}(j)}, r_{X_{1}(j)}, \ldots, r_{X_{p}(j)} , r_{X_{p}(j)}, 0 \right)^{T};\]
\noindent we can build the matrices of order $(2p+1) \times m:$

 \[\mathbf{X^{c}}=[\mathbf{x^{c}_{1}} \, \mathbf{x^{c}_{2}} \, \ldots,  \mathbf{x^{c}_{m}}]  \qquad \textrm{and} \qquad \mathbf{X^{r}}=[\mathbf{x^{r}_{1}} \, \mathbf{x^{r}_{2}} \, \ldots \, \mathbf{x^{r}_{m}}] \]

With these matrices, the minimization problem in (\ref{eq2.11}) may be rewritten in matricial form as follows:

\begin{equation}
\begin{array}{lll}
{\mathbf \min} &  & \|\mathbf{y^c}-(\mathbf{X^{c}})^{T}\mathbf{b}\|^2+ \frac{1}{3}\|\mathbf{y^r}-(\mathbf{X^{r}})^{T}\mathbf{b}\|^2\\
& & \\
  \mathbf {s. t.} & & \alpha_{k},\beta_{k} \geq 0,   k\in\{1,2,\ldots,p\}\\
  & & \gamma \in \mathbb{R}
\end{array}
 \label{eq2.11AA}
 \end{equation}\medskip

But, as the parameters for the centers and half-ranges may not be obtained independently, we may rewrite the optimization problem (\ref{eq2.11AA}) as a least square problem:

\begin{equation}
\begin{array}{lll}
\mathbf{\min} &  & \displaystyle \left \|\left[\begin{array}{c} \mathbf{y^c}\\ \frac{1}{\sqrt{3}}\mathbf{y^r}\end{array}\right]-\left[\begin{array}{c} (\mathbf{X^{c}})^{T}\\ \frac{1}{\sqrt{3}}(\mathbf{X^{r}})^{T}\end{array}\right] \mathbf{b}\right\|^2=\|\mathbf{Y}-\mathbf{X}\mathbf{b}\|^2\\
\\
& & \\
  \mathbf{ s. t.} & & \alpha_{k},\beta_{k} \geq 0,   k\in\{1,2,\ldots,p\}\\
  & & \gamma \in \mathbb{R}
\end{array}
\label{eq2.11AAA}
 \end{equation}\medskip

Several methods may be found in the literature to solve the constrained least squares problem (\ref{eq2.11AAA}) and, therefore the constrained quadratic optimization problem (\ref{eq2.11}).

As the quadratic function to optimize is convex and the feasible region too, it may be ensured that the vectors that verify the \textit{Kuhn Tucker conditions} are the vectors where the function reaches the global minimum, i.e. are the optimal solutions. In cases when the objective function is strictly convex we can ensure that the optimal solution is unique.

Let $\left( \alpha_{1}^{*}, \beta_{1}^{*},\cdots, \alpha_{n}^{*},\beta_{n}^{*} ,\gamma^{*} \right)$ be an optimal solution of the optimization problem in (\ref{eq2.11}). Dias and Brito \cite{dibr12} proved that the mean of the predicted histogram-valued variable $\overline{\widehat{Y}}$ is given by:

\begin{equation} \label{eq2.11B}
\overline{\widehat{Y}}=\displaystyle\sum\limits_{k=1}^p \left(\alpha_{k}^{*}-\beta_{k}^{*}\right) \overline{X_{k}} + \gamma^{*}.
\end{equation}

As the quantile function
$\Psi_{X_k(j)}^{-1}(t)$ and the respective symmetric $\Psi_{X_k(j)}^{-1}(1-t)$ with $t \in [0,1]$ are both in the \textit{DSD Model}, it is important to analyze the behavior of the model in the situations where these functions are collinear.
\textit{Proposition \ref{p2.1}} below allows deducing the collinearity conditions.

\begin{proposition}\label{p2.1}
The quantile functions $\Psi_{X_{k}(j)}^{-1}(t)=c_{X_{k}(j)}+r_{X_{k}(j)}(2t-1)$ and $-\Psi_{{X}(j)}^{-1}(1-t)=-c_{X_{k}(j)}+r_{X_{k}(j)}(2t-1)$
with $ 0 \leq t \leq 1$ that represent the intervals $I_{X_{k}(j)}$ and $-I_{X_{k}(j)},$
respectively, are collinear if the interval $I_{X_{k}(j)}$ has $c_{X_{k}(j)}=0$, which means that
the interval is symmetric, or $r_{X_{k}(j)}=0,$ which means that the interval is reduced to a real number (degenerate interval).
\end{proposition}

\textbf{Proof:}
The quantile functions $\Psi_{X_{k}(j)}^{-1}(t)$ and $-\Psi_{{X_{k}}(j)}^{-1}(1-t)$ with $ 0 \leq t \leq 1$
are collinear if there exists a real number $\lambda \neq 0$ such that
$-\Psi_{{X_{k}}(j)}^{-1}(1-t)=\lambda \Psi_{X_{k}(j)}^{-1}(t),$ with $t \in \left[0,1\right].$
\[\begin{array}{l}
-\Psi_{{X_{k}}(j)}^{-1}(1-t)=\lambda \Psi_{X_{k}(j)}^{-1}(t) \Longleftrightarrow -c_{X_{k}(j)}+r_{X_{k}(j)}(2t-1)=\lambda\left(c_{X_{k}(j)}+r_{X_{k}(j)}(2t-1)\right) \\ \Rightarrow  \left(c_{X_{k}(j)}=0 \,  \wedge
 \, \lambda=1 \,  \wedge \, r_{X_{k}(j)} \in \mathbb R \right) \, \vee \, \left( r_{X_{k}(j)}=0 \, \wedge \,  \lambda=-1 \, \wedge \, c_{X_{k}(j)} \in \mathbb R \right)
 \end{array}.\]

Therefore two quantile functions are collinear when the interval $I_{X_{k}(j)}$ is symmetric that is $I_{X_{k}(j)}=\left[-r_{X_{k}(j)};r_{X_{k}(j)}\right]$ or degenerate, i.e., $I_{X_{k}(j)}=c_{X_{k}(j)}$ $\qquad \Box$

The \textit{DSD Model} can nevertheless be applied when the quantile function $\Psi_{X_k(j)}^{-1}(t)$ and the respective symmetric are collinear. However, Equation (\ref{eq2.7}) is reduced to the classical linear regression model, between the centers, in Equation (\ref{eq2.9}), when the all intervals of the explicative interval-valued variables are degenerate and between the half ranges, in Equation (\ref{eq2.10}), when all intervals of the explicative interval-valued variables are symmetric.

When the collinearity between the interval-valued variable and respective symmetric is verified, the optimization problem has an optimal solution but it is not unique because, in this situation the quadratic function to optimize is not strictly convex (the columns of $\mathbf{X}$ in Equation (\ref{eq2.11AA}) are linearly dependent). However all values of the parameters when the global minimum is attained allow obtaining the same model, that in these cases isa classical model between the centers or the half ranges.

As the  \textit{DSD model} for interval-valued variables is a particular case of the model defined by Dias and Brito \cite{dibr12}
for histogram-valued variables, the optimal solution of the quadratic optimization problem
for interval-valued variables with non-negative constraints verifies the \textit{Kuhn Tucker conditions.}
It is therefore possible to prove the following decomposition \cite{dibr12}:

$$
\displaystyle\sum\limits_{j=1}^m D_{M}^{2} \left(\Psi_{Y(j)}^{-1}(t),\overline{Y}\right)=\displaystyle\sum\limits_{j=1}^m  D_{M}^{2} \left(\Psi_{Y(j)}^{-1}(t),\Psi_{\widehat{Y}(j)}^{-1}(t)\right)+\displaystyle\sum\limits_{j=1}^m  D_{M}^{2} \left(\Psi_{\widehat{Y}(j)}^{-1}(t),\overline{Y}\right).
$$

This decomposition allows defining the goodness-of-fit measure for the proposed model for interval-valued variables.

\begin{definition}\label{def2.5}
Consider the observed and predicted ranges of values of the interval-valued variable
$Y$ and $\widehat{Y}$ represented, respectively, by their quantile functions
$\Psi_{Y(j)}$ and $\Psi_{\widehat{Y}(j)}^{-1}.$ Consider also the symbolic mean of the interval-valued variable $Y,$ given by
$\overline{Y}=\frac{1}{m} \displaystyle \sum_{j=1}^{m} c_{Y(j)}$ \cite{bidi03}. The goodness-of-fit measure is given by
\[
\Omega=\frac{\displaystyle\sum\limits_{j=1}^m D_{M}^{2} \left(\Psi_{\widehat{Y}(j)}^{-1}(t),\overline{Y}\right)}{\displaystyle\sum\limits_{j=1}^m D_{M}^{2} \left(\Psi_{Y(j)}^{-1}(t),\overline{Y}\right)}=\frac{\displaystyle\sum\limits_{j=1}^m \left(c_{\widehat{Y}(j)}-\overline{Y}\right)^2+\frac{1}{3}r_{\widehat{Y}(j)}^2}{\displaystyle\sum\limits_{j=1}^m. \left(c_{Y(j)}-\overline{Y}\right)^2+\frac{1}{3}r_{Y(j)}^2}
\]
\end{definition}

As in classical linear regression, where the coefficient of determination $R^2$ ranges from 0 to 1,
the goodness-of-fit measure, $\Omega$, also ranges between 0 and 1.

\subsection{The \textit{DSD Model} is a generalization of the classical linear regression model}\label{s2.3}

Symbolic variables, introduced in Symbolic Data Analysis, are a generalization of classical variables,
and the statistical concepts and methods defined for these variables should also generalize the classical ones.
As we will see below, the \textit{DSD linear regression model} defined for histogram-valued variables \cite{dibr12}
and its present particularization for interval-valued variables, may be written for classical variables since their values are degenerate intervals (the upper and lower bounds are identical).

\begin{proposition}\label{p2.2}
The expression that allows predicting the values that the response variable takes in a classical linear regression model is a particular case
of the one obtained by the \textit{DSD linear regression model} for interval-valued variables given in (\ref{eq2.7}), if we consider intervals where the upper and lower bounds of the intervals are the same.
\end{proposition}

\textbf{Proof:}
Consider the observations of the explicative classical variables $X_{k},$
with $k \in \left\{1,2,\ldots,p\right\}$ and the observations of the response classical variables $Y.$
For each unit $j,$ the observed values of the variables $X_{k}$ are real numbers
 $b_{X_{k}(j)}$ that may be represented by the interval $[b_{X_{k}(j)},b_{X_{k}(j)}]$
 or by the quantile function $\Psi_{X_{k}}^{-1}(t)=b_{X_{k}(j)}$  (that in this case is a constant function).
 For each unit $j,$ the predicted value of the classical variable  $Y,$
 is the real number $\widehat{y}(j)$ that may similarly be represented by an interval or a quantile function.

Equation (\ref{eq2.7}) allows predicting the values of variable $Y,$ as follows:

\[ \widehat{y}(j) = \gamma+\sum_{k=1}^{p}\left(\alpha_{k}-\beta_{k}\right) b_{X_{k}(j)} \]

with $\alpha_{k},\beta_{k} \geq 0,$  $k \in \left\{1,2,\ldots,p \right\}$ and $\gamma \in \mathbb{R}.$

As $\alpha_{k},\beta_{k} \geq 0,$ $\alpha_{k}-\beta_{k}$ is a real number.
If we consider $\lambda_{k}=\alpha_{k}-\beta_{k}$ we have the classical linear regression model

\[\widehat{y}(j)=\gamma+\sum_{k=1}^{p} \lambda_{k} b_{X_{k}(j)}\]
with $\lambda_{k} \in \mathbb{R}$ and $k \in \left\{1,2,\ldots,p \right\}.$ $\qquad \Box$

As we have referred before, in a situation of degenerate intervals, the function to optimize is not strictly convex, and therefore  more than one optimal solution exists. However, for all parameters $\alpha_{k}$ and $\beta_{k}$ we obtain the same parameter $\lambda_{k}.$
Since no constraint is imposed on this parameter, we have in this case a classical linear regression model.

Also, the goodness-of-fit measure for interval-valued variables is a generalization
of the coefficient of determination $R^2$ of classical variables.
To obtain this result, it is first necessary to prove that follow proposition.

\begin{proposition}\label{p2.3}
The Mallows distance between intervals reduced to real numbers is the
Euclidean distance between two real numbers.
\end{proposition}

\textbf{Proof:}
Consider two intervals $I_{X}$ and $I_{Y}$ with equal bounds,
$I_{X}=[b_1,b_1]$ and $I_{Y}=[b_2,b_2]$ with $b_1,b_2 \in \mathbb{R}$;
those intervals may be represented by the quantile functions $\Psi_{X}^{-1}(t)=b_1$ and
$\Psi_{X}^{-1}(t)=b_2$ for $0 \leq t\leq 1.$

The Mallows distance in \textit{ Definition \ref{def2.2}} applied to these particular intervals $I_{X}$
and $I_{Y}$ whose centers are $b_1$ and $b_2$, respectively, and both have range $0$, is:

\[ D^{2}_M(\Psi_{X}^{-1},\Psi_{Y}^{-1})=(b_1-b_2)^2\]

So, we obtain the squared Euclidean distance between two unidimensional points.$\qquad \Box$

To conclude the previous result, we just need to state the following straightforward proposition:

\begin{proposition}\label{p2.4}
The goodness-of-fit measure in \textit{Definition \ref{def2.5}} particularized to degenerated intervals
is the coefficient of determination
$R^2$ of the classical linear regression model.
\end{proposition}

Therefore, it may be said that the
\textit{DSD Model} under uniformity is a theoretical generalization of the classical linear regression model.

\subsection{The single \textit{DSD Regression Model} for interval-valued variables} \label{s2.4}

Using \textit{Definition \ref{def2.3}} for the special case of only one explicative variable,
the predicted quantile function $\Psi_{\widehat{Y}(j)}^{-1}$
for each unit of the predicted interval-valued variable is given by
\begin{equation}\label{eq2.12}
\Psi_{\widehat{Y}(j)}^{-1}(t)=\gamma+\left(\alpha - \beta\right) c_{X(j)} +\left( \alpha+\beta \right) r_{X(j)} (2t-1) , \quad 0\leq t \leq 1.
\end{equation}
The corresponding predicted interval $I_{\widehat{Y}(j)}$ is the following:
\begin{equation}\label{eq2.13}
I_{\widehat{Y}(j)}=\left[\alpha\underline{I}_{X(j)}-\beta\overline{I}_{X(j)}+\gamma,
\alpha \overline{I}_{X(j)}-\beta \underline{I}_{X(j)}+\gamma \right]
\end{equation}

\begin{proposition}\label{p2.5}
Consider the interval-valued variable $\widehat{Y}$ predicted by the \textit{DSD Model} from the interval-valued variable $X.$ From this relationship we may conclude:
\begin{enumerate}
  \item The centers of the predicted intervals are in a classical linear relation with the centers of the observed intervals of the variable $X.$
  \item For each unit $j,$ the ratio between the half ranges of the intervals $\widehat{Y}(j)$ and $X(j)$ is constant.
\end{enumerate}
\end{proposition}

\textbf{Proof:}
 
From the \textit{DSD Model} we obtain the relationship between the centers and half ranges of the interval-valued variables in \textit{Equations (\ref{eq2.9})}  and \textit{(\ref{eq2.10})}. Particularizing these equations to one explicative variable,
we obtain for each unit $j,$  with $j=\left\{1,2,\ldots,m\right\}$ the following:

\[c_{\widehat{Y}(j)}=\left(\alpha-\beta\right) c_{X(j)}+\gamma\]
and
\[r_{\widehat{Y}(j)}=\left(\alpha+\beta\right) r_{X(j)}\Longleftrightarrow \frac{r_{\widehat{Y}(j)}}{r_{X(j)}}=\alpha+\beta\]

So, when two interval-valued variables are in perfect linear relationship,
the centers of the intervals are in a perfect classical linear relationship
and the ratio of the ranges of the intervals is constant and equal for all units.$\qquad \Box$

In this situation, when we predict one interval-valued variable from only one interval-valued variable,
it is straightforward to obtain the expressions of the parameters $\alpha,$ $\beta$ and $\gamma$ of the \textit{DSD Model}.
To find these expressions it is necessary to solve the quadratic optimization problem
with non-negative constrains for the parameters $\alpha$ and $\beta,$
as described in (\ref{eq2.11}), but now considering only one explicative variable.
The minimization problem is in this case as follows:

\begin{equation}
\begin{array}{lll}
\mathbf {\min} &  & f(\alpha, \beta, \gamma)  = \displaystyle \sum\limits_{j=1}^m\left[\left( c_{Y(j)}-\left(\alpha-\beta\right) c_{X(j)}-\gamma \right)^2+\frac{1}{3}\left( r_{Y(j)}-\left(\alpha+\beta\right) r_{X(j)} \right)^2\right]\\
& & \\
  \mathbf {s. t.} & & g_{1}(\alpha, \beta, \gamma)=-\alpha \leq 0\\
  & & g_{2}(\alpha, \beta, \gamma)=-\beta \leq 0 \\
  & & \gamma \in \mathbb{R}
\end{array}
\label{eq2.14}
 \end{equation}\medskip

\begin{proposition}\label{p2.6}
Consider the minimization problem in Equation (\ref{eq2.14}). When the function to minimize is strictly convex, the optimal solution of this problem, i.e., the values estimated for the parameters of the \textit{DSD Model} when the objective function reaches the minimum value, are given by:
 \begin{itemize}
   \item If {\scriptsize $\displaystyle\sum\limits_{j=1}^m\frac{1}{3}r_{X(j)}r_{Y(j)}\displaystyle \sum\limits_{j=1}^m\left(c_{X(j)}-\overline{X}\right)^2>\displaystyle \sum\limits_{j=1}^m\left(c_{Y(j)}-\overline{Y}\right)c_{X(j)}\displaystyle \sum\limits_{j=1}^m\frac{1}{3}r^2_{X(j)}$ then $\alpha^{*},\beta^{*} \neq 0.$}\medskip

\noindent In this case,
      {\scriptsize \[\alpha^{*}= \frac{ \displaystyle \sum\limits_{j=1}^m\left(c_{Y(j)}-\overline{Y}\right)\left(c_{X(j)}-\overline{X}\right) \displaystyle \sum\limits_{j=1}^m\frac{1}{3}r^2_{X(j)}+ \displaystyle \sum\limits_{j=1}^m\frac{1}{3}r_{X(j)}r_{Y(j)} \displaystyle \sum\limits_{j=1}^m\left(c_{X(j)}-\overline{X}\right)^2}{2 \displaystyle \sum\limits_{j=1}^m\left(c_{X(j)}-\overline{X}\right)^2 \displaystyle \sum\limits_{j=1}^m\frac{1}{3}r^2_{X(j)}}\]}

 {\scriptsize\[\beta^{*}= \frac{ -\displaystyle \sum\limits_{j=1}^m\left(c_{Y(j)}-\overline{Y}\right)\left(c_{X(j)}-\overline{X}\right) \displaystyle \sum\limits_{j=1}^m\frac{1}{3}r^2_{X(j)}+ \displaystyle \sum\limits_{j=1}^m\frac{1}{3}r_{X(j)}r_{Y(j)} \displaystyle \sum\limits_{j=1}^m\left(c_{X(j)}-\overline{X}\right)^2}{2 \displaystyle \sum\limits_{j=1}^m\left(c_{X(j)}-\overline{X}\right)^2 \displaystyle \sum\limits_{j=1}^m\frac{1}{3}r^2_{X(j)}}\]}

 {\scriptsize\[\gamma^{*}= \overline{Y}-\left(\alpha^{*}-\beta^{*}\right)\overline{X}.\]}

\vspace{1cm}

   \item If  {\scriptsize$\displaystyle\sum\limits_{j=1}^m\frac{1}{3}r_{X(j)}r_{Y(j)}\displaystyle \sum\limits_{j=1}^m\left(c_{X(j)}-\overline{X}\right)^2<\displaystyle \sum\limits_{j=1}^m\left(c_{Y(j)}-\overline{Y}\right)c_{X(j)}\displaystyle \sum\limits_{j=1}^m\frac{1}{3}r^2_{X(j)},$}  then  {\scriptsize$\alpha^{*}=0$ $\vee$ $\beta^{*}=0.$} \medskip
  \vspace{0.5cm}

\noindent In this case,
       \begin{itemize}
         \item  If {\scriptsize$\displaystyle\sum\limits_{j=1}^m\frac{1}{3}r_{X(j)}r_{Y(j)}> \displaystyle \sum\limits_{j=1}^m\left(c_{Y(j)}-\overline{Y}\right)c_{X(j)}$} then

            {\scriptsize $\alpha^{*}=0; \quad
              \beta^{*}=\frac{\displaystyle\sum\limits_{j=1}^m\frac{1}{3}r_{X(j)}r_{Y(j)}-\displaystyle \sum\limits_{j=1}^m\left(c_{Y(j)}-\overline{Y}\right)\left(c_{X(j)}-\overline{X}\right)}{\displaystyle \sum\limits_{j=1}^m\left(c_{X(j)}-\overline{X}\right)^2+\displaystyle \sum\limits_{j=1}^m\frac{1}{3}r^2_{X(j)}};
               \gamma^{*}= \overline{Y}+\beta^{*}\overline{X}.$}
  \vspace{0.5cm}
         \item  If {\scriptsize $\displaystyle\sum\limits_{j=1}^m\frac{1}{3}r_{X(j)}r_{Y(j)}< \displaystyle \sum\limits_{j=1}^m\left(c_{Y(j)}-\overline{Y}\right)c_{X(j)}$} then

            {\scriptsize $\alpha^{*}=\frac{\displaystyle\sum\limits_{j=1}^m\frac{1}{3}r_{X(j)}r_{Y(j)}+\displaystyle \sum\limits_{j=1}^m\left(c_{Y(j)}-\overline{Y}\right)\left(c_{X(j)}-\overline{X}\right)}{\displaystyle \sum\limits_{j=1}^m\left(c_{X(j)}-\overline{X}\right)^2+\displaystyle \sum\limits_{j=1}^m\frac{1}{3}r^2_{X(j)}};\quad
             \beta^{*}=0; \quad \gamma^{*}= \overline{Y}-\alpha^{*}\overline{X}.$}
       \end{itemize}

       \vspace{1cm}

   \item  If \scriptsize{ $\displaystyle\sum\limits_{j=1}^m\frac{1}{3}r_{X(j)}r_{Y(j)}=\displaystyle \sum\limits_{j=1}^m\left(c_{Y(j)}-\overline{Y}\right)c_{X(j)}$ then $\alpha^{*}=0;$ $\beta^{*}=0$ and $\gamma^{*}=\overline{Y}.$}
 \end{itemize}
\end{proposition}
     \vspace{1cm}
     
\textbf{Proof:}
The proof is given in \textit{Appendix A}. $\qquad \Box$

\section{Simulation studies} \label{s3}

The simulation studies that we will now present have two main goals. The first study aims at identifying the error function characteristics that are needed to disturb the linear regression in different given ways. In the second study, we want to evaluate empirically the behavior of the parameter estimation of the \textit{DSD Model} applied to interval-valued variables, when the  explicative and response variables present different levels of linearity.

\subsection{Building symbolic simulated data tables} \label{s3.1}

To build the symbolic simulated data tables it is necessary to generate the observations of the interval-valued variables $X_{k},\, k=\left\{1,\ldots,p\right\}$ and $Y,$ where $Y$ is the variable to be modelized from $X_{k}$ by the \textit{DSD Model}. The process to obtain these data tables is similar to the one used in the simulation study for histogram-valued variables in Dias and Brito \cite{dibr12}. To obtain the $m$ observations associated to a interval-valued variable $X_{k}$, we start by uniformly simulating  5000 real values corresponding to each unit. For each observation, we select the minimum and maximum of these values and build an interval associated to each unit. For the explicative variables $X_{k},$ we consider three levels of variability:
\begin{itemize}
  \item Low variability - when the intervals associated to the variable $X_{k}$ have similar small half ranges;
  \item High variability - when the intervals associated to the variable $X_{k}$ have similar large half ranges;
  \item Mixed variability - when we have a mixture of intervals associated to the variable $X_{k}$ with variable half ranges.
\end{itemize}

Afterwards, the intervals that are the observations of the interval-valued variable $Y$ are obtained considering the \textit{DSD Model} for particular values of the parameters and the error function $\varepsilon_{j}(t).$
So, the values of the interval-valued variable  $Y,$ for each unit $j$ are obtained by
       \[\Psi_{Y^{*}(j)}^{-1}(t)=\gamma+\sum_{k=1}^{p}\alpha_{k}\Psi_{X_{k}(j)}^{-1}(t)-\sum_{k=1}^{p}\beta_{k}\Psi_{X_{k}(j)}^{-1}(1-t)+\varepsilon_{j}(t)\]
with
\[\varepsilon_{j}(t)=a_{(j)}+\left(2t-1\right) b_{(j)} \quad t \in [0,1]\]

Each quantile function $\Psi_{Y^{*}(j)}^{-1}(t)$ is randomly disturbed by an error function $\varepsilon_{j}(t)$ for different values of $a_{(j)}$ and  $b_{(j)}.$  The values of $b_{(j)}$ cannot be larger than the respective value of the half range $r_{Y^{*}(j)},$ else for this unit $j$ the half range $r_{Y(j)}$ would be negative.

To perform the simulation study, symbolic data tables that illustrate different situations were created according to a selected factorial design. For each situation considered, 1000 data tables were generated.The values of in the tables of the \textit{Appendixes B} and \textit{C}, are the mean of 1000 values together with the respective standard deviation values $s.$

\subsection{Simulation study I} \label{s3.2}
In the first simulation study, the goal is to analyze the behavior of the error function and see if is it possible to establish a relationship between the error function and the goodness-of-fit measures. To analyze the behavior of the error function, we consider intervals (the observations associated to the explicative and response variables) that have low variability, high variability or a mixture of intervals with variable half ranges.
The following goodness-of-fit measures are considered in this study:

\begin{itemize}
	\item $\overline{\Omega},$ where $\Omega$ is the measure deduced from the \textit{DSD Model} (see \textit{Subsection \ref{s2.2}});

	\item Root-mean-square error $(RMSE_M)$, a measure defined using the Mallows distance (also used in the \textit{DSD Model}), proposed by Irpino and Verde \cite{irve12}; it is defined by
 \[RMSE_{M}=\sqrt{ \frac{\displaystyle \sum_{j=1}^{m}\displaystyle \int_{0}^{1}\left(\Psi_{\widehat{Y}(j)}^{-1}(t)-\Psi_{Y(j)}^{-1}(t)\right)^2dt}{m}} \]
\end{itemize} \medskip

\textit{Factorial design}

In this study a full factorial design was employed, with the following factors:

\begin{itemize}
    \item Sample size: m=10;100.
    \item Number of explicative interval-valued variables $p=1.$
    \item Levels of variability in the explicative variable $X.$ (The distribution of the values in microdata is Uniform).
           \begin{description}
            \item[i)] Low variability - $X(j)\sim \mathcal{U}(\delta_1(j),\delta_2(j))$ are randomly generated considering for each $j \in \left\{1,\ldots,m\right\},$ $\delta_1(j)\sim \mathcal{U}(-2,0)$ and $\delta_2(j)\sim \mathcal{U}(4,6)$;
            \item[ii)] High variability - $X(j)\sim \mathcal{U}(\delta_3(j),\delta_4(j))$ are randomly generated considering for each $j \in \left\{1,\ldots,m\right\},$ $\delta_3(j)\sim \mathcal{U}(-14,-12)$ and $\delta_4(j)\sim \mathcal{U}(16,18)$;
           \item[iii)] Mixture with variable half ranges - $X(j)\sim \mathcal{U}(\delta_5(j),\delta_6(j))$ are randomly generated considering for each $j \in \left\{1,\ldots,m\right\},$ several options:
               \begin{itemize}
                 \item $\delta_5(j)\sim \mathcal{U}(-2,0)$ and $\delta_6(j)\sim \mathcal{U}(0,2);$
                 \item $\delta_5(j)\sim \mathcal{U}(-1,1)$ and $\delta_6(j)\sim \mathcal{U}(2,4);$
                 \item $\delta_5(j)\sim \mathcal{U}(-3,-1)$ and $\delta_6(j)\sim \mathcal{U}(9,11);$
                 \item $\delta_5(j)\sim \mathcal{U}(-11,-9)$ and $\delta_6(j)\sim \mathcal{U}(29,31);$
                 \item $\delta_5(j)\sim \mathcal{U}(-1,1)$ and $\delta_6(j)\sim \mathcal{U}(19,21).$
               \end{itemize}
            \end{description}
    \item Parameters of the \textit{DSD Model}. The selection of the parameters influences the levels of variability in the response variable $Y.$
     \begin{description}
        \item[i)] $\alpha=2;$ $\beta=1;$ $\gamma=-1$ (generate intervals with low (high) variability when the intervals of the explicative variable have low (high) variability);
        \item[ii)] $\alpha=6;$ $\beta=0;$ $\gamma=2$ (generate intervals with moderate/high variability when the intervals of the explicative variable have low or high variability);
        \item[iii)]  $\alpha=2;$ $\beta=8;$ $\gamma=3$ (generate intervals with high variability when the intervals of the explicative variable have low or high variability).
         \end{description}
\item The error function $\varepsilon_{j}(t)=a_{(j)}+(2t-1)b_{(j)},$ with $t \in [0,1]$ is defined considering:
       \begin{description}
        \item[i)] Different levels of variability for the values $a_{(j)}.$
        The values of $a_{(j)}$ are randomly (uniformly) generated in $\mathcal{U}(-s_{a},s_{a})$ with $s_{a}=\left\{0,2,5,10,20,40,80,120,180\right\}.$
        \item[ii)] Different levels of variability for the values $b_{(j)}.$
        The values of $b_{(j)}$ are randomly (uniformly) generated in $\mathcal{U}_{s_{b}}=\mathcal{U}(-s_{b},s_{b})$ with $s_{b}=\left\{0,1,2,3,4,5,6, \right.$ $ \left. 10,20,40,80,120\right\}.$ As the value of $b_{(j)}$ cannot be larger than the respective minimum value of the half range $r_{Y^{*}(j)},$ in all situations when $mr=\displaystyle\min_{j \in \{1,\ldots,m\}}\left\{r_{Y^{*}(j)}\right\}$ is lower than $s_{b}$ we consider $\mathcal{U}_{s_{b}}=\mathcal{U}\left(-mr,mr\right).$
       \end{description}
       The selection of the values $s_{a}$ and $s_{b}$ is done according
       to the size of the values in the intervals associated to the response variable $Y.$
       For this simulation study, to choose the highest value of $s_{a},$
       we consider that $a_{(j)}$ must be outside the interval $\left[c_{Y^{*}(j)}-r_{Y^{*}(j)},c_{Y^{*}(j)}+r_{Y^{*}(j)}\right].$
       For the value $s_{b},$ the last chosen value is close to $mr,$ since for higher values results are similar.
\end{itemize}\medskip

\textit{Results and conclusions}

\begin{figure}
\begin{center}
\includegraphics[width=1\textwidth]{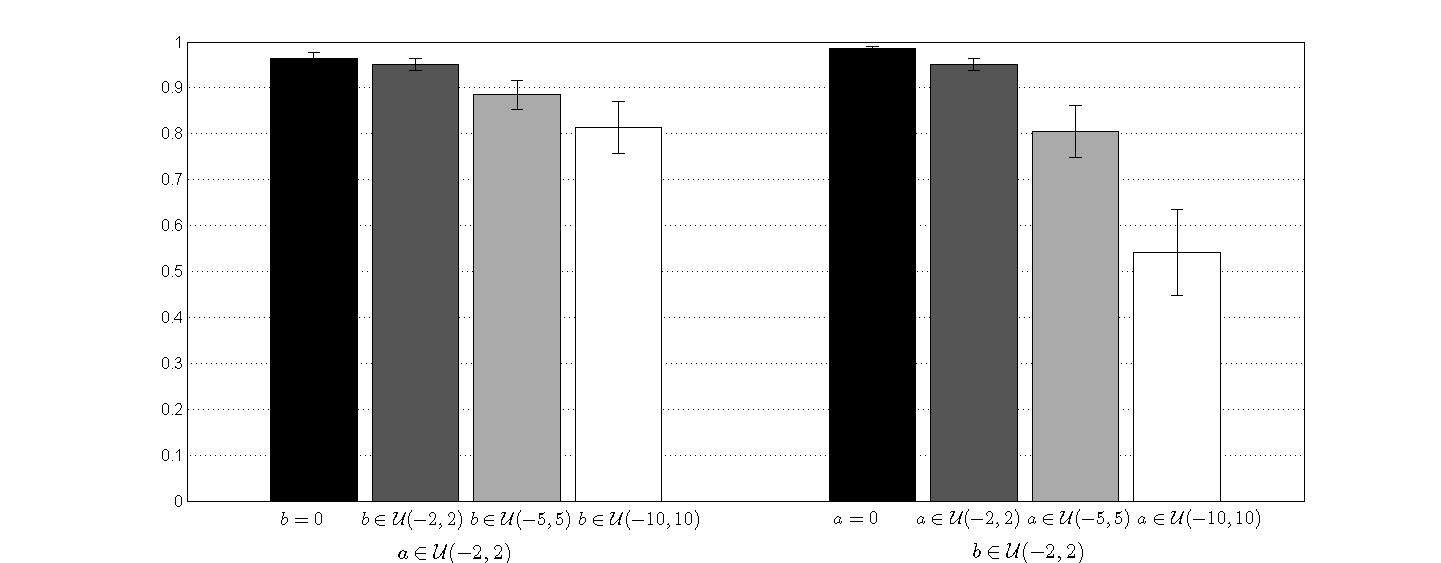}
\caption{Mean values of $\Omega$ and the respective standard deviation for different error functions.}
\label{fig1S}
\end{center}
\end{figure}

\begin{figure}
\begin{center}
\includegraphics[width=1\textwidth]{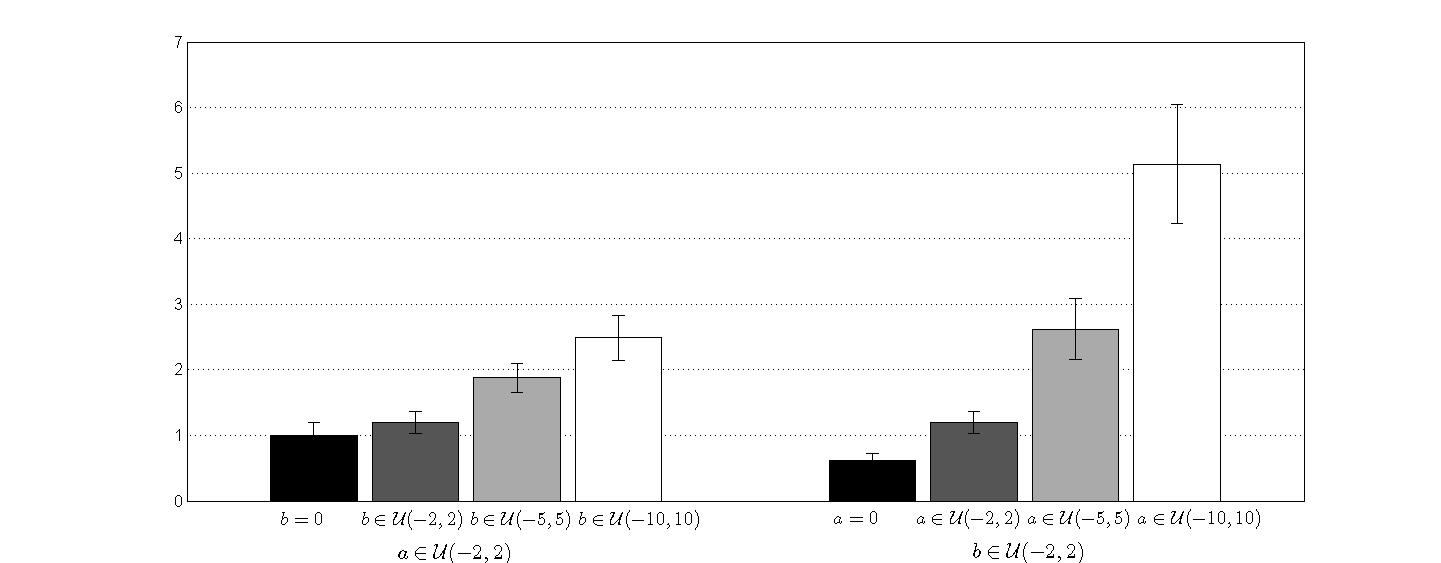}
\caption{Mean values of $RMSE_M$ and the respective standard deviation for different error functions.}
\label{fig2S}
\end{center}
\end{figure}

The tables with the results of the study may be found in \textit{Appendix B}. From \textit{Tables \ref{table1SA2}} to \textit{\ref{table3SA2}} we present the means of the goodness-of-fit measures $\Omega$ and the means of the $RMSE_M$ when the interval-valued variable $X$ presents low variability and the interval-valued variable $Y$ was generated by the \textit{DSD Model} considering the three selections for the parameters. The variability in $Y$ is lower when the intervals of $Y$ are generated by the model $\alpha=2;$ $\beta=1;$ $\gamma=-1$ (\textit{Table \ref{table1SA2}}); moderate when the intervals of $Y$ are generated by the model $\alpha=6;$ $\beta=0;$ $\gamma=2$ (\textit{Table \ref{table2SA2}}) and higher when the intervals of $Y$ are generated by the model $\alpha=2;$ $\beta=8;$ $\gamma=3$ (\textit{Table \ref{table3SA2}}). To analyze the behavior of the error function and the impact of the values $a$ and  $b$ in the disturbance of the linear relation between interval-valued variables we considered several possible options for $b$; several values of $a$ were associated with each $b$. From \textit{Table \ref{table4SA2}} to \textit{\ref{table6SA2}} and from \textit{Table \ref{table7SA2} to \ref{table9SA2}}, we present the results of the similar studies applied to a interval-valued variable $X$ that presents high variability or different variabilities.
As concerns  the influence of the values $a$ and $b$ that compose the error function that disturbs the function, we may observe that when the variability in all intervals of the interval-valued variable $X$ is low or high (\textit{Tables \ref{table1SA2}} to \textit{\ref{table6SA2}}), the linearity between the data is more affected by the values of $a$ then the values of $b.$ In all situations, when we consider the same disturbance for the values of $a,$ the increase in disturbance of the values of $b$ affects less the linear relation between the variables than when we considered the same disturbance for the values of $b$ and increase the disturbance of the values of $a.$ \textit{Figures \ref{fig1S}} and \textit{\ref{fig2S}} illustrate the situation in \textit{Table \ref{table1SA2}}, where  $X$ has low variability and the parameters of the \textit{DSD Model} are $\alpha=2; \beta=1;$ and  $\gamma=-1.$

It is important to underline that the simulation study imposes a higher limit for the selected values of $b_{j}$ in the error function. This limitation prevents the analysis of the behavior of the models when the values of $b_{j}$ are high, according to the size of the values of the half ranges  $r_{Y^{*}(j)}$ or when we have a mixture of very different half-ranges (\textit{Tables \ref{table7SA2}} to \textit{\ref{table9SA2}}). When these situations occur, the disturbance of the perfect linear regression, as previously described, and that we use to generate the symbolic data tables, is affected almost only by the values of $a.$

Analyzing in detail the obtained results enables concluding that to obtain a model with low linearity, the value of $a_{j}$ should not be in the interval $\left[\underline{I}_{Y^{*}(j)},\overline{I}_{Y^{*}(j)}\right].$  Considering this information, it is possible to suitably select the values of $a_{j}$ to disturb the linear relationship between interval-valued variables when the intervals in all observations have similar half ranges. Consequently, higher values of $a_{j}$ are necessary in the error function $\varepsilon_{j}$ to disturb the linear relationships when the half range of the intervals of the response variable is large. However, when we have a mixture of half ranges in the explicative variables, this choice is more difficult.

In this study we considered two measures to assess the goodness-of-fit: the coefficient of determination $\Omega$ and the root-mean-square error $RMSE_M.$ According to the obtained results we may conclude that the $RMSE_M$ is not a relative measure. This measure takes into account the size of the values in intervals and therefore the magnitude of the values that compose the error function must take into account the size of the values that compose the intervals, when the goal is to disturb the perfect linear regression in a similar way. For interval-valued variables, even when the values of the intervals have very different sizes, we can have similar results of the measure $RMSE_M$ when we disturb the perfect linear relationship with similar error functions (similar values are selected for the values $a_{j}$ and $b_{j}$ to compose the error function $\varepsilon_{j}$). However, the respective values of $\Omega$ may be very different. For example, from \textit{\textit{Tables \ref{table1SA2}} to \textit{\ref{table4SA2}}}, when the error function considers $a \in \mathcal{U}(-20,20)$ and $b \in \mathcal{U}(-10,10),$ the mean value of $RMSE_M$ is always around $11$ whereas the respective mean value of $\Omega$ are very different for the different situations. The measure $\Omega$ evaluates the quality of the linear relation independently of the magnitude of the values whereas to interpret the values of the measure $RMSE_M$ we have to take into consideration the size of the values in the intervals.

\subsection{Simulation study II} \label{s3.1.1}

In the second simulation study, the goal is to analyze the behavior of the parameters' estimation and the performance of the \textit{DSD Model} considering two levels of linearity between the interval-valued variables. For all situations, the observations of the interval-valued variables are generated from micro data with Uniform distribution. In addition to the goodness-of-fit measures $\Omega$ and $RMSE_M$ considered in \textit{Simulation Study I}, we compute the lower and the upper bound root-mean-square ($RMSE_{L}$ and $RMSE_{U},$ respectively) that Lima Neto and De Carvalho \cite{netocar10, netocar08} use to study the performance of their linear regression models. These measures are defined as follows:
 \[RMSE_{L}=\sqrt{ \frac{1}{m}\displaystyle \sum_{j=1}^{m}(\underline{I}(j)-\widehat{\underline{I}}(j))^2}\]
 \[ RMSE_{U}=\sqrt{ \frac{1}{m} \displaystyle \sum_{j=1}^{m}(\overline{I}(j)-\widehat{\overline{I}}(j))^2}\]

with $\left[\underline{I}(j),\overline{I}(j)\right[$ and $\left[\widehat{\underline{I}}(j),\widehat{\overline{I}}(j)\right[$ the observed and predicted intervals, for each unit $j.$ \medskip

\textit{Factorial design}

In this study a full factorial design was employed, with the following factors:

\begin{itemize}
    \item Number of explicative interval-valued variables: $p=1$ and $p=3.$
     \item Parameters of the \textit{DSD Model}.
     \begin{description}
       \item[$\circ$] For $p=1:$
     \begin{description}
        \item[i)] $\alpha=2;$ $\beta=1;$ $\gamma=-1;$ ($\alpha$ and $\beta$ are close)
        \item[ii)]  $\alpha=6;$ $\beta=0;$ $\gamma=2;$ ($\alpha$ is higher than $\beta$)
         \item[iii)]  $\alpha=2;$ $\beta=8;$ $\gamma=3;$ ($\alpha$ is lower than $\beta$)
         \end{description}
        \item[$\circ$] For $p=3:$
        \begin{description}
        \item[i)]  $\alpha_1=2;$ $\beta_1=1;$ $\alpha_2=0.5;$ $\beta_2=3;$ $\alpha_3=1.5;$ $\beta_3=1;$ $\gamma=-1;$ (the values of $\alpha$ and $\beta$ are close)
        \item[ii)]  $\alpha_1=6;$ $\beta_1=0;$ $\alpha_2=2;$ $\beta_2=8;$ $\alpha_3=10;$ $\beta_3=5;$ $\gamma=3;$ (the values of $\alpha$ and $\beta$ are apart)
         \end{description}
       \end{description}
    \item Levels of variability in explicative variables $X_{k}.$
           \begin{description}
            \item[i)] Low variability - $X_{k}(j)\sim \mathcal{U}(\delta_1(j),\delta_2(j))$ are randomly generated considering for each $j \in \left\{1,\ldots,m\right\}$ and $k \in \left\{1,2,3\right\}:$
                \begin{itemize}
                  \item $k=1$ - $\delta_1(j)\sim \mathcal{U}(-2,0)$ and $\delta_2(j)\sim \mathcal{U}(4,6)$;
                  \item $k=2$ - $\delta_1(j)\sim \mathcal{U}(1,3)$ and $\delta_2(j)\sim \mathcal{U}(3,5)$;
                  \item $k=3$ - $\delta_1(j)\sim \mathcal{U}(4,6)$ and $\delta_2(j)\sim \mathcal{U}(9,11)$;
                \end{itemize}
            \item[ii)] High variability - $X_{k}(j)\sim \mathcal{U}(\delta_3(j),\delta_4(j))$ are randomly generated considering for each $j \in \left\{1,\ldots,m\right\}$ and $k \in \left\{1,2,3\right\}:$
                 \begin{itemize}
                  \item $k=1$ - $\delta_3(j)\sim \mathcal{U}(-14,-12)$ and $\delta_4(j)\sim \mathcal{U}(16,18)$;
                  \item $k=2$ - $\delta_3(j)\sim \mathcal{U}(1,3)$ and $\delta_4(j)\sim \mathcal{U}(25,27)$;
                  \item $k=3$ - $\delta_3(j)\sim \mathcal{U}(-16,-14)$ and $\delta_4(j)\sim \mathcal{U}(-1,1)$;
                \end{itemize}
                \item[iii)] Variable half ranges - $X_{k}(j)\sim \mathcal{U}(\delta_5(j),\delta_6(j))$ are randomly generated considering for each $j \in \left\{1,\ldots,m\right\}$  and $k \in \left\{1,2,3\right\},$ the several options:
                    \begin{itemize}
                 \item $\delta_5(j)\sim \mathcal{U}(-2,0)$ and $\delta_6(j)\sim \mathcal{U}(0,2);$
                 \item $\delta_5(j)\sim \mathcal{U}(-1,1)$ and $\delta_6(j)\sim \mathcal{U}(2,4);$
                 \item $\delta_5(j)\sim \mathcal{U}(-3,-1)$ and $\delta_6(j)\sim \mathcal{U}(9,11);$
                 \item $\delta_5(j)\sim \mathcal{U}(-11,-9)$ and $\delta_6(j)\sim \mathcal{U}(29,31);$
                 \item $\delta_5(j)\sim \mathcal{U}(-1,1)$ and $\delta_6(j)\sim \mathcal{U}(19,21).$
               \end{itemize}
            \end{description}
            \item Two levels of linearity are considered. For each $j \in \left\{1,\ldots,m\right\},$ the values of $a(j)$ and $b(j)$ are randomly generated as follows:
       \begin{description}
        \item[i)] Low linearity - $a(j)\sim \mathcal{U}(-\frac{ml+mu}{2},\frac{ml+mu}{2})$ and $b(j)\sim \mathcal{U}(-mr,mr).$
        \item[ii)]  High linearity - $a(j)\sim \mathcal{U}(-\frac{1}{8}\frac{ml+mu}{2},\frac{1}{8}\frac{ml+mu}{2})$ and $b(j)\sim \mathcal{U}(-\frac{1}{8}mr,\frac{1}{8}mr).$
        \end{description}
          where $ml=\left|\displaystyle\min_{j \in \{1,\ldots,m\}}\left\{\underline{I}_{Y^{*}(j)}\right\}\right|;$ $mu=\left|\displaystyle\max_{j \in \{1,\ldots,m\}}\left\{\overline{I}_{Y^{*}(j)}\right\}\right|$ and $mr=\displaystyle\min_{j \in \{1,\ldots,m\}}\left\{r_{Y^{*}(j)}\right\}$
        \item Sample size: m=10; 30; 100; 250.
\end{itemize}\medskip

\textit{Results and conclusions}

The tables with the results of the study can be found in \textit{Appendix C}. From \textit{Tables \ref{table1SA3}} to \textit{\ref{table3SA3}} the results obtained for the parameters estimated and the goodness-of-fit measures, with $p=1$ for the three selected values of $\alpha,$ $\beta$ and $\gamma.$ In \textit{Tables \ref{table4SA3}} to \textit{\ref{table7SA3}} similar results are presented for the considered cases where $p=3.$  Based on the obtained results, presented in \textit{Appendix C}, we can see that the behavior of  the parameters' estimation is independent of the number of explicative variables in the model and the parameters selected for the model. In each of these situations, three levels of variability in the explicative variables $X_{k}$ were considered each of them with two levels of linearity, and two types of behavior were observed.
\begin{description}
  \item[$\circ$] When the linearity between the variables is high and the diversity of the half ranges of the intervals of  $X_{k}$ is low or we have variable half ranges, the estimated parameters are close to the initial parameter values. However, for high levels of linearity, when the  variability of the half ranges of the intervals is high and mainly when the sample size is small, the estimated parameters are more distant from the initial parameters. This difference is larger in the independent parameter.
  \item[$\circ$] When the level of linearity between the variables is low, many of the estimated parameters were distant from the original ones. These cases, observed mainly when the number of observations is low, are not surprising because other models may exist that adjust better the interval data.
\end{description}

\begin{figure}[h!]
\begin{center}
\includegraphics[width=0.75\textwidth]{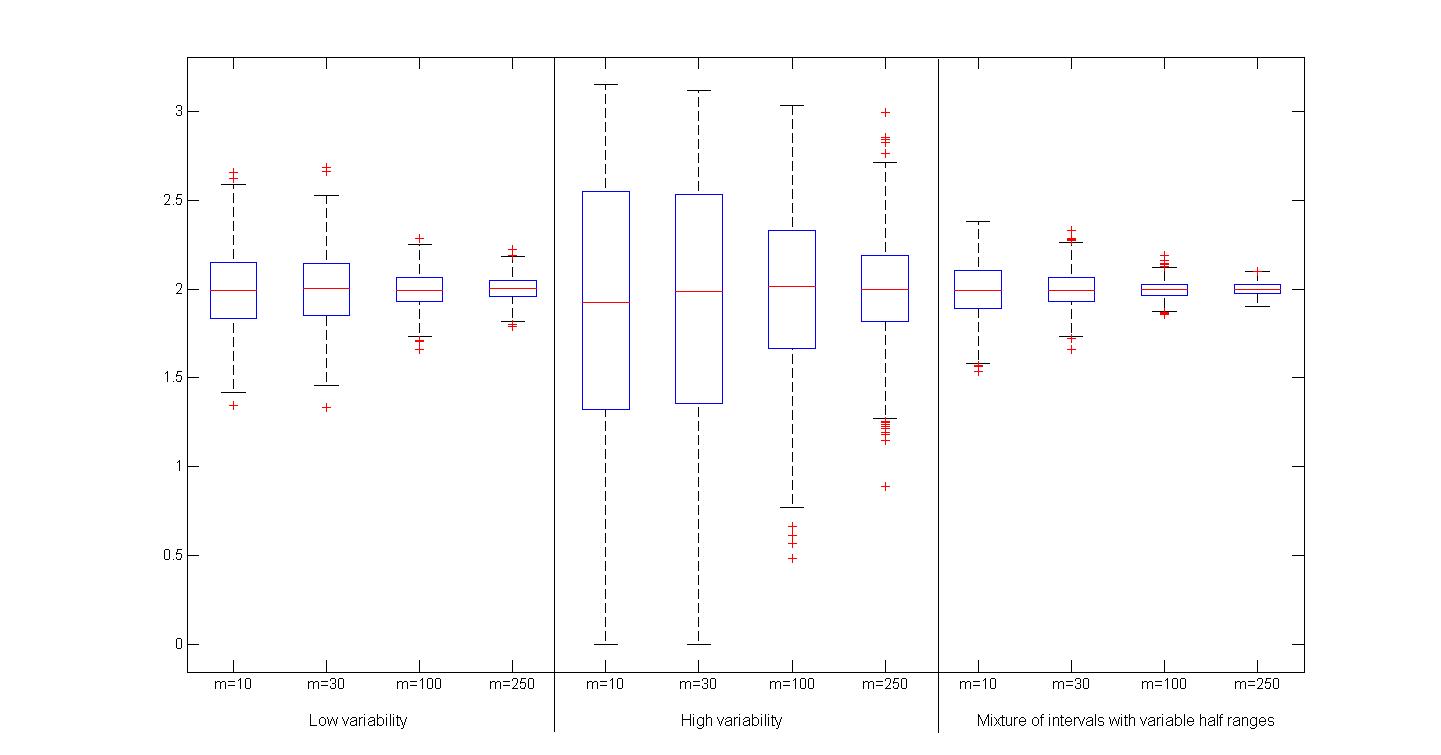}
\caption{Boxplot for the estimated parameter $\widehat{\alpha}$ for a high level of linearity and when the original parameters are $\alpha=2; \beta=1; \gamma=-1$.}
\label{fig3S}
\end{center}
\end{figure}

According to this, the analysis of the behavior of the $MSE$ and the mean of the estimated parameters is essentially applicable in situations where the level of linearity is high. For almost all these cases, we observe that the values of the $MSE$ decrease and tend to zero as the number of observations increases and the mean of the estimated parameters becomes very close to the respective parameters of the model. For situations where the half ranges of the intervals of $Y$ is larger (which occurs when the variability of $X$ is high or when the values of the parameters are far apart), the independent parameter has a high value of $MSE$ and a high standard deviation associated to the mean value. As such, intervals with large half ranges in the response variable cause more instability in the \textit{DSD Model} and therefore the parameters' estimation is more unstable, essentially on the independent parameter.  In the boxplots presented in \textit{Figures \ref{fig3S}} to \textit{\ref{fig5S}} we may observe the behavior described above for the situations where $p=1$ and the original values of the parameters are $\alpha=2; \beta=1; \gamma=-1.$

\begin{figure}[h!]
\begin{center}
\includegraphics[width=0.75\textwidth]{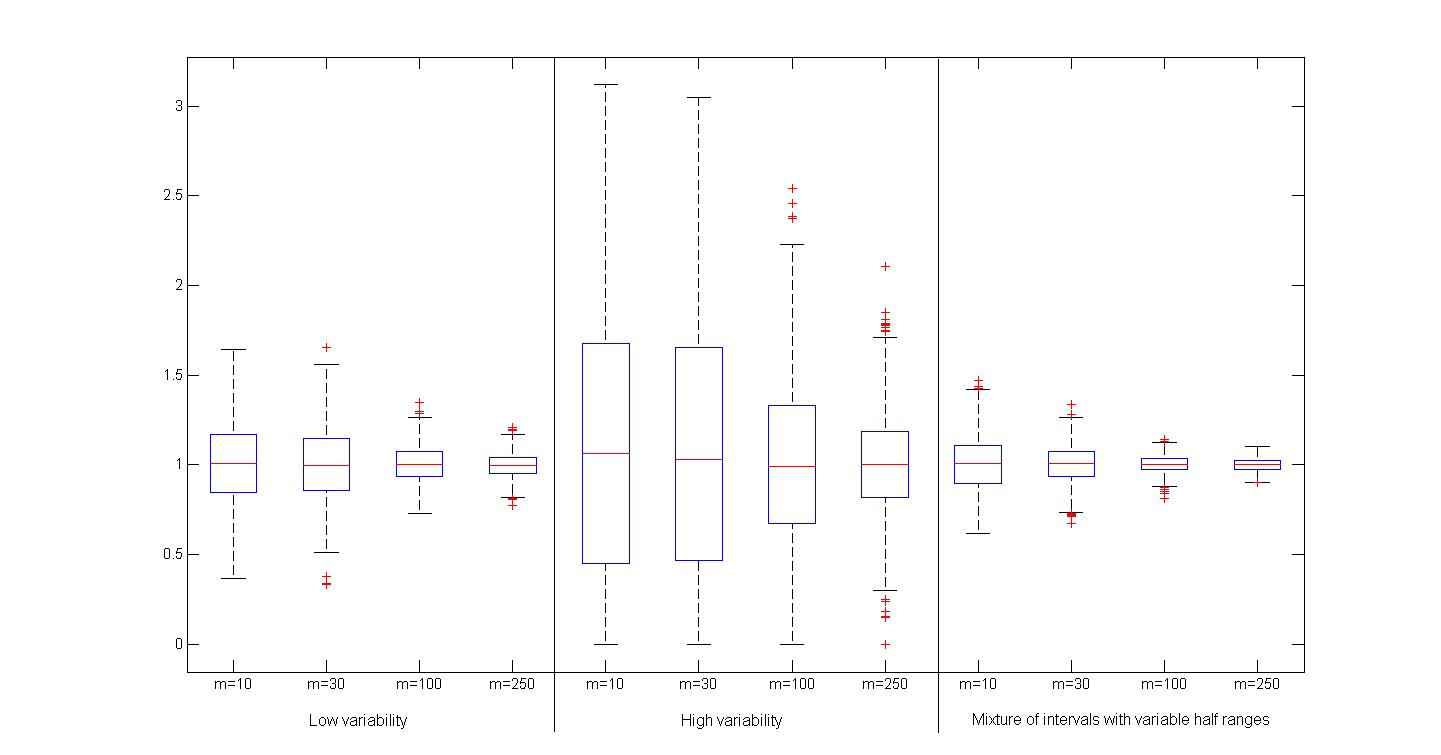}
\caption{Boxplot for the estimated parameter $\widehat{\beta}$ for a high level of linearity and when the original parameters are $\alpha=2; \beta=1; \gamma=-1$.}
\label{fig4S}
\end{center}
\end{figure}

\begin{figure}[h!]
\begin{center}
\includegraphics[width=0.75\textwidth]{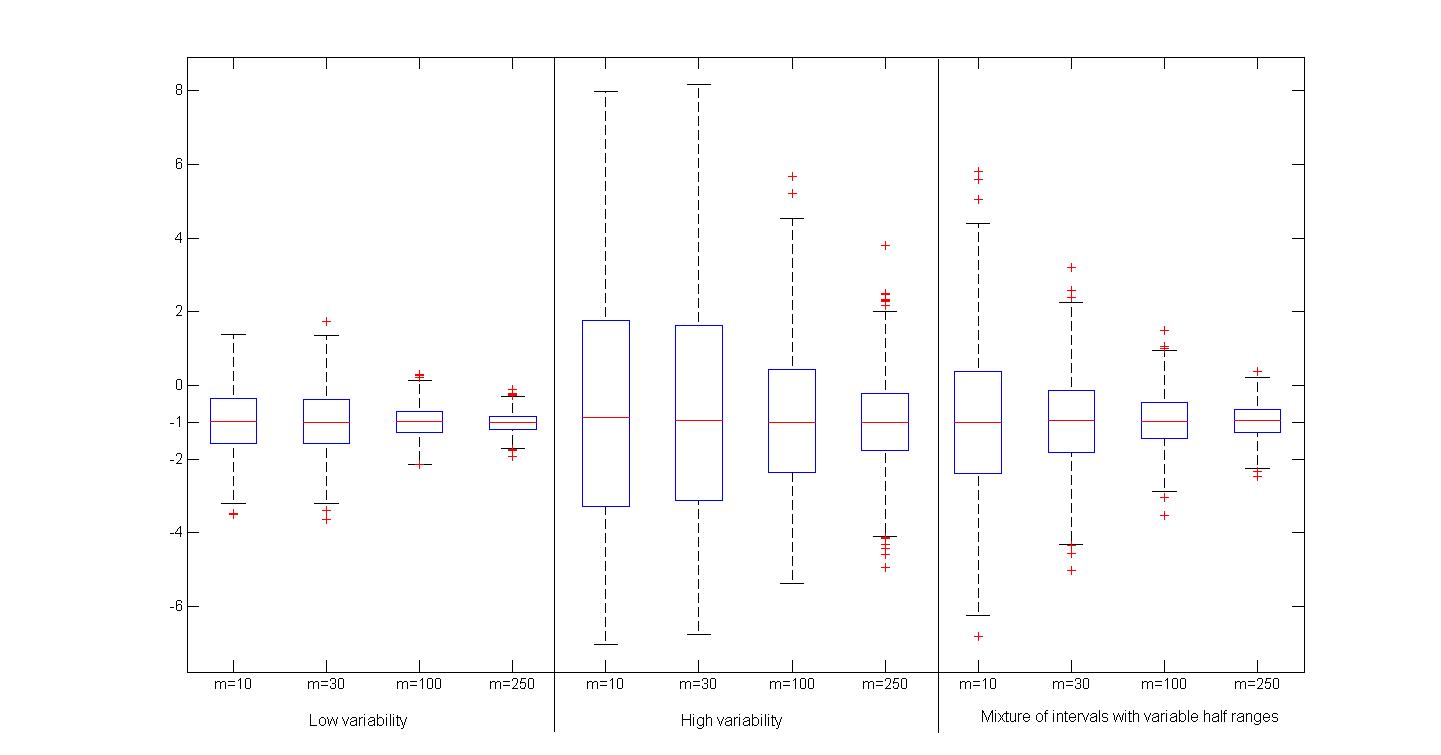}
\caption{Boxplot for the estimated parameter $\widehat{\gamma}$ for a high level of linearity and when the original parameters are $\alpha=2; \beta=1; \gamma=-1$.}
\label{fig5S}
\end{center}
\end{figure}

Based on this simulation study, we can also assess the behavior of the coefficient of determination associated to the \textit{DSD Model} and the values of the root-mean-square errors.
The values obtained for $\Omega$ show that this value provides a good evaluation for the level of linearity. The models slightly disturbed present values of $\Omega$ close to one. On the other hand, when the error function applied to the model causes a high disturbance in the linear relation, the values of $\Omega$ are closer to zero. Furthermore, the mean values of $\Omega$ are consistent with the respective values of the measures $RMSE_M;$ $RMSE_L$ and $RMSE_U.$ In general, as expected, in each situation and to the respective level of variability of the explicative variables, the highest values of $\Omega$ correspond to the lowest values of $RMSE_M.$
The values that compose the error function $\varepsilon_{j}$ are obtained considering the same criterion in all situations, but when the explicative variables include a mixture of intervals with different half ranges, the values that we obtain for $\Omega$ are lower than the ones obtained in other situations. This happens because as we have a variety of intervals, the error functions will not affect all intervals in the same way.

\section{Applied examples} \label{s4}

\subsection{The relation between time of unemployment and years of employment} \label{s4.2}

The 2008 Portuguese Labour Force Survey provides individual information about the people that live in Portugal.
The original data table that we analyzed contains, among others, demographic variables (such as gender, marital status,
age, level of education, employer...) and geographical location (region, city,...).
In this study we are interested in analyzing if the time of unemployment (in months) is related to
the time (in years) that people have worked previously. However, we are not interested in performing this study for each individual,
as it may be of greater interest to determine what happens in certain categories, such as young women who live in North of Portugal.
Since each of these categories consists of several individuals, the observed 'value' is no longer a single point but an interval.
So, in this case, the symbolic data table is built considering that the units (higher units) are classes of individuals
obtained by crossing $gender{\times}region{\times}age{\times}education$. Here, there are two genders (female (F), male (M)), four regions (north (N), Center (C), Lisbon and Tagus Valley (L), South (S)); three age groups (15 to 24 (A1), 25 to 44 (A2), 45 to 64 (A3)) and three levels of education (basic education (B), secondary education (S) and graduate (G)). In total we have $2{\times}4{\times}3{\times}3=72$ possible classes (categories). The time of unemployment and the time of work before unemployment are now interval-valued symbolic variables.

\textit{Table \ref{table1E2}} represents the symbolic table that results from the original data table, for the variables $X$ (time of employment before unemployment) and  $Y$ (time of unemployment).

\begin{table}[h!]
  \centering
  {\scriptsize
\begin{tabular}{lcc|lcc|lcc} \toprule
 Units & Y & X & Units & Y & X & Units & Y & X \\
 \hline
  $F{\times}C{\times}A1{\times}B$ & $\left[3;49\right]$ & $\left[0;4\right]$ & $F{\times}N{\times}A3{\times}S$ & $\left[0;123\right]$ & $\left[23;35\right]$ &$M{\times}L{\times}A3{\times}B$ & $\left[1;244\right]$ & $\left[22;57\right]$\\
  $F{\times}C{\times}A1{\times}S$ & $\left[1;6\right]$ & $\left[0;2\right]$ & $F{\times}S{\times}A1{\times}B$ & $\left[1;52\right]$ & $\left[1;7\right]$ &$M{\times}L{\times}A3{\times}S$ & $\left[2;65\right]$ & $\left[25;50\right]$\\
  $F{\times}C{\times}A2{\times}B$ & $\left[2;147\right]$ & $\left[2;34\right]$& $F{\times}S{\times}A1{\times}S$ & $\left[1;36\right]$ & $\left[0;9\right]$ &$M{\times}L{\times}A3{\times}G$ & $\left[7;44\right]$ & $\left[28;40\right]$\\
  $F{\times}C{\times}A2{\times}S$ & $\left[3;61\right]$ & $\left[5;22\right]$&$F{\times}S{\times}A1{\times}G$ & $\left[1;13\right]$ & $\left[0;1\right]$ &$M{\times}N{\times}A1{\times}B$ & $\left[1;33\right]$ & $\left[0;18\right]$\\
  $F{\times}C{\times}A2{\times}G$ & $\left[4;16\right]$ & $\left[0;15\right]$ &$F{\times}S{\times}A2{\times}B$ & $\left[1;101\right]$ & $\left[0;33\right]$ &$M{\times}N{\times}A1{\times}S$ & $\left[1;15\right]$ & $\left[1;4\right]$\\
  $F{\times}C{\times}A3{\times}B$ & $\left[1;108\right]$ & $\left[23;47\right]$&$F{\times}S{\times}A2{\times}S$ & $\left[0;96\right]$ & $\left[0;25\right]$ &$M{\times}N{\times}A2{\times}B$ & $\left[1;97\right]$ & $\left[1;35\right]$\\
  $F{\times}L{\times}A1{\times}B$ & $\left[1;18\right]$ & $\left[1;7\right]$&$F{\times}S{\times}A2{\times}G$ & $\left[1;21\right]$ & $\left[1;27\right]$ &$M{\times}N{\times}A2{\times}S$ & $\left[1;46\right]$ & $\left[0;21\right]$\\
  $F{\times}L{\times}A1{\times}S$ & $\left[1;19\right]$ & $\left[1;11\right]$& $F{\times}S{\times}A3{\times}B$ & $\left[1;265\right]$ & $\left[8;52\right]$ &$M{\times}N{\times}A2{\times}G$ & $\left[2;100\right]$ & $\left[2;14\right]$\\
  $F{\times}L{\times}A2{\times}B$ & $\left[0;156\right]$ & $\left[3;34\right]$&$F{\times}S{\times}A3{\times}S$ & $\left[3;26\right]$ & $\left[20;37\right]$ &$M{\times}N{\times}A3{\times}B$ & $\left[0;159\right]$ & $\left[15;52\right]$\\
  $F{\times}L{\times}A2{\times}S$ & $\left[2;69\right]$ & $\left[3;25\right]$&$M{\times}C{\times}A1{\times}B$  & $\left[3;6\right]$ & $\left[0;8\right]$ &$M{\times}N{\times}A3{\times}S$ & $\left[9;35\right]$ & $\left[20;40\right]$\\
  $F{\times}L{\times}A2{\times}G$ & $\left[0;63\right]$ & $\left[0;22\right]$&$M{\times}C{\times}A1{\times}S$ & $\left[2;3\right]$ & $\left[0;4\right]$ &$M{\times}N{\times}A3{\times}G$ & $\left[9;19\right]$ & $\left[31;36\right]$\\
  $F{\times}L{\times}A3{\times}B$ & $\left[1;320\right]$ & $\left[29;58\right]$&$M{\times}C{\times}A2{\times}B$ & $\left[2;97\right]$ & $\left[10;28\right]$ &$M{\times}S{\times}A1{\times}B$ & $\left[1;35\right]$ & $\left[0;10\right]$\\
   $F{\times}L{\times}A3{\times}S$ & $\left[2;162\right]$ & $\left[22;36\right]$&$M{\times}C{\times}A2{\times}G$ & $\left[7;13\right]$ & $\left[4;10\right]$ &$M{\times}S{\times}A1{\times}S$ & $\left[4;63\right]$ & $\left[1;6\right]$\\
    $F{\times}L{\times}A3{\times}G$ & $\left[8;27\right]$ & $\left[12;32\right]$&$M{\times}C{\times}A3{\times}B$ & $\left[4;98\right]$ & $\left[30;51\right]$ &$M{\times}S{\times}A2{\times}B$ & $\left[0;157\right]$ & $\left[4;35\right]$\\
    $F{\times}N{\times}A1{\times}B$ & $\left[1;61\right]$ & $\left[0;9\right]$&$M{\times}C{\times}A3{\times}S$ & $\left[20;38\right]$ & $\left[25;39\right]$ &$M{\times}S{\times}A2{\times}S$ & $\left[1;21\right]$ & $\left[7;24\right]$\\
     $F{\times}N{\times}A1{\times}S$ & $\left[0;10\right]$ & $\left[0;3\right]$&$M{\times}L{\times}A1{\times}B$ & $\left[2;20\right]$ & $\left[0;9\right]$ &$M{\times}N{\times}A2{\times}G$ & $\left[4;18\right]$ & $\left[5;20\right]$\\
     $F{\times}N{\times}A2{\times}B$ & $\left[1;325\right]$ & $\left[6;32\right]$&$M{\times}L{\times}A1{\times}S$ & $\left[4;14\right]$ & $\left[1;9\right]$ &$M{\times}S{\times}A3{\times}B$ & $\left[1;274\right]$ & $\left[26;56\right]$\\
       $F{\times}N{\times}A2{\times}S$ & $\left[2;88\right]$ & $\left[2;25\right]$&$M{\times}L{\times}A2{\times}B$& $\left[1;194\right]$ & $\left[0;31\right]$ &$M{\times}S{\times}A3{\times}S$ & $\left[11;26\right]$ & $\left[28;42\right]$\\
       $F{\times}N{\times}A2{\times}G$ & $\left[2;80\right]$ & $\left[1;25\right]$&$M{\times}L{\times}A2{\times}S$ & $\left[4;133\right]$ & $\left[3;23\right]$ & &  & \\
       $F{\times}N{\times}A3{\times}B$ & $\left[1;372\right]$ & $\left[11;57\right]$&$M{\times}L{\times}A2{\times}G$& $\left[6;65\right]$ & $\left[4;16\right]$ & &  & \\
\toprule
\end{tabular}}
\caption{Symbolic data table where the two variables, time of activity before unemployment and time of unemployment are interval-valued variables.}
\label{table1E2}
\end{table}

The main goal of this study is to analyze the linear relationship between the interval-valued variables: logarithm of the time of unemployment, $LNY,$ $(LNY=LN(Y+2)),$ and time of activity before the unemployment $X$, considering as observed units (higher units) the classes of individuals previously described.

We predicted the quantile function representing the interval taken by the interval-valued variable $LNY$ from the \textit{DSD Model}, and obtained:
$$\Psi_{\widehat{LNY}(j)}^{-1}(t)=2.2277+0.0779\Psi_{X(j)}^{-1}(t)-0.0503\Psi_{X(j)}^{-1}(1-t)$$

In this case, the predicted interval for each unit $j,$ is given by
\[\left[0.0276c_{X(j)}-0.1282r_{X(j)}+2.2277, 0.0276c_{X(j)}+0.1282r_{X(j)}+2.2277\right].\]

\begin{figure}[h!]
\begin{center}
\subfigure[Observed intervals for $LNY.$] {\label{figA}
\resizebox*{7cm}{!}{\includegraphics{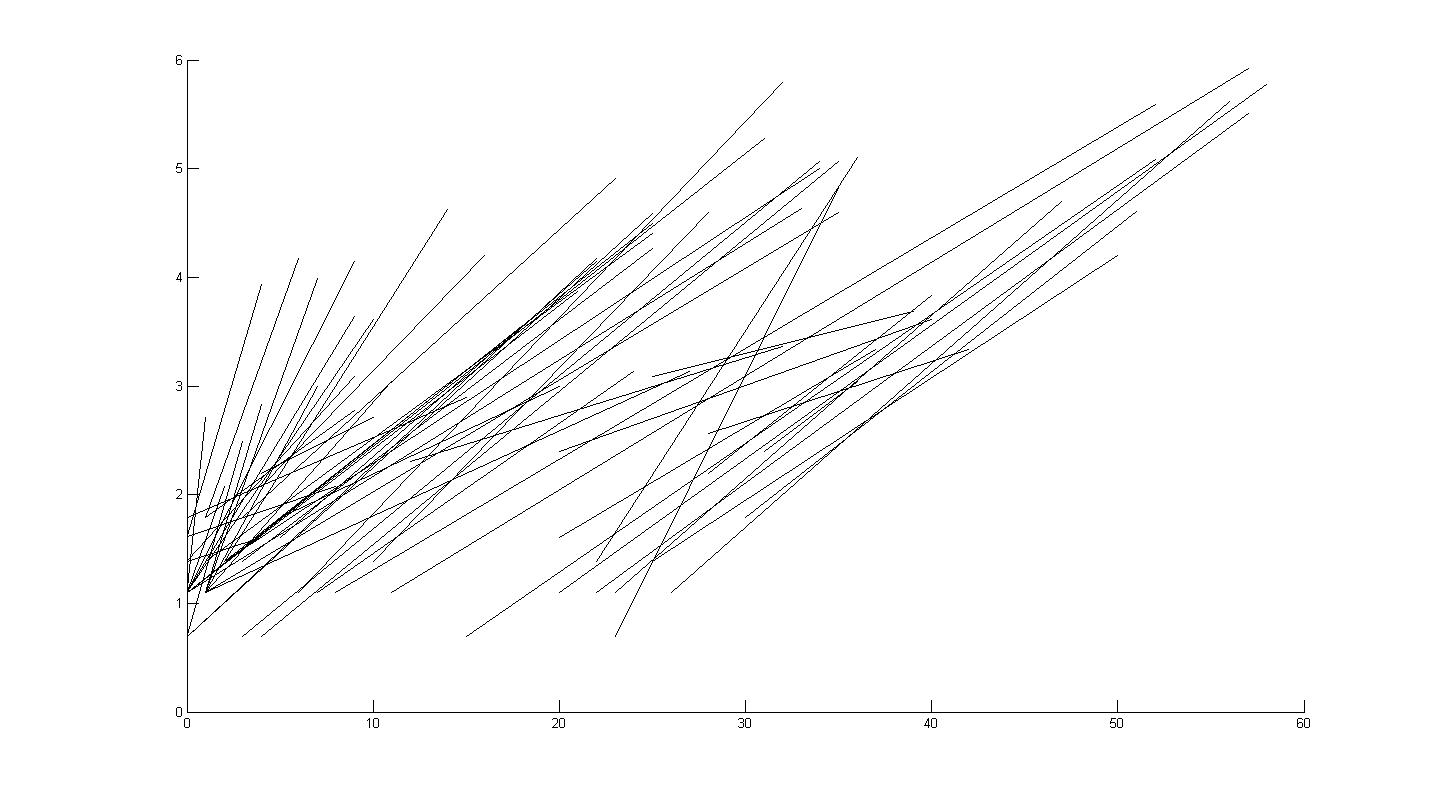}}}
\hspace{5pt}
\subfigure[Predicted intervals for $LNY$.]{\label{figB}
\resizebox*{7cm}{!}{\includegraphics{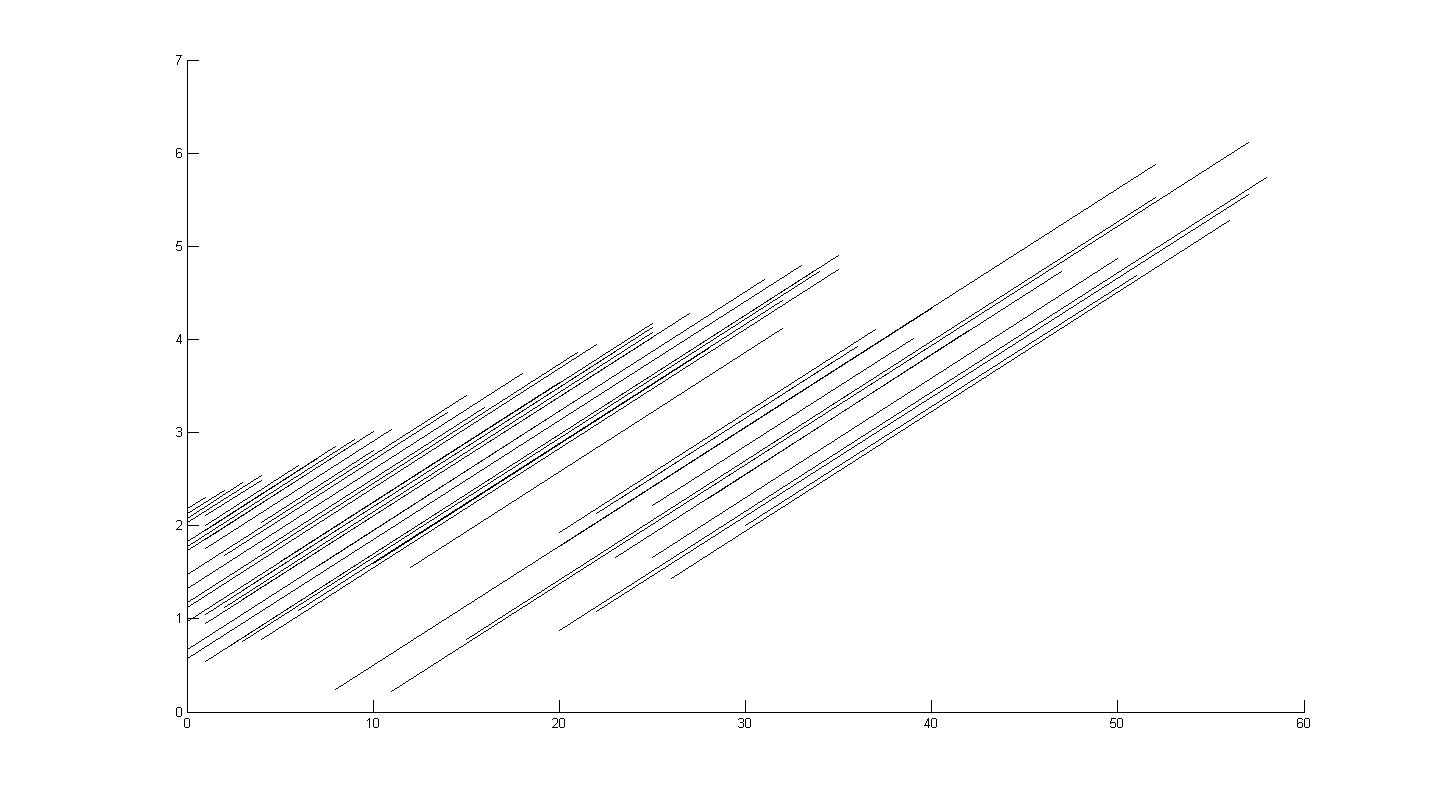}}}
\caption{Scatter plot considering the observed intervals for the interval-valued variables $X$ and $LNY$ or the predicted intervals.}%
\label{fig3E2}
\end{center}
\end{figure}

As we interpreted in \textit{Subsection \ref{s2.2}}, the interval-valued variables $X$ and $LNY$
have a linear relation that tends to be direct, because the value estimated for the parameter $\alpha=0.0779$
is slightly greater than $\beta=0.0503.$ For the set of classes of individuals to which the data refer, when the symbolic mean of time of activity before the unemployment increases one year, the symbolic mean of the $LNY$ (in months) increases $0.0276.$ However, the relationship described by the \textit{DSD Model} is not very strong. The value of the goodness-of-fit measure $\Omega$ deduced to the model is for these data $0.7715.$
The scatter plot of these data can be observed in \textit{Figure \ref{figA}}. However, as we have a large number of units, the scatter plot that represents the observed intervals of both variables by a rectangle is very hard to interpret and we chose to represent the diagonals of the rectangle.

As we have said in \textit{Subsection \ref{s2.2}}, the perfect linear regression by the \textit{DSD Model} between two interval valued variables induces a perfect linear regression between the centers of the intervals and also induces that the ratio of the ranges of the intervals is constant and equal for all observations. These behaviors can be illustrated by the scatter plot in \textit{Figure \ref{figB}}, that considers the intervals observed to the variable $X$ and the predicted intervals by the \textit{DSD Model} to the variable $LNY.$

The purpose of this example is not only to illustrate the \textit{DSD Model}, but also to compare the results with other models already proposed \cite{netocar10, netocar08, bidi07, bidi02, bidi00}. In \textit{Table \ref{table3E2}} we present the models and the \textit{Root Mean Square Error} generally used as measures of goodness of fit.
\begin{table}[h!]
{\scriptsize
  \centering
\begin{tabular}{ccccc} \toprule
   Models & Expressions that allow predicting the intervals of $LNY$ for each $j$ &$RMSE_{L}$ & $RMSE_{U}$ & $RMSE_{M}$ \\  \hline
  \multirow{2}{*}{\textit{DSD}} &  $\Psi_{\widehat{LNY}(j)}^{-1}(t)=2.2277+0.0779\Psi_{X(j)}^{-1}(t)-0.0503\Psi_{X(j)}^{-1}(1-t)$  & \multirow{2}{*}{0.5745} &\multirow{2}{*}{0.6710} & \multirow{2}{*}{0.4679}  \\
   & $\widehat{c}_{{LNY}(j)}=2.2277+0.0276c_{X(j)}$ and  $\widehat{r}_{{LNY}(j)}=0.1282r_{X(j)}$ & & &  \\  \hline
  \textit{CM} &  $\widehat{c}_{{LNY}(j)}=2.2277+0.0276c_{X(j)}$ & 1.1622 & 1.3146 & 0.7759 \\ \hline
  \textit{Billard 2007} &  $\widehat{c}_{{LNY}(j)}=1.9009+0.0468c_{X(j)}$ & 1.1504 & 1.0365  & 0.7255 \\  \hline
  \multirow{2}{*}{\textit{MinMax}} &  $\widehat{\underline{I}}_{{LNY}(j)}=1.2236+0.0206\underline{I}_{{LNY}(j)}$ & \multirow{2}{*}{0.4725} &\multirow{2}{*} {0.7329} & \multirow{2}{*} {0.4621} \\
  & $\widehat{\overline{I}}_{{LNY}(j)}=2.8704+0.0436\overline{I}_{{LNY}(j)}$ & & & \\  \hline
 \multirow{2}{*}{\textit{CRM }} &  $\widehat{c}_{{LNY}(j)}=2.2277+0.0276c_{X(j)}$ & \multirow{2}{*}{0.4458} & \multirow{2}{*}{0.6541} & \multirow{2}{*}{0.4397} \\
 & $\widehat{r}_{{LNY}(j)}=1.0642+0.0855r_{X(j)}$  & & & \\  \hline
  \multirow{2}{*}{\textit{CCRM }} &  $\widehat{c}_{{LNY}(j)}=2.2277+0.0276c_{X(j)}$ & \multirow{2}{*}{0.4458} & \multirow{2}{*}{0.6541} & \multirow{2}{*}{0.4397} \\
  &  $\widehat{r}_{{LNY}(j)}=1.0642+0.0855r_{X(j)}$ & & & \\ \toprule
\end{tabular}}
\caption{Comparison of the performance between linear regression model for interval-valued variables.}
\label{table3E2}
\end{table}

In this example the \textit{CRM} and \textit{CCRM} are the same because in the \textit{CRM} the parameters estimated for the half ranges are all non-negative, the constrains imposed in \textit{CCRM} to these parameters are met. We can also observe that the linear regression induced by the \textit{DSD Model} relative to the centers of the intervals is obtained by the models where a linear regression between the centers is considered.
The results of the \textit{Root Mean Square Error (RMSE)} allow comparing the predicted and the observed intervals of the response variable $LNY.$ These measures are not deduced from the model, therefore they may serve as independent comparison measures. Observing the values of the \textit{RMSE}, we can conclude that the \textit{DSD Model} and \textit{CRM (and CCRM)} have similar results, that is not surprising because the linear regression between the centers is the same. It is important to underline that the goal of the work developed in this paper is not propose a model that provides better results than the previous models. The \textit{DSD Model} for interval-valued variable emerges from the particularization of a more general model, the \textit{DSD Linear Regression Model} for histogram-valued variables. The advantage of the \textit{DSD Model} when applied to interval-valued variables is that it allows taking into consideration a distribution within the intervals.

\subsection{Predicted burned area of forest fires, in the northeast region of Portugal,} \label{s4.3}

This study considers forest fire data from the Montesinho natural park, in the northeast region of Portugal. The original data can be found in \cite{basedados08} and details are described in \cite{cortez07}. For this study we selected the response variable \textit{area} (the burned area of the forest (in ha)) and three explicative variables: \textit{temp} (temperature in Celsius degrees); \textit{wind} (wind speed in km/h); \textit{rh}  (relative humidity in percentage). As in the classical study \cite{cortez07}, the response variable \textit{area} was transformed with a $ln(x+1)$ function and we represent it as \textit{LNarea}.
To build the symbolic data (macrodata) we aggregated the information by months. The units (higher units) of this study are the months and the observations of the variables \textit{temp, wind, rh} and \textit{LNarea} associated to each month were organized in intervals. To build these macrodata we considered only the months and the records in which forest fires occurred. For this reason January and November were eliminated. The symbolic data considered in this example is represented in \textit{Table \ref{table1E3}}.

\begin{table}[h!]
  \centering
\begin{tabular}{ccccc} \toprule
  Months & LNarea (Y) & temp  & wind  & rh   \\
\hline

 Feb & $\left[0.74;3.97\right]$  & $\left[4.6;12.4\right]$ & $\left[0.9;9.4\right]$&  $\left[35;82\right]$ \\

  Mar & $\left[0.67;3.63\right]$ & $\left[5.3;17\right]$ & $\left[0.9;9.4\right]$& $\left[26;70\right]$ \\

  Apr & $\left[1.47;4.13\right]$  & $\left[5.8;13.7\right]$&  $\left[3.1;9.4\right]$&  $\left[33;64\right]$ \\

  May & $\left[3.58;3.58\right]$  & $\left[18;18\right]$ & $\left[4;4\right]$&  $\left[40;40\right]$ \\

  June & $\left[0.64;4.27\right]$  &$\left[14.3;28\right]$& $\left[1.8;9.4\right]$&  $\left[34;79\right]$ \\

  July & $\left[0.31;5.63\right]$ &$\left[11.2;33.3\right]$& $\left[0.4;8.9\right]$&  $\left[22;88\right]$ \\

  Aug & $\left[0.09;6.62\right]$  &$\left[11.2;33.3\right]$& $\left[0.4;8.9\right]$&  $\left[22;88\right]$\\

  Sep & $\left[0.29;7\right]$  &$\left[10.1;29.6\right]$& $\left[0.9;7.6\right]$&  $\left[15;78\right]$ \\

  Oct & $\left[1.9;3.9\right]$  &$\left[16.1;20.2\right]$& $\left[2.7;4.5\right]$&  $\left[25;45\right]$ \\

  Dec & $\left[1.9;3.2\right]$ &$\left[2.2;5.1\right]$& $\left[4.9;8.5\right]$&  $\left[21;61\right]$ \\
\toprule
\end{tabular}
\caption{Burned area data, where the four variables \textit{LNarea, temp, wind} and \textit{rh} are now interval-valued variables.}
\label{table1E3}
\end{table}

Considering the conditions described above, the model that allows predicting the intervals of \textit{LNarea} from the intervals of the explicative variables $temp, wind$ and $rh$ for each month $j$ is as follows:

\begin{center}
\begin{equation}\label{eq1_ex2}
 \Psi_{\widehat{LNarea}(j)}^{-1}(t) = 1.8637+0.0224\Psi_{temp(j)}^{-1}(t)-0.0215\Psi_{temp(j)}^{-1}(1-t)-0.0143\Psi_{rh(j)}^{-1}(1-t)
\end{equation}
\end{center}
with $t \in \left[0,1\right]$.

The goodness-of-fit measure associated to this situation is $\Omega=0.9202,$ that shows that this linear regression model describes well the relationship between the interval-valued variables. So, if we know the forecast for the temperature, wind and relative humidity for one month, it is possible to predict the minimum and maximum of  area of burned area of the forest.

\begin{figure}[h!]
\begin{center}
\includegraphics[width=0.9\textwidth]{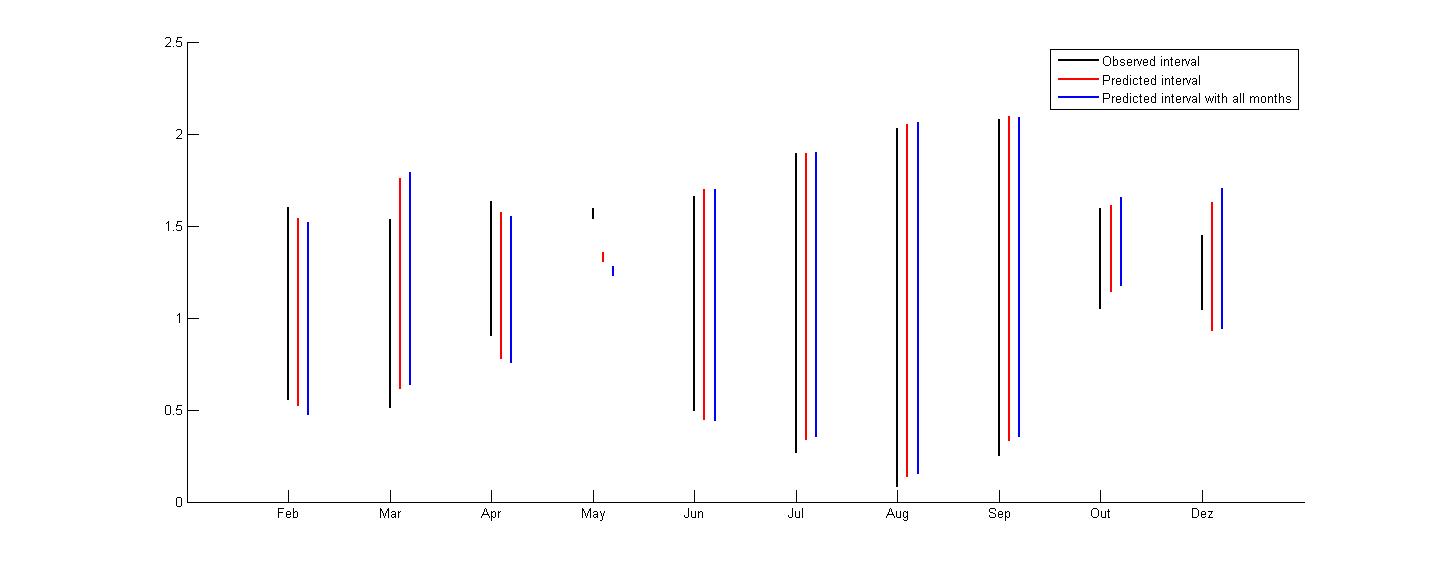}
\caption{Observed and predicted intervals for the interval-valued variable \textit{LNarea}, for each month.}
\label{fig1E3}
\end{center}
\end{figure}

As we obtain a good behavior of the model in this situation, in the next study, we will compare the observed and predicted intervals associated to the variable \textit{LNarea} for each month $j,$ $j \in \left\{ \right.$ February, March, April, May, June, July, August, September, October, December$\left. \right\}.$ We will consider that when we predict the interval of hectares of burned area
(\textit{LNarea}) for  month $j$ this month will not be considered in  building the model. The results are represented in \textit{Figure \ref{fig1E3}}. The months are represented in the $x-$axis and the intervals, observed and predicted in the two ways described above, are represented in the $y-$axis.

Observing \textit{Figure \ref{fig1E3}} we can say that the prediction of the intervals is in general quite good, slightly worse for May and December. Comparing the predictions obtained by the \textit{DSD Model} in (\ref{eq1_ex2}) with those obtained by other models,
we can observe that when the month value is not used in the estimation of the parameters of the model, small differences are generally observed.

The expressions of the models proposed by Lima Neto and De Carvalho \cite{netocar10} and Billard and Diday \cite{bidi00} that allow predicting the intervals of values of burned area of forest fires are as follows in \textit{Table \ref{table2E3}}.

\begin{table}[h!]
{\footnotesize
  \centering
\begin{tabular}{cc} \toprule
   Models & Expressions that allow predicting the intervals of $LNarea$ for each $j$ \\  \hline
  \multirow{2}{*}{\textit{DSD}} &  $\Psi_{\widehat{LNarea}(j)}^{-1}(t) = 1.8637+0.0224\Psi_{temp(j)}^{-1}(t)-0.0215\Psi_{temp(j)}^{-1}(1-t)-0.0143\Psi_{rh(j)}^{-1}(1-t)$ \\
   & $\widehat{c}_{{LNarea}(j)}=1.8637+0.0009c_{temp(j)}-0.0143c_{rh(j)}$  \\
   & $\widehat{r}_{{LNarea}(j)}=0.0439r_{temp(j)}+0.0143r_{rh(j)}$ \\  \hline
  \textit{CM} &  $\widehat{c}_{{LNarea}(j)}=1.9163+0.0015c_{temp(j)}+0.0027c_{wind(j)}-0.0158c_{rh(j)}$  \\  \hline
  \multirow{2}{*}{\textit{MinMax}} &  $\widehat{\underline{I}}_{{LNarea}(j)}=1.1559+0.0123\underline{I}_{{temp}(j)}-0.0379\underline{I}_{{wind}(j)}+0.0085\underline{I}_{{rh}(j)}$ \\
  & $\widehat{\overline{I}}_{{LNarea}(j)}=-0.3930+0.01127\overline{I}_{{temp}(j)}+0.2372\overline{I}_{{wind}(j)}+0.0168\overline{I}_{{rh}(j)} $  \\  \hline
 \multirow{2}{*}{\textit{CRM }} &  $\widehat{c}_{{LNarea}(j)}=1.9163+0.0015c_{temp(j)}+0.0027c_{wind(j)}-0.0159c_{rh(j)} $  \\
 & $\widehat{r}_{{LNarea}(j)}=0.0091+0.0652r_{temp(j)}-0.0072r_{wind(j)}+0.0089r_{rh(j)} $   \\  \hline
  \multirow{2}{*}{\textit{CCRM }} &  $\widehat{c}_{{LNarea}(j)}=1.9163+0.0015c_{temp(j)}+0.0027c_{wind(j)}-0.0159c_{rh(j)}$  \\
  &  $\widehat{r}_{{LNarea}(j)}=0.0037+0.0651r_{temp(j)}+0.0081r_{rh(j)}$  \\ \toprule
\end{tabular}}
\caption{Comparison of the performance between linear regression model for interval-valued variables.}
\label{table2E3}
\end{table}

In this case, as one of the estimated parameters of the model associated to the half ranges in  \textit{CRM} is negative,
the expression for the half ranges in \textit{CCRM} is already different.

\begin{table}[h!]
  \centering
\begin{tabular}{ccccc} \toprule
   Models &$RMSE_{L}$ & $RMSE_{U}$ & $RMSE_{M}$ \\ \hline
  \textit{DSD} & 0.1106 & 0.1222 & 0.1066 \\ \hline
  \textit{CM} &  0.3076 & 0.2676 & 0.1856\\ \hline
  \textit{MinMax}  & 0.1481 & 0.0940 & 0.1044 \\ \hline
  \textit{CRM } & 0.1030 & 0.1161 & 0.1038 \\ \hline
  \textit{CCRM } & 0.1034 & 0.1159 & 0.1038 \\ \toprule
\end{tabular}
\caption{Comparison of the performance of different linear regression models for the burned-area interval data.}
\label{table3E3}
\end{table}

In \textit{Table \ref{table3E3}} we present also the \textit{RMSE} for the models previously proposed \cite{netocar10, bidi00} and for the \textit{DSD Model.} As we observed in Example \ref{s4.2}, the results of the \textit{RMSE} calculated for the \textit{CRM, CCRM} and \textit{DSD Model} are again very similar.

\begin{figure}
        \centering
        \begin{subfigure}
                \centering
               \includegraphics[width=0.75\textwidth]{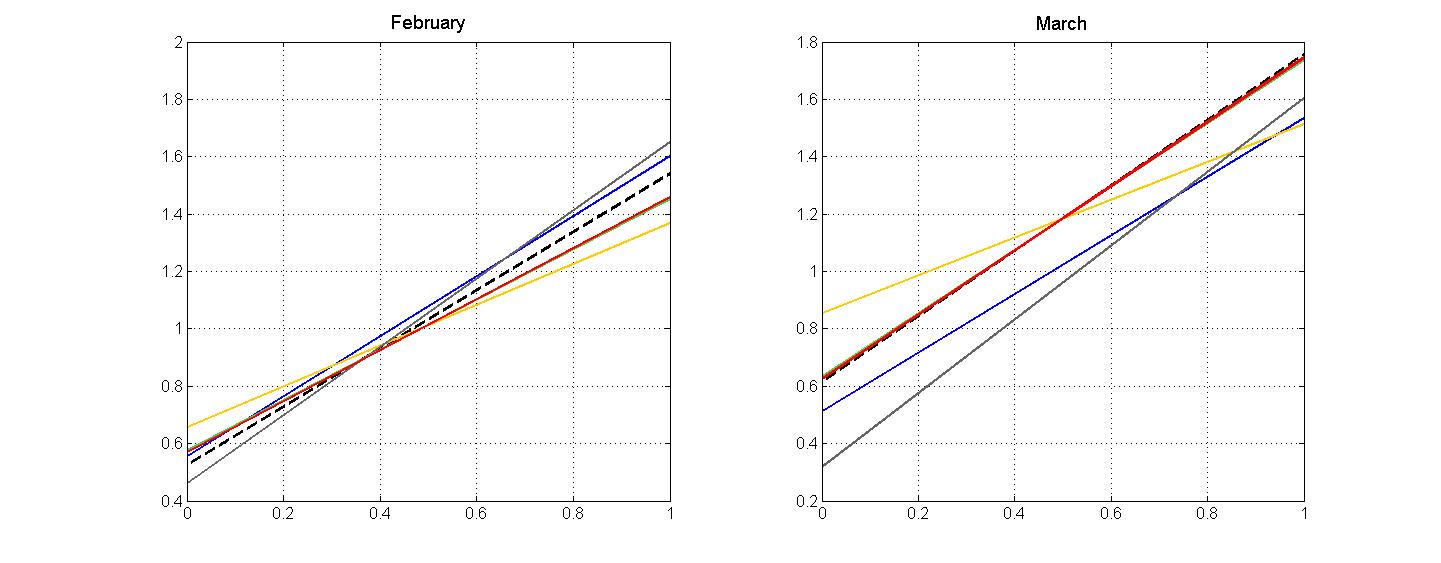}
        \end{subfigure}%
        \begin{subfigure}
                \centering
                \includegraphics[width=0.75\textwidth]{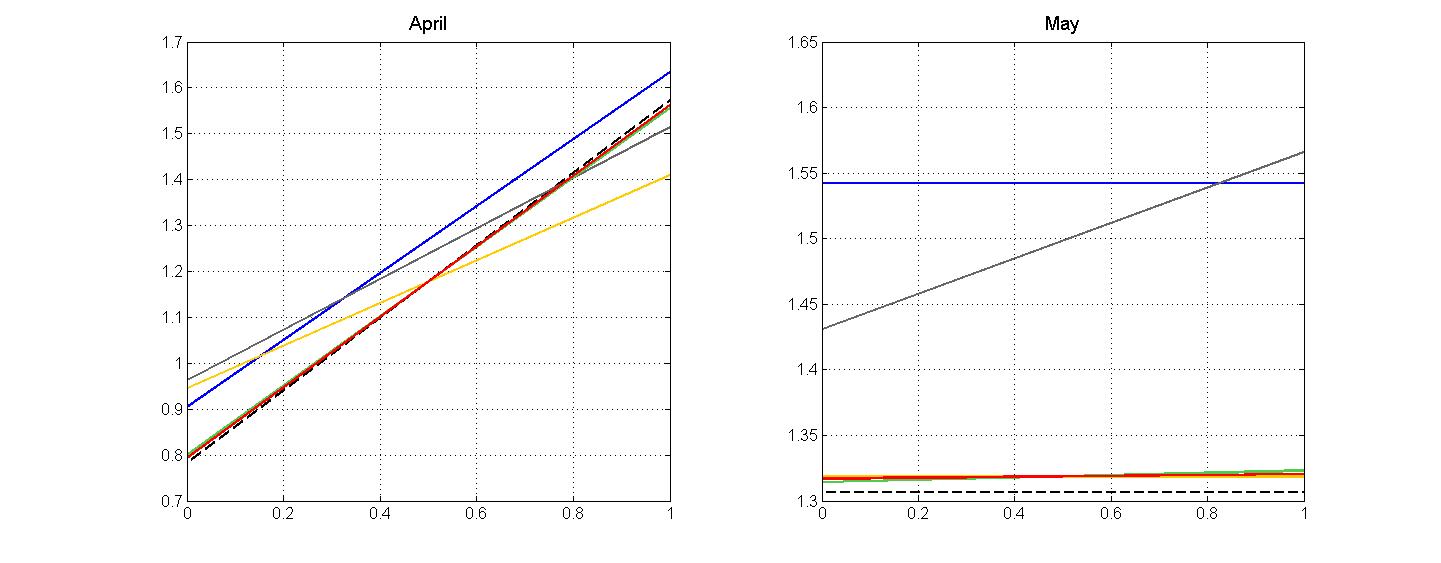}
        \end{subfigure}
        \begin{subfigure}
                \centering
                \includegraphics[width=0.75\textwidth]{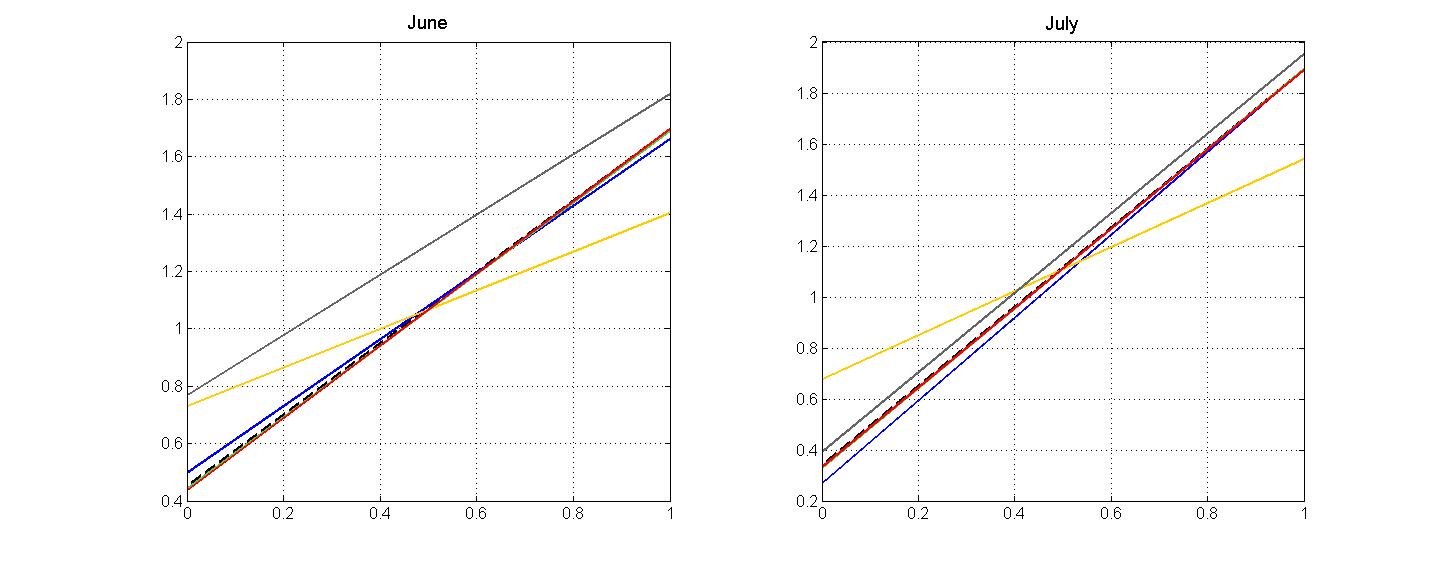}
        \end{subfigure}
        \begin{subfigure}
                \centering
                \includegraphics[width=0.75\textwidth]{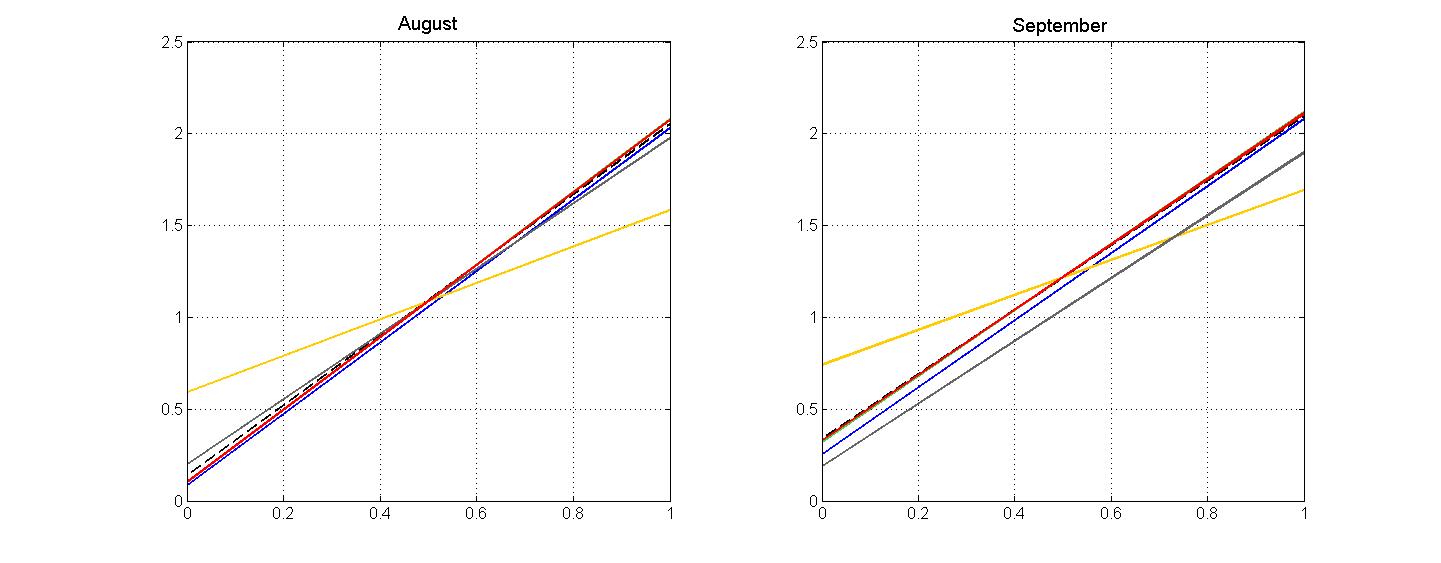}
        \end{subfigure}
        \begin{subfigure}
                \centering
                \includegraphics[width=0.75\textwidth]{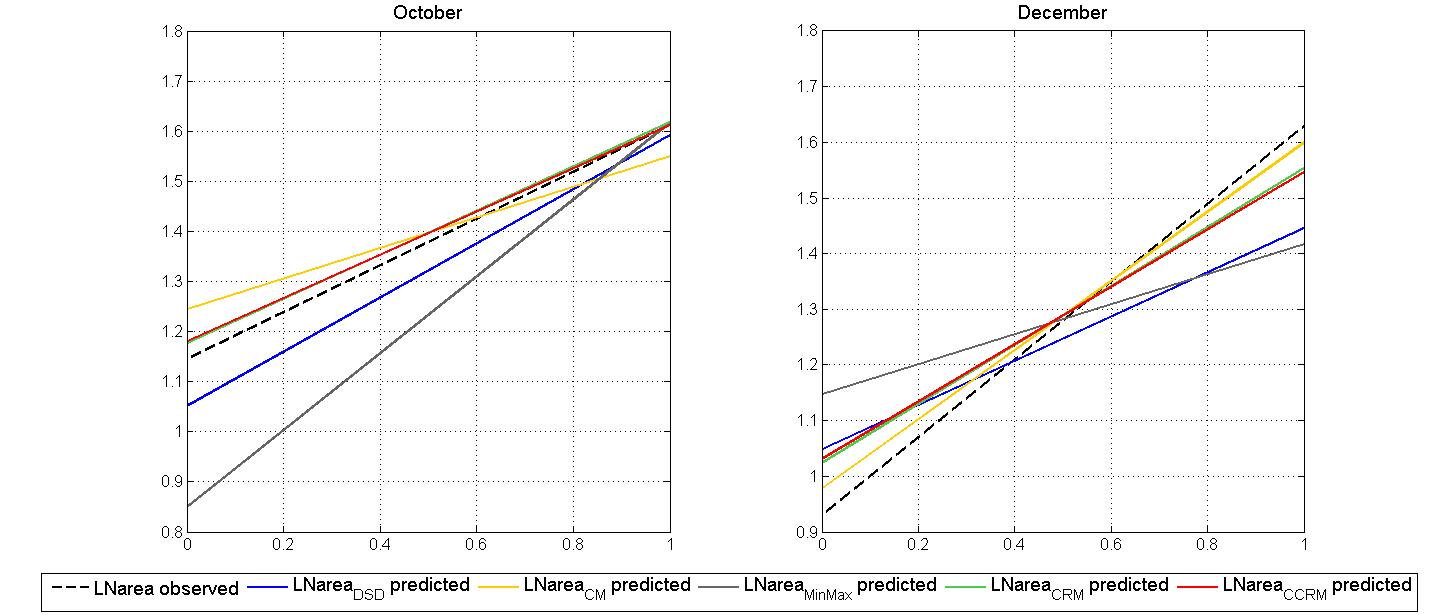}
        \end{subfigure}
        \caption{Observed and predicted intervals for the \textit{LNarea} in all months, predicted with several models.}\label{fig2E3}
\end{figure}

In \textit{Figure \ref{fig2E3}} we may compare the predicted intervals of the values of burned area of forest fires in all months considering the linear regression models \textit{CM, MinMax, CRM, CCRM} and \textit{DSD}.

\section{Conclusion} \label{s5}

An interval-valued variable is a particular case of a histogram-valued variable if for all observations we only have one interval with weight equal to one. A classical variable is a particular case of an interval-valued variable, when to all observations corresponds a degenerate interval (an interval where the lower and upper bounds are the same). Because of this link between histogram, interval and classical variables it was logical that the \textit{DSD Model} for histogram-valued variables could be particularized to interval-valued variables that in turn could be particularized to real values, as we have observed in this paper.

The main advantages of the \textit{DSD Model} are that it defines a linear relationship between one response variable and $n$ explicative variables without decomposing the intervals in their bounds or centers and half ranges. In fact, this model, as it uses the quantile function to represent the intervals, allows working with the intervals and consider the distributions within intervals. In this paper we assume the Uniform distribution in all intervals that are the ``values" observed for the interval-valued variables. For these conditions, the \textit{DSD Model} induces a relation between the half ranges and a relation between the centers of the intervals where the respective estimated parameters are not independent; in the case of the half ranges, this relation is always direct, similarly to what occurs in the \textit{Constrained Center and Range Method}.

The \textit{DSD Model} has the potential of taking into consideration the distribution in the intervals associated to the observations of the interval-valued variables. As such, it is possible to adapt the proposed model to interval-valued variables with other distributions.
For example, the \textit{DSD Model} may be developed considering a triangular distribution in the intervals.
As in most studies of \textit{Symbolic Data Analysis}, it is considered that the values in the intervals are uniformly distributed, all descriptive statistics would also have to be also redefined.

Furthermore, a generalization of the \textit{DSD Model} is currently under development with the aim of obtaining a more flexible model. In this new approach, applied both to histogram-valued variables and interval-valued variables, the independent parameter is a quantile function instead of a real number.

As a future research perspective, other models and methods in Symbolic Data Analysis based on linear relationships between interval-valued variables, such as logistic regression, may now be developed using this approach.

\vspace{2cm}

\textbf{Acknowledgements}

This work is partly funded by the ERDF – European Regional Development Fund through the
COMPETE Programme (operational programme for competitiveness) and by National Funds through the
FCT – Funda\c{c}\~{a}o para a Ci\^{e}ncia e a Tecnologia (Portuguese Foundation for Science and Technology)
within project «FCOMP - 01-0124-FEDER-022701».

\pagebreak

\newpage
\appendix

\section*{\large Appendix A. Proof of Proposition \ref{p2.6}.}\label{ap1}
\addcontentsline{toc}{chapter}{Appendix A: Proof of Proposition \ref{p2.6}}

\vspace{1cm}

Before prooving Proposition \ref{p2.6} it is necessary to consider two theorems \cite{win94} and to define the function to optimize in matricial form.

\begin{theorem}\label{t1}
Consider the minimization problem in (\ref{eq2.14}). If $b^{*}=(\alpha^{*}, \beta^{*}, \gamma^{*})$ is an optimal solution of this problem, $b^{*}$ must satisfy the constrains of the optimization problem and the Kuhn Tucker conditions:
\begin{itemize}
  \item Constrains: $-\alpha \leq 0$ and $-\beta \leq 0$
    \item Kuhn Tucker conditions:
    \begin{enumerate}
  \item $\frac{\partial f (b^{*})}{\partial \alpha}-\lambda_{1}=0$
 \item $\frac{\partial f (b^{*})}{\partial \beta} -\lambda_{2}=0$
   \item $\frac{\partial f (b^{*})}{\partial \gamma}=0$
  \item $\lambda_{1} \alpha^{*} =0$
  \item $\lambda_{2} \beta^{*} =0$
\item $\lambda_{1}, \lambda_{2} \geq 0.$
 \end{enumerate}
\end{itemize}
\end{theorem}\medskip

\begin{theorem}\label{t2}
Consider the minimization problem in (\ref{eq2.14}). If $f(\alpha, \beta, \gamma),$ $g_{1}(\alpha, \beta, \gamma)$ and $g_{2}(\alpha, \beta, \gamma)$ are convex functions, then any vector that satisfies the hypotheses of \textit{Theorem \ref{t1}} is an optimal solution of the optimization problem in (\ref{eq2.14}).
\end{theorem}\medskip

The function $f(\alpha, \beta, \gamma) = \displaystyle \sum\limits_{j=1}^m \left[\left( c_{Y(j)}-\left(\alpha-\beta\right) c_{X(j)}-\gamma \right)^2+\frac{1}{3}\left( r_{Y(j)}-\left(\alpha+\beta\right) r_{X(j)} \right)^2\right]$ to be optimized in problem (\ref{eq2.14}) may be rewritten as follows:

\begin{center}
\begin{equation}
f(\alpha, \beta, \gamma) = \mathbf{b}^{T}\mathbf{Hb}+\mathbf{q}^{T}\mathbf{b+d}
\end{equation}
\end{center}
where the matrices and vectores involved are the following:
\begin{itemize}
  \item  $\mathbf{H}$ is the hessian matrix, a symmetric matrix of order $3,$

  \[\mathbf{ H}=\left[  \begin{array}{ccc}
     \displaystyle \sum\limits_{j=1}^m c_{X(j)}^2+\frac{1}{3}r_{X(j)}^2 & \displaystyle \sum\limits_{j=1}^m -c_{X(j)}^2+\frac{1}{3}r_{X(j)}^2 & \displaystyle \sum\limits_{j=1}^m c_{X(j)} \\
     \displaystyle \sum\limits_{j=1}^m -c_{X(j)}^2+\frac{1}{3}r_{X(j)}^2 & \displaystyle \sum\limits_{j=1}^m c_{X(j)}^2+\frac{1}{3}r_{X(j)}^2 & \displaystyle \sum\limits_{j=1}^m -c_{X(j)} \\
  \displaystyle \sum\limits_{j=1}^m c_{X(j)} & \displaystyle \sum\limits_{j=1}^m -c_{X(j)} & m
    \end{array}
  \right]\]

  \item $\mathbf{q}$ is the column vector  of independent terms, \[\mathbf{ q}=\left[
                           column                                     \begin{array}{c}
                                                                \displaystyle \sum\limits_{j=1}^m -2c_{Y(j)}c_{X(j)}-\frac{2}{3}r_{Y(j)}r_{X(j)}\\
                                                                 \displaystyle \sum\limits_{j=1}^m 2c_{Y(j)}c_{X(j)}-\frac{2}{3}r_{Y(j)}r_{X(j)}\\
                                                                   \displaystyle \sum\limits_{j=1}^m -2c_{Y(j)}
                                                                   \end{array}\right]\]
  \item $\mathbf{b}$ is the column vector  of the parameters $ \mathbf{b}=\left[ \alpha \quad  \beta \quad  \gamma \right]^T$
  \item $d$ is the real value $d=\displaystyle \sum\limits_{j=1}^m \displaystyle\sum\limits_{i=1}^n c_{Y(j)}^{2}+\frac{1}{3}r_{Y(j)}^{2}.$
\end{itemize}

\vspace{1cm}

\noindent \textbf{Proof of Proposition \ref{p2.6}:}

\textbf{Proof:}
Consider the optimization problem in (\ref{eq2.14}) where:
 \begin{description}
   \item[a)] the functions $g_{1}(\alpha, \beta, \gamma)$ and $g_{2}(\alpha, \beta, \gamma)$ that define the non-negative constrains are convex, so the feasible region of the optimization problem is a convex set;\medskip

   \item[b)] $f(\alpha, \beta, \gamma)$ is a convex function because $\mathbf{H}$ is positive semi-definite. Consider the matriz $\mathbf{X}$ defined in Equation (\ref{eq2.11AAA}) but now only for one explicative variable. In this particular case, we have:

       \[\mathbf{X}=\left[\begin{array}{ccc}
                        c_{X(1)} & -c_{X(1)} & 1\\
                        \vdots & \vdots & \vdots \\
                        c_{X(m)} & -c_{X(m)} & 1 \\
                        \frac{1}{\sqrt{3}}r_{X(1)} & \frac{1}{\sqrt{3}}r_{X(1)} & 0 \\
                       \vdots & \vdots & \vdots \\
                        \frac{1}{\sqrt{3}}r_{X(m)} & \frac{1}{\sqrt{3}}r_{X(m)} & 0
                      \end{array}\right]\]

   As $\mathbf{H}=\mathbf{X}^{T}\mathbf{X},$  $\mathbf{H}$ is positive semi-definite.\medskip

   \item[c)] the intervals of the explicative variable $X$ are not all degenerate ($r_{X_{j}}\neq 0$) or symmetric ($c_{X_{j}}\neq 0$). In this situation, the columns of $\mathbf{X}$ are linearly independent, so  $\mathbf{H}$ is positive definite and consequently the function $f(\alpha, \beta, \gamma)$ is strictly convex. When the objective function is strictly convex the optimal solution is unique.
 \end{description}\medskip

 As the optimization problem in (\ref{eq2.14}) verifies the conditions of  \textit{Theorem \ref{t2}}, it is possible to find the expressions of the parameters for the linear regression model in \ref{def2.5}.
 Considering the \textit{Kuhn Tucker conditions} (4) and (5) we have:

    \[\lambda_{1} \alpha^{*} =0 \: \wedge \: \lambda_{2} \beta^{*} =0 \; \Leftrightarrow \left(\lambda_{1}=0 \: \vee  \: \alpha^{*} =0 \right) \: \wedge \: \left(\lambda_{2} =0 \: \vee  \: \beta^{*} =0 \right).\]

 So, we may consider four situations.

 \begin{description}
   \item[I] Suppose  $\alpha^{*}=\beta^{*} =0.$ The system formed by the \textit{Kuhn Tucker conditions} is

  \[ \left\{\begin{array}{c}
              \frac{\partial f (b^{*})}{\partial \alpha}-\lambda_{1}=0 \\
             \frac{\partial f (b^{*})}{\partial \beta} -\lambda_{2}=0 \\
            \frac{\partial f (b^{*})}{\partial \gamma}=0 \\
            \end{array}\right.\]

  Solving this system we prove that in this situation \[\displaystyle\sum\limits_{j=1}^m\frac{1}{3}r_{X(j)}r_{Y(j)}=\displaystyle \sum\limits_{j=1}^m\left(c_{Y(j)}-\overline{Y}\right)c_{X(j)}.\]

\noindent So, in this case $\alpha^{*}=0;$ $\beta^{*}=0$ and $\gamma^{*}=\overline{Y}.$

 \item[II] Suppose  $\alpha^{*}=0$ and $\lambda_{2}=0.$ Considering that in this case, $\lambda_{1}\geq 0$ and $\beta^{*} > 0,$ we have:

 \[ \left\{\begin{array}{l}
              \frac{\partial f (b^{*})}{\partial \alpha}-\lambda_{1}=0 \\
            \frac{\partial f (b^{*})}{\partial \beta}=0 \\
            \frac{\partial f (b^{*})}{\partial \gamma}=0 \\
            \end{array}\right.\]

\noindent from which we conclude that:

       {\scriptsize     \[ \left\{\begin{array}{l}
            \displaystyle\sum\limits_{j=1}^m\frac{1}{3}r_{X(j)}r_{Y(j)}\displaystyle \sum\limits_{j=1}^m\left(c_{X(j)}-\overline{X}\right)^2 \leq \displaystyle \sum\limits_{j=1}^m\left(c_{Y(j)}-\overline{Y}\right)c_{X(j)}\displaystyle \sum\limits_{j=1}^m\frac{1}{3}r^2_{X(j)}\\
            \beta^{*}=\frac{\displaystyle\sum\limits_{j=1}^m\frac{1}{3}r_{X(j)}r_{Y(j)}-\displaystyle \sum\limits_{j=1}^m\left(c_{Y(j)}-\overline{Y}\right)\left(c_{X(j)}-\overline{X}\right)}{\displaystyle \sum\limits_{j=1}^m\left(c_{X(j)}-\overline{X}\right)^2+\displaystyle \sum\limits_{j=1}^m\frac{1}{3}r^2_{X(j)}} \quad {\rm if} \quad
            \displaystyle\sum\limits_{j=1}^m\frac{1}{3}r_{X(j)}r_{Y(j)} > \displaystyle \sum\limits_{j=1}^m\left(c_{Y(j)}-\overline{Y}\right)c_{X(j)}\\
            \gamma^{*}= \overline{Y}+\beta^{*}\overline{X}\\
                 \end{array}\right.
            \]}

\item[III] Suppose  $\lambda_{1}=0$ and $\beta^{*}=0.$  Considering that in this case, $\lambda_{2}\geq 0$ and $\alpha^{*} > 0,$ we have:

  \[ \left\{
            \begin{array}{l}
              \frac{\partial f (b^{*})}{\partial \alpha}=0 \\
            \frac{\partial f (b^{*})}{\partial \beta}-\lambda_{2}=0 \\
            \frac{\partial f (b^{*})}{\partial \gamma}=0 \\
            \end{array}\right.\]

  \noindent from which we conclude that:

       {\scriptsize     \[ \left\{\begin{array}{l}
            \alpha^{*}=\frac{\displaystyle\sum\limits_{j=1}^m\frac{1}{3}r_{X(j)}r_{Y(j)}+\displaystyle \sum\limits_{j=1}^m\left(c_{Y(j)}-\overline{Y}\right)\left(c_{X(j)}-\overline{X}\right)}{\displaystyle \sum\limits_{j=1}^m\left(c_{X(j)}-\overline{X}\right)^2+\displaystyle \sum\limits_{j=1}^m\frac{1}{3}r^2_{X(j)}} \quad {\rm if} \quad \displaystyle\sum\limits_{j=1}^m\frac{1}{3}r_{X(j)}r_{Y(j)} < \displaystyle \sum\limits_{j=1}^m\left(c_{Y(j)}-\overline{Y}\right)c_{X(j)}\\
            \displaystyle\sum\limits_{j=1}^m\frac{1}{3}r_{X(j)}r_{Y(j)}\displaystyle \sum\limits_{j=1}^m\left(c_{X(j)}-\overline{X}\right)^2 \leq \displaystyle \sum\limits_{j=1}^m\left(c_{Y(j)}-\overline{Y}\right)c_{X(j)}\displaystyle \sum\limits_{j=1}^m\frac{1}{3}r^2_{X(j)}\\
            \gamma^{*}= \overline{Y}-\alpha^{*}\overline{X}\\
                 \end{array}\right.
            \]}

   \item[IV] Suppose  $\lambda_{1}=0$ and $\lambda_{2}=0.$ Then,

   \[\left\{
            \begin{array}{c}
              \frac{\partial f (b^{*})}{\partial \alpha}=0 \\
            \frac{\partial f (b^{*})}{\partial \beta}=0 \\
            \frac{\partial f (b^{*})}{\partial \gamma}=0 \\
            \end{array}\right.\]

 From this system we conclude that:

 {\scriptsize     \[ \left\{\begin{array}{l}
 \alpha^{*}= \frac{ \displaystyle \sum\limits_{j=1}^m\left(c_{Y(j)}-\overline{Y}\right)\left(c_{X(j)}-\overline{X}\right) \displaystyle \sum\limits_{j=1}^m\frac{1}{3}r^2_{X(j)}+ \displaystyle \sum\limits_{j=1}^m\frac{1}{3}r_{X(j)}r_{Y(j)} \displaystyle \sum\limits_{j=1}^m\left(c_{X(j)}-\overline{X}\right)^2}{2 \displaystyle \sum\limits_{j=1}^m\left(c_{X(j)}-\overline{X}\right)^2 \displaystyle \sum\limits_{j=1}^m\frac{1}{3}r^2_{X(j)}}\\
 \beta^{*}= \frac{ -\displaystyle \sum\limits_{j=1}^m\left(c_{Y(j)}-\overline{Y}\right)\left(c_{X(j)}-\overline{X}\right) \displaystyle \sum\limits_{j=1}^m\frac{1}{3}r^2_{X(j)}+ \displaystyle \sum\limits_{j=1}^m\frac{1}{3}r_{X(j)}r_{Y(j)} \displaystyle \sum\limits_{j=1}^m\left(c_{X(j)}-\overline{X}\right)^2}{2 \displaystyle \sum\limits_{j=1}^m\left(c_{X(j)}-\overline{X}\right)^2 \displaystyle \sum\limits_{j=1}^m\frac{1}{3}r^2_{X(j)}}\\
\gamma^{*}= \overline{Y}-\left(\alpha^{*}-\beta^{*}\right)\overline{X}\\
\end{array}\right.\]}

As in this case, $\alpha^{*} > 0$ and $\beta^{*} > 0,$ the expressions of $\alpha^{*}$ and  $\beta^{*}$ are non-negative only if
 $$\displaystyle\sum\limits_{j=1}^m\frac{1}{3}r_{X(j)}r_{Y(j)}\displaystyle \sum\limits_{j=1}^m\left(c_{X(j)}-\overline{X}\right)^2>\displaystyle \sum\limits_{j=1}^m\left(c_{Y(j)}-\overline{Y}\right)c_{X(j)}\displaystyle \sum\limits_{j=1}^m\frac{1}{3}r^2_{X(j)} \qquad \Box$$

\end{description}

\pagebreak

\newpage

\begin{landscape}

\section*{\large Appendix B: Simulation results of the study presented in \textit{Subsection \ref{s3.2}}.} \label{ap2}
\addcontentsline{toc}{chapter}{Appendix C: Simulation results of the study presented in \textit{Subsection \ref{s3.2}}.}

\vspace{2cm}
\renewcommand{\arraystretch}{1.5}
\begin{table}[h!]
\begin{center}
{\scriptsize
\begin{tabular}{p{2cm}|p{0.4cm}|c|c|c|c|c|c|c|c}
 \toprule
    &  & \multicolumn{2}{|c|}{$b=0$}& \multicolumn{2}{|c|}{$b \in \mathcal{U}(-2,2)$} & \multicolumn{2}{|c|}{$b \in \mathcal{U}(-5,5)$} & \multicolumn{2}{|c|}{$b \in \mathcal{U}(-10,10)$} \\ \hline
   & $n$ & $\overline{\Omega} \, (s)$ & $\overline{RMSE}_{M} \, (s)$ & $\overline{\Omega} \, (s)$ & $\overline{RMSE}_{M} \, (s) $ & $\overline{\Omega}\, (s)$&  $\overline{RMSE}_{M} \, (s)$ & $\overline{\Omega} \, (s)$ & $\overline{RMSE}_{M} \, (s)$   \\ \toprule
   \multirow{2}{2cm}{$a =0$} & 10 & 1.0000 (1.5773E-14) & 4.5604E-5 (5.6949E-19) & 0.9858 (0.0045) & 0.6252 (0.1022) & 0.9162 (0.0287) & 1.5598 (0.2592) & 0.8381 (0.0577) & 2.2502 (0.3753) \\
   & 100 & 1.0000 (2.3326E-14) & 2.8418E-6 (6.7373E-20) & 0.9846 (0.0015)& 0.6625 (0.0304) & 0.9112 (0.0089) & 1.6549 (0.0732) & 0.8504 (0.0154) & 2.2246 (0.0973)\\ \hline
  \multirow{2}{2cm}{$a \in \mathcal{U}(-2,2)$} & 10 &  0.9640 (0.0126) & 1.0060 (0.1906) & 0.9505 (0.0135) & 1.1953 (0.1673) & 0.8847 (0.0314) & 1.8760 (0.2246) & 0.8131 (0.0563) & 2.4888 (0.3456)\\
   & 100 & 0.9557 (0.0039) & 1.1426 (0.0534) & 0.9417 (0.0041)& 1.3212 (0.0467) & 0.8740 (0.0104)& 2.0133 (0.0691) & 0.8181 (0.0161) & 2.4973 (0.0906) \\ \hline
   \multirow{2}{2cm}{$a \in \mathcal{U}(-5,5)$} & 10 & 0.8144 (0.0565) & 2.5400 (0.4815) & 0.8046 (0.0570) & 2.6179 (0.4631) & 0.7582 (0.0621) & 2.9846 (0.4261)&0.7035 (0.0824) & 3.4178 (0.4423) \\
   & 100 & 0.7764 (0.0159) & 2.8529 (0.1294) & 0.7666 (0.0164)&2.9330 (0.1283) & 0.7205 (0.0192) & 3.3086 (0.1194) & 0.6828 (0.0232) & 3.6201 (0.1208) \\
\end{tabular}}
\caption{Results of the \textit{DSD Model} with $\alpha=2$ $\beta=1$ $\gamma=-1$ for the case of low variability in the intervals of variable $X$. }
\label{table1SA2}
\end{center}
\end{table}

\vspace{1cm}

\begin{table}[h!]
\begin{center}
{\scriptsize
\begin{tabular}{p{2cm}|p{0.4cm}|c|c|c|c|c|c|c|c}
  \toprule
    &  & \multicolumn{2}{|c|}{$b=0$}& \multicolumn{2}{|c|}{$b \in \mathcal{U}(-2,2)$} & \multicolumn{2}{|c|}{$b \in \mathcal{U}(-5,5)$} & \multicolumn{2}{|c|}{$b \in \mathcal{U}(-10,10)$} \\ \hline
   & $n$ & $\overline{\Omega} \, (s)$ & $\overline{RMSE}_{M} \, (s)$ & $\overline{\Omega} \, (s)$ & $\overline{RMSE}_{M} \, (s) $ & $\overline{\Omega}\, (s)$&  $\overline{RMSE}_{M} \, (s)$ & $\overline{\Omega} \, (s)$ & $\overline{RMSE}_{M} \, (s)$   \\ \toprule
   \multirow{2}{2cm}{$a =0$} & 10 & 1.0000 (2.1549E-14) & 2.4683E-6 (3.5169E-20) & 0.9967 (0.0011) & 0.6210 (0.1058) & 0.9792 (0.0067) &1.5732(0.2606)  & 0.9229 (0.0260) & 3.1039 (0.5143) \\
   & 100 & 1.0000 (1.7217E-14) & 9.5283E-8 (3.9725E-22) & 0.9963 (3.4504E-4) & 0.6635 (0.0308) &0.9773 (0.0021) & 1.6556 (0.0764) & 0.9144 (0.0887) & 3.3198 (0.1501) \\ \hline
  \multirow{2}{2cm}{$a \in \mathcal{U}(-5,5)$} & 10 &   0.9451 (0.0178) & 2.6139 (0.4752) & 0.9425 (0.0171) & 2.6838 (0.4465) &0.9218 (0.0194) & 3.0767 (0.4260) & 0.8757 (0.0322)& 4.0883 (0.4998) \\
   & 100 & 0.9352 (0.0053) & 2.8598 (0.1269) & 0.9316 (0.0053) & 2.9439 (0.1225) & 0.9150 (0.0059) & 3.3104 (0.1146) & 0.8604 (0.0107) & 4.3762 (0.1438)  \\ \hline
   \multirow{2}{2cm}{$a \in \mathcal{U}(-10,10)$} & 10 & 0.8156 (0.0521) & 5.2187 (0.9551) & 0.8101 (0.0519) & 5.3306 (0.9411) & 0.8014 (0.0520) & 5.4522 (0.8819)   & 0.7582 (0.0595) & 6.1447 (0.8242) \\
   & 100 & 0.7824 (0.0154) & 5.7397 (0.2623) & 0.7809 (0.0158) & 5.7613 (0.2598) &0.7689 (0.0159) & 5.9653 (0.2517) & 0.7298 (0.0195) & 6.6188 (0.2432) \\
\end{tabular}}
\caption{Results of the \textit{DSD Model} with $\alpha=6$ $\beta=0$ $\gamma=2$  for the case of low variability in the intervals of variable $X$.  }
\label{table2SA2}
\end{center}
\end{table}

\begin{table}[h!]
\begin{center}
{\scriptsize
\begin{tabular}{p{2cm}|p{0.4cm}|c|c|c|c|c|c|c|c}
  \toprule
    &  & \multicolumn{2}{|c|}{$b=0$}& \multicolumn{2}{|c|}{$b \in \mathcal{U}(-5,5)$} & \multicolumn{2}{|c|}{$b \in \mathcal{U}(-10,10)$} & \multicolumn{2}{|c|}{$b \in \mathcal{U}(-20,20)$} \\ \hline
   & $n$ & $\overline{\Omega} \, (s)$ & $\overline{RMSE}_{M} \, (s)$ & $\overline{\Omega} \, (s)$ & $\overline{RMSE}_{M} \, (s) $ & $\overline{\Omega}\, (s)$&  $\overline{RMSE}_{M} \, (s)$ & $\overline{\Omega} \, (s)$ & $\overline{RMSE}_{M} \, (s)$   \\ \toprule
   \multirow{2}{2cm}{$a =0$} & 10 & 1.0000 (9.9970E-16) & 1.3681E-4 (1.3017E-18) & 0.9922 (0.0025) & 1.5559 (0.2568) & 0.9693 (0.0099)& 3.1197 (0.5075)&  0.8878 (0.0381)& 6.1721 (1.0576)\\
   & 100 & 1.0000 (1.5218E-14) & 1.3870E-5 (1.8814E-19) & 0.9914 (7.6563E-4) & 1.6598 (0.0727) &0.9665 (0.0032)& 3.3110 (0.1518) & 0.8782 (0.0123) & 6.6258 (0.2967) \\ \hline
  \multirow{2}{2cm}{$a \in \mathcal{U}(-5,5)$} & 10 &  0.9794 (0.0074) & 2.5396 (0.4899) & 0.9718 (0.0077) & 3.0010 (0.4137) & 0.9495 (0.0131) & 4.0508 (0.5214) & 0.8692 (0.0379) & 6.7761 (0.9245) \\
   & 100 & 0.9750 (0.0022) & 2.8527 (0.1287) & 0.9668 (0.0023)& 3.3016 (0.1173) & 0.9433 (0.0041) & 4.3675 (0.1443) & 0.8581 (0.0133) & 7.2226 (0.2862) \\ \hline
   \multirow{2}{2cm}{$a \in \mathcal{U}(-10,10)$} & 10 & 0.9229 (0.0248) & 5.1006 (0.9176) & 0.9162 (0.0264) & 5.3303 (0.9299) & 0.8970 (0.0272)& 5.9936 (0.8222) & 0.8254 (0.0462) & 8.0706 (0.9828)  \\
   & 100 & 0.9068 (0.0076) & 5.7126 (0.2587) & 0.8995 (0.0081)& 5.9558 (0.2553) &0.8789 (0.0086) & 6.6070 (0.2374) & 0.8055 (0.0157)& 8.7512 (0.2833) \\
\end{tabular}}
\caption{Results of the \textit{DSD Model} with $\alpha=2$ $\beta=8$ $\gamma=3$ for the case of low variability in the intervals of variable $X$. }
\label{table3SA2}
\end{center}
\end{table}

\vspace{1cm}

\begin{table}[h!]
\begin{center}
{\scriptsize
\begin{tabular}{p{2cm}|p{0.4cm}|c|c|c|c|c|c|c|c}
  \toprule
    &  & \multicolumn{2}{|c|}{$b=0$}& \multicolumn{2}{|c|}{$b \in \mathcal{U}(-10,10)$} & \multicolumn{2}{|c|}{$b \in \mathcal{U}(-20,20)$} & \multicolumn{2}{|c|}{$b \in \mathcal{U}(-40,40)$} \\ \hline
   & $n$ & $\overline{\Omega} \, (s)$ & $\overline{RMSE}_{M} \, (s)$ & $\overline{\Omega} \, (s)$ & $\overline{RMSE}_{M} \, (s) $ & $\overline{\Omega}\, (s)$&  $\overline{RMSE}_{M} \, (s)$ & $\overline{\Omega} \, (s)$ & $\overline{RMSE}_{M} \, (s)$   \\ \toprule
   \multirow{2}{2cm}{$a =0$} & 10 & 1.0000 (2.549E-14) & 5.0256E-6 (8.7288E-20)  & 0.9855 (0.0048) & 3.0977 (0.5255) & 0.9439 (0.0190) & 6.1711 (1.0613)  & 0.8023 (0.0659)& 12.4165 (1.9913)  \\
   & 100 & 1.0000 (3.3323E-15) & 1.4212E-6 (1.1017E-20) & 0.9838 (0.0015)& 3.3177 (0.1551) & 0.9384 (0.0058) & 6.6285 (0.2955) & 0.7910 (0.0228)& 13.2528 (0.6163) \\ \hline
  \multirow{2}{2cm}{$a \in \mathcal{U}(-10,10)$} & 10 &  0.9615 (0.0127) & 5.1233 (0.9187)  & 0.9472 (0.0134) & 6.0701 (0.7913)& 0.9086 (0.0233) & 8.1289 (0.9790)& 0.7820 (0.0714) & 13.5523 (1.9721)\\
   & 100 & 0.9535 (0.0040) & 5.7202 (0.2606) & 0.9387 (0.0044) & 6.6179 (0.2392) & 0.8971 (0.0077) & 8.7618 (0.2866) & 0.7611 (0.0226)& 14.4461 (0.5581) \\ \hline
   \multirow{2}{2cm}{$a \in \mathcal{U}(-20,20)$} & 10 & 0.8604 (0.0402) & 10.4207 (1.8047) &0.8481 (0.0416) & 10.9671 (1.7686) & 0.8166 (0.0474) & 12.2167 (1.6930) &0.7085 (0.0826)& 16.37308 (1.9637) \\
   & 100 & 0.8373 (0.0127) & 11.4216 (0.5348) & 0.8260 (0.0123) & 11.8970 (0.4823) & 0.7933 (0.0141) & 13.2180 (0.4709) & 0.6848 (0.0266) & 17.5246 (0.5641)\\
\end{tabular}}
\caption{Results of the \textit{DSD Model} with $\alpha=2$ $\beta=1$ $\gamma=-1$ for the case of high variability  in the intervals of variable $X$.  }
\label{table4SA2}
\end{center}
\end{table}

\vspace{1cm}

\begin{table}[h!]
\begin{center}
{\scriptsize
\begin{tabular}{p{2cm}|p{0.4cm}|c|c|c|c|c|c|c|c}
  \toprule
    &  & \multicolumn{2}{|c|}{$b=0$}& \multicolumn{2}{|c|}{$b \in \mathcal{U}(-20,20)$} & \multicolumn{2}{|c|}{$b \in \mathcal{U}(-40,40)$} & \multicolumn{2}{|c|}{$b \in \mathcal{U}(-80,80)$} \\ \hline
   & $n$ & $\overline{\Omega} \, (s)$ & $\overline{RMSE}_{M} \, (s)$ & $\overline{\Omega} \, (s)$ & $\overline{RMSE}_{M} \, (s) $ & $\overline{\Omega}\, (s)$&  $\overline{RMSE}_{M} \, (s)$ & $\overline{\Omega} \, (s)$ & $\overline{RMSE}_{M} \, (s)$   \\ \toprule
   \multirow{2}{2cm}{$a =0$} & 10 & 1.0000 (7.1090E-15) & 3.5413E-5 (6.7797E-20) & 0.9854 (0.0046) & 6.2264 (1.0222)   & 0.9436 (0.0177) & 12.4846 (2.0005) &0.8038 (0.0713) & 24.8492 (4.2778)\\
   & 100 & 1.0000 (1.7550E-14) & 2.8423E-6 (3.8559E-20) & 0.9839 (0.0015) & 6.6174 (0.3021) & 0.9385 (0.0060)& 13.2516 (0.6071) & 0.7914 (0.0227)& 26.4916 (1.2190) \\ \hline
  \multirow{2}{2cm}{$a \in \mathcal{U}(-20,20)$} & 10 &  0.9593 (0.0127) & 10.5898 (1.8003) & 0.9452 (0.0144) & 12.3940 (1.6926)& 0.9066 (0.0242) & 16.4188 (2.0539)  & 0.7760 (0.0718) & 27.0897 (3.8528)\\
   & 100 & 0.9536 (0.0041) & 11.4310 (0.5313) & 0.9390 (0.0044) & 13.2119 (0.4791) & 0.8974 (0.0076)& 17.5174 (0.5465) & 0.7627 (0.0227)& 28.8749 (1.1240)  \\ \hline
   \multirow{2}{2cm}{$a \in \mathcal{U}(-40,40)$} & 10 & 0.8566 (0.0408) & 21.2015 (3.6407) & 0.8467 (0.0399) & 22.0670 (3.3662)   & 0.8159 (0.0455) & 24.5339 (3.2660)&0.7034 (0.0794) & 33.1065 (3.6605) \\
   & 100 &0.8361 (0.0123) & 22.9573 (1.0286) & 0.8251 (0.0125)&23.8750 (1.0036) & 0.7932 (0.0147) & 26.4868 (0.9985) & 0.6869 (0.0253) & 35.0127 (1.0882)  \\
\end{tabular}}
\caption{Results of the \textit{DSD Model} with $\alpha=6$ $\beta=0$ $\gamma=2$ for the case of high variability in the intervals of variable $X$. }
\label{table5SA2}
\end{center}
\end{table}

\vspace{1cm}

\begin{table}[h!]
\begin{center}
{\scriptsize
\begin{tabular}{p{2cm}|p{0.4cm}|c|c|c|c|c|c|c|c}
  \toprule
    &  & \multicolumn{2}{|c|}{$b=0$}& \multicolumn{2}{|c|}{$b \in \mathcal{U}(-40,40)$} & \multicolumn{2}{|c|}{$b \in \mathcal{U}(-80,80)$} & \multicolumn{2}{|c|}{$b \in \mathcal{U}(-120,120)$} \\ \hline
   & $n$ & $\overline{\Omega} \, (s)$ & $\overline{RMSE}_{M} \, (s)$ & $\overline{\Omega} \, (s)$ & $\overline{RMSE}_{M} \, (s) $ & $\overline{\Omega}\, (s)$&  $\overline{RMSE}_{M} \, (s)$ & $\overline{\Omega} \, (s)$ & $\overline{RMSE}_{M} \, (s)$   \\ \toprule
   \multirow{2}{2cm}{$a =0$} & 10 & 1.0000 (1.8439E-14) & 5.6135E-5 (1.1119E-18) & 0.9791 (0.0068)& 12.4398 (2.0347) & 0.9197 (0.0270) & 24.8927 (4.3585) &  0.8341 (0.0580) & 37.5643 (6.0387) \\
   & 100 & 1.0000 (1.3107E-14) & 9.3047E-6 (9.6610E-20) &  0.9770 (0.0021) & 13.2434 (0.5995) & 0.9134 (0.0085)& 26.5086 (1.1887) & 0.8242 (0.0186)& 39.7911 (1.7756) \\ \hline
  \multirow{2}{2cm}{$a \in \mathcal{U}(-20,20)$} & 10 &  0.9859 (0.0050)& 10.1933 (1.9191) &   0.9657 (0.0084) & 16.1862 (1.9646)  & 0.9060 (0.0274) & 27.2311 (3.8488) & 0.8253 (0.0591) & 38.4246 (5.9630)\\
   & 100 & 0.9827 (0.0016) & 11.4494 (0.5259) & 0.9605 (0.0027) & 17.4940 (0.5711) & 0.8994 (0.0091) & 28.8380 (1.1226) & 0.8131 (0.0190)& 41.2507 (1.6972)  \\ \hline
   \multirow{2}{2cm}{$a \in \mathcal{U}(-40,40)$} & 10 & 0.9438 (0.0176)& 20.8909 (3.6073)&  0.9261 (0.0190) & 24.2228 (3.2109) & 0.8725 (0.0340) & 32.5674 (3.9407)  & 0.7930 (0.0631) & 43.1086 (5.6312)\\
   & 100 & 0.9348 (0.0054) & 22.8082 (1.0105) & 0.9144 (0.0065) & 26.4097 (1.0093) & 0.8584 (0.0108) & 35.0370 (1.1680) & 0.7801 (0.0204)& 45.7994 (1.6846) \\
\end{tabular}}
\caption{Results of the \textit{DSD Model} with $\alpha=2$ $\beta=8$ $\gamma=3$ for the case of high variability in the intervals of $X$. }
\label{table6SA2}
\end{center}
\end{table}

\begin{table}[h!]
\begin{center}
{\scriptsize
\begin{tabular}{p{2cm}|p{0.4cm}|c|c|c|c|c|c}
 \toprule
    &  & \multicolumn{2}{|c|}{$b=0$}& \multicolumn{2}{|c|}{$b \in \mathcal{U}(-1,1)$} & \multicolumn{2}{|c|}{$b \in \mathcal{U}(-2,2)$} \\ \hline
   & $n$ & $\overline{\Omega} \, (s)$ & $\overline{RMSE}_{M} \, (s)$ & $\overline{\Omega} \, (s)$ & $\overline{RMSE}_{M} \, (s) $ & $\overline{\Omega}\, (s)$&  $\overline{RMSE}_{M} \, (s)$   \\ \toprule
   \multirow{2}{2cm}{$a =0$} & 10 & 1.0000 (2.1771E-14) & 3.4708E-6 (7.0339E-20) & 0.9997 (0.0001) & 0.3147 0.0522 & 0.9992 (0.0003) & 0.5365 (0.0873)   \\
   & 100 & 1.0000 (1.3218E-14) & 3.7144E-7 (6.1970E-21)& 0.9998 (2.1279E-5) & 0.3317 (0.015)& 0.9995 (4.6331E-5) & 0.4780 (0.0226) \\ \hline
  \multirow{2}{2cm}{$a \in \mathcal{U}(-5,5)$} & 10 & 0.9814 (0.0066) & 2.5499 (0.4793) & 0.9813 (0.0065) & 2.5582 (0.4769) & 0.9806 (0.0063) & 2.6067 (0.4400) \\
   & 100 & 0.9828 (0.0015) & 2.8522 (0.1281) & 0.9825 (0.0016) & 2.8754 (0.1341) & 0.9823 (0.0016) & 2.8911 (0.1308)\\ \hline
   \multirow{2}{2cm}{$a \in \mathcal{U}(-10,10)$} & 10 & 0.9303 (0.0238) &  5.0894 (0.9514) & 0.9292 (0.024) & 5.1327 (0.9787) & 0.9298 (0.0239) & 5.1025 (0.9632)  \\
   & 100 & 0.9345 (0.0057) & 5.7116 (0.2626) & 0.9341 (0.0057) & 5.7331 (0.2622) & 0.9339 (0.0056) & 5.738 (0.2590) \\ \hline
  \multirow{2}{2cm}{$a\in \mathcal{U}(-20,20)$} & 10 & 0.7789 (0.0670) & 10.1157 (1.9085) & 0.7766 (0.0665) & 10.1611 (1.9311) &  0.7728 (0.0645) & 10.3056 (1.8866)  \\
   & 100 & 0.7815 (0.0162) & 11.4235 (0.5251)& 0.7812 (0.0161) & 11.4383 (0.5278) & 0.7813 (0.0161) & 11.4286 (0.5256) \\ \hline
   \multirow{2}{2cm}{$a \in \mathcal{U}(-40,40)$} & 10 & 0.4945 (0.1034) & 20.4073 (3.5533) & 0.5001 (0.1108) & 20.1758 (3.7656) & 0.5026 (0.1172) & 20.1149 (3.9868)  \\
   & 100 & 0.4755 (0.0270) & 22.8224 (1.0857) & 0.4747 (0.0243) & 22.8285 (0.9893) & 0.4735 (0.0255) & 22.8831 (1.0547) \\ \toprule
\end{tabular}}
\caption{Results of the \textit{DSD Model} with $\alpha=2$ $\beta=1$ $\gamma=-1$ for the case of a mixture of intervals in $X$ with variable half ranges.}
\label{table7SA2}
\end{center}
\end{table}

\vspace{1cm}

\begin{table}[h!]
\begin{center}
{\scriptsize
\begin{tabular}{p{2cm}|p{0.4cm}|c|c|c|c|c|c}
  \toprule
    &  & \multicolumn{2}{|c|}{$b=0$}& \multicolumn{2}{|c|}{$b \in \mathcal{U}(-2,2)$} & \multicolumn{2}{|c|}{$b \in \mathcal{U}(-4,4)$} \\ \hline
   & $n$ & $\overline{\Omega} \, (s)$ & $\overline{RMSE}_{M} \, (s)$ & $\overline{\Omega} \, (s)$ & $\overline{RMSE}_{M} \, (s) $ & $\overline{\Omega}\, (s)$&  $\overline{RMSE}_{M} \, (s)$   \\ \toprule
   \multirow{2}{2cm}{$a =0$} & 10 & 1.0000 (2.0216E-14) & 1.0259E-5 (1.8475E-19)  &  0.9998 (0.0001) & 0.6344 (0.1000) & 0.9994 (0.0002) & 1.0838 (0.1725) \\
   & 100 & 1.0000 (2.1771E-14) & 4.8294E-7 (4.3432E-21) & 0.9998 (1.5717E-5) & 0.6627 (0.0297) & 0.9996 (3.1573E-5) & 0.9591 (0.0413)\\ \hline
  \multirow{2}{2cm}{$a \in \mathcal{U}(-20,20)$} & 10 & 0.9502 (0.0166) & 10.3065 (1.9296) & 0.9491 (0.0169) & 10.4514 (1.9103) & 0.9504 (0.0148) & 10.3518 (1.7058) \\
   & 100 & 0.9502 (0.0044) & 11.4452 (0.5244) & 0.9502 (0.0045) & 11.4421 (0.5309) & 0.9498 (0.0043) & 11.4967 (0.5095)\\ \hline
   \multirow{2}{2cm}{$a \in \mathcal{U}(-40,40)$} & 10 & 0.8277 (0.0510) & 20.7545 (3.6153) & 0.8247 (0.0513) & 20.9247 (3.6069) &  0.8275 (0.0495) & 20.7101 (3.5881) \\
   & 100 & 0.8274 (0.0145) & 22.8651 (1.0635) & 0.8275 (0.0147) & 22.8731 (1.0626) & 0.8275 (0.0143) & 22.8633 (1.0365)\\ \hline
  \multirow{2}{2cm}{$a\in \mathcal{U}(-80,80)$} & 10 &  0.5584 (0.1084) & 41.4547 (7.4647) & 0.5600 (0.1044) & 41.4495 (7.4732) &  0.5579 (0.1063) & 41.2382 (7.3919) \\
   & 100 & 0.5448 (0.0319) & 45.7474 (2.1530) & 0.5447 (0.0311) & 45.7854 (2.0515) & 0.5456 (0.0322) & 45.7837 (2.1359)\\ \hline
   \multirow{2}{2cm}{$a \in \mathcal{U}(-120,120)$} & 10 & 0.3749 (0.1164) & 62.5107 (10.9749) & 0.3770 (0.1182) & 62.1691 (11.0598) &  0.3836 (0.1208) & 61.9878 (11.3465)\\
   & 100 & 0.3494 (0.0378) & 68.6348 (3.2748) & 0.3490 (0.0373) & 68.6286 (3.1566) & 0.3475 (0.0364) & 68.6923 (3.0647)\\ \toprule
\end{tabular}}
\caption{Results of the \textit{DSD Model} with $\alpha=6$ $\beta=0$ $\gamma=2$ for the case of a mixture of intervals in $X$ with variable half ranges.}
\label{table8SA2}
\end{center}
\end{table}

\begin{table}[h!]
\begin{center}
{\scriptsize
\begin{tabular}{p{2cm}|p{0.4cm}|c|c|c|c|c|c}
\toprule
    &  & \multicolumn{2}{|c|}{$b=0$}& \multicolumn{2}{|c|}{$b \in \mathcal{U}(-3,3)$} & \multicolumn{2}{|c|}{$b \in \mathcal{U}(-6,6)$} \\ \hline
   & $n$ & $\overline{\Omega} \, (s)$ & $\overline{RMSE}_{M} \, (s)$ & $\overline{\Omega} \, (s)$ & $\overline{RMSE}_{M} \, (s) $ & $\overline{\Omega}\, (s)$&  $\overline{RMSE}_{M} \, (s)$   \\ \toprule
   \multirow{2}{2cm}{$a =0$} & 10 & 1.0000 (2.0660E-14) &  6.4510E-6 (1.2373E-19)& 0.9998 (0.0001) & 0.9433 (0.1518) & 0.9993 (0.0002) & 1.7813 (0.2949) \\
   & 100 & 1.0000 (2.6881E-14) & 1.4868E-6 (1.3347E-20) & 0.9998 (1.5744E-5) & 0.9965 (0.0449) & 0.9996 (3.9619E-05) & 1.5916 (0.0707)\\ \hline
  \multirow{2}{2cm}{$a \in \mathcal{U}(-20,20)$} & 10 & 0.9771 (0.0079) & 10.0727 (1.8686) & 0.9765 (0.0083) & 10.2002 (1.8992) &  0.9761 (0.0083) & 10.2856 (1.8941) \\
   & 100 & 0.9775 (0.0020) & 11.4170 (0.5217) & 0.9774 (0.0020) & 11.4554 (0.5212) & 0.9771 (0.0020) & 11.5137 (0.5241)\\ \hline
   \multirow{2}{2cm}{$a \in \mathcal{U}(-40,40)$} & 10 & 0.9145 (0.0296) & 20.1803 (3.8982) & 0.9133 (0.0290) & 20.3594 (3.8103) &  0.9136 (0.0294) & 20.2686 (3.8420)\\
   & 100 & 0.9158 (0.0073) & 22.8335 (1.0420) & 0.9153 (0.0070) & 22.9138 (1.0009) & 0.9151 (0.0070) & 22.9342 (1.0078)\\ \hline
  \multirow{2}{2cm}{$a\in \mathcal{U}(-80,80)$} & 10 & 0.7354 (0.0779) & 40.4491 (7.6280) & 0.7301 (0.0765) & 40.9811 (7.3988) & 0.7311 (0.0791) & 40.7989 (7.5507) \\
   & 100 & 0.7318 (0.0199) & 45.6359 (2.0838)& 0.7304 (0.0198) & 45.7955 (2.1034)& 0.7310 (0.0204) & 45.7559 (2.1003)\\ \hline
   \multirow{2}{2cm}{$a \in \mathcal{U}(-160,160)$} & 10 & 0.4400 (0.1200) & 81.4531 (15.2694) & 0.4312 (0.1135) & 82.7159 (14.8059) &  0.4379 (0.1166) & 81.8070 (14.7164)  \\
   & 100 &0.4100 (0.0320) & 91.0520 (4.4318) & 0.4074 (0.0305) & 91.3262 (4.1702) & 0.4096 (0.0316) & 91.4414 (4.2867)\\ \toprule
\end{tabular}}
\caption{Results of the \textit{DSD Model} with $\alpha=2$ $\beta=8$ $\gamma=3$ for the case of a mixture of intervals in $X$ with variable half ranges.}
\label{table9SA2}
\end{center}
\end{table}

\end{landscape}

\newpage

\begin{landscape}

\section*{\large Appendix C: Simulation results of the study presented in \textit{Subsection \ref{s3.1.1}}.} \label{ap3}
\addcontentsline{toc}{chapter}{Appendix C: Simulation results of the study presented in \textit{Subsection \ref{s3.1.1}}.}

\vspace{1cm}

\renewcommand{\arraystretch}{1.25}
\begin{table}[h!]
\begin{center}
{\scriptsize
\begin{tabular}{p{1.3cm}|p{1cm}|c|c|c|c|c|c|c|c|c|c|c}
\toprule
   \multirow{2} {1.3cm} {\textit{Degree of variability}} & \multirow{2} {1cm} {\textit{Degree of linearity }} & \multirow{2} {*} {$n$} & \multicolumn{6}{|c|}{\textit{Parameters estimated}} & \multicolumn{4}{|c|}{\textit{Goodness of fit measures}} \\ \cline{4-13}
      & & & $\overline{\widehat{\alpha}} \, (s)$ & MSE($\alpha$)  & $\overline{\widehat{\beta}} \, (s)$  & MSE($\beta$)&  $\overline{\widehat{\gamma}} \, (s)$ & MSE($\gamma$)&  $\overline{\Omega} \, (s)$ & $\overline{RMSE}_{M} \, (s)$ & $\overline{RMSE}_{L} \, (s)$ & $\overline{RMSE}_{U} \, (s)$ \\ \toprule
\multirow{12} {1.1cm}{Low variability} & \multirow{4} {1cm}{High linearity} & 10 & 1.9948 (0.2254) &  0.0508 & 1.0049 (0.2251) & 0.0506 & -0.9794 (0.8684) & 0.7538 &  0.9796 (0.0061) & 0.7525 (0.1161) & 0.8435 (0.1801) & 0.8416 (0.1760) \\
                                   &                                 & 30 & 2.0013 (0.2105) & 0.0443 & 0.9984 (0.2106) & 0.0443 & -1.0095 (0.8727) & 0.7609 & 0.9763 (0.0037) & 0.8202 (0.0658) & 0.9089 (0.0970) & 0.8988 (0.0985) \\
                                   &                                 & 100 & 1.9976 (0.1015) & 0.0103 & 1.0026 (0.1015) & 0.0103 & -0.9885 (0.4122) & 0.1698 & 0.9736 (0.0022) & 0.8728 (0.0366) & 0.9583 (0.0536) & 0.9542 (0.0549) \\
                                   &                                 & 250 & 2.0027 (0.0679) & 0.0046 & 0.9975 (0.0681) & 0.0046 & -1.0106 (0.2760) & 0.0762 & 0.9729 ( 0.0013) & 0.8813  (0.0221) & 0.9567 (0.0328) & 0.9552 (0.0329) \\ \cline{2-13}
                                   & \multirow{4} {1cm} {Low linearity}
                                                                        & 10 & 1.7733 (1.2370) & 1.5801 & 1.3057 (1.2318) & 1.6093 & -0.0553 (5.0654) & 26.5254 & 0.4544 (0.1056) & 6.0933 (0.9154) & 6.8473 (1.4886) & 6.8037 (1.3653) \\
\end{tabular}}
\caption{Results of the \textit{DSD Model} with $\alpha=2$ $\beta=1$ $\gamma=-1$ in different conditions.}
\label{table1SA3}
\end{center}
\end{table}

\renewcommand{\arraystretch}{1.25}
\begin{table}[h!]
\begin{center}
{\scriptsize
\begin{tabular}{p{1.3cm}|p{1cm}|c|c|c|c|c|c|c|c|c|c|c}
  \toprule
   \multirow{2} {1.3cm} {\textit{Degree of variability}} & \multirow{2} {1cm} {\textit{Degree of linearity }} & \multirow{2} {*} {$n$} & \multicolumn{6}{|c|}{\textit{Parameters estimated}} & \multicolumn{4}{|c|}{\textit{Goodness of fit measures}} \\ \cline{4-13}
      & & & $\overline{\widehat{\alpha}} \, (s)$ & MSE($\alpha$)  & $\overline{\widehat{\beta}} \, (s)$  & MSE($\beta$)&  $\overline{\widehat{\gamma}} \, (s)$ & MSE($\gamma$)&  $\overline{\Omega} \, (s)$ & $\overline{RMSE}_{M} \, (s)$ & $\overline{RMSE}_{L} \, (s)$ & $\overline{RMSE}_{U} \, (s)$ \\ \toprule
\multirow{12} {1.1cm}{Low variability} & \multirow{4} {1cm}{High linearity} & 10 & 2.0008 (0.8607) & 0.7401 & 7.9985 (0.8668) & 0.7506 & 2.9979 (3.3514) & 11.2207 & 0.9800 (0.0062) & 2.5118 (0.4063) & 2.8222 (0.6041) & 2.8179 (0.6030)\\
                                   &                                 & 30 & 1.9837 (0.6671) & 0.4448 & 8.0138 (0.6676) & 0.4454 & 3.0667 (2.7846) & 7.7504 & 0.9758 (0.0037) & 2.7783 (0.2210) & 3.0555 (0.3222) & 3.0527 (0.3190)\\
                                   &                                 & 100 & 2.0017 (0.3340) & 0.1114 & 7.9978 (0.3359) & 0.1127 & 2.9925 (1.3601) & 1.8481 & 0.9738 (0.0022) & 2.9180 (0.1244) & 3.1892 (0.1787) & 3.2028 (0.1898)\\
                                   &                                 & 250 &  2.0010 (0.2191) & 0.0480 & 8.0005 (0.2209) & 0.04870 & 3.0010 (0.8959) & 0.8018 & 0.9731 (0.0013) & 2.9446 (0.0739) & 3.1900 (0.1087) & 3.1952 (0.1089)\\ \cline{2-13}
                                   & \multirow{4} {1cm} {Low linearity}
                                                                        & 10 & 3.2795 (3.7377) & 15.5933 & 6.9963 (4.0266) & 17.2044 & -1.0177 (15.7588) & 264.2341 & 0.4430 (0.1078) & 20.8759 (3.2957) & 23.3040 (4.8818) & 23.2727 (4.8518)\\
\end{tabular}}
\caption{Results of the \textit{DSD Model} with $\alpha=2$ $\beta=8$ $\gamma=3$ in different conditions.}
\label{table2SA3}
\end{center}
\end{table}

\renewcommand{\arraystretch}{1.25}
\begin{table}[h!]
\begin{center}
{\scriptsize
\begin{tabular}{p{1.3cm}|p{1cm}|c|c|c|c|c|c|c|c|c|c|c}
\toprule
   \multirow{2} {1.3cm} {\textit{Degree of variability}} & \multirow{2} {1cm} {\textit{Degree of linearity }} & \multirow{2} {*} {$n$} & \multicolumn{6}{|c|}{\textit{Parameters estimated}} & \multicolumn{4}{|c|}{\textit{Goodness of fit measures}} \\ \cline{4-13}
      & & & $\overline{\widehat{\alpha}} \, (s)$ & MSE($\alpha$)  & $\overline{\widehat{\beta}} \, (s)$  & MSE($\beta$)&  $\overline{\widehat{\gamma}} \, (s)$ & MSE($\gamma$)&  $\overline{\Omega} \, (s)$ & $\overline{RMSE}_{M} \, (s)$ & $\overline{RMSE}_{L} \, (s)$ & $\overline{RMSE}_{U} \, (s)$ \\ \toprule
\multirow{12} {1.1cm}{Low variability} & \multirow{4} {1cm}{High linearity} & 10 & 5.8225 (0.3250) & 0.1371 & 0.2093 (0.2937) & 0.1300 & 2.7258 (1.2647) & 2.1246 & 0.9776 (0.0063) & 1.6442 (0.2471) & 1.8256 (0.3754) & 1.8139 (0.3507)\\
                                   &                                 & 30 & 5.8499 (0.2552) & 0.0876 & 0.1643 (0.2431) & 0.0860 & 2.6382 (1.0667) & 1.5439 & 0.9747 (0.0039) & 1.7263 (0.1370) & 1.8920 (0.1987) & 1.8849 (0.2031)\\
                                   &                                 & 100 & 5.9278 (0.1273) & 0.0214 & 0.0801 (0.1184) & 0.0204 & 2.2963 (0.5193) & 0.3572 & 0.9733 (0.0021) & 1.8011 (0.0736) & 1.9667 (0.1082) & 1.9616 (0.1094) \\
                                   &                                 & 250 & 5.9539 (0.0820) & 0.0088 & 0.0517 (0.0759) & 0.0084 & 2.1958 (0.3307) & 0.1476 & 0.9729 (0.0014) & 1.8027 (0.0462) & 1.9507 (0.0667) & 1.9480 (0.0678) \\ \cline{2-13}
                                   & \multirow{4} {1cm} {Low linearity}
                                                                        & 10 & 4.7773 (2.3476) & 7.0008 & 1.5180 (2.1009) & 6.7135 & 7.3110 (9.3862) & 116.2198 & 0.4336 (0.1030) & 13.1140 (2.1595) & 14.5021 (3.0595) & 14.5375 (3.0036)\\
\end{tabular}}
\caption{Results of the \textit{DSD Model} with $\alpha=6$ $\beta=0$ $\gamma=2$ in different conditions.}
\label{table3SA3}
\end{center}
\end{table}

\end{landscape}

\newpage

\begin{landscape}
\vspace{10cm}
\renewcommand{\arraystretch}{1.2}
\begin{table}
\begin{center}
{\scriptsize
\begin{tabular}{p{1.2cm}|p{1cm}|p{0.25cm}|c|@{}c@{}|c|@{}c@{}|c|@{}c@{}|c|@{}c@{}|c|@{}c@{}|c|@{}c@{}}
\toprule
   \multirow{2} {1.2cm} {\textit{Degree of variability}}  & \multirow{2} {1cm} {\textit{Degree of linearity }} & \multirow{2} {0.25cm} {$n$}&  \multicolumn{12}{|c|}{\textit{Parameters estimated}} \\ \cline{4-15}
      & & & $\overline{\widehat{\alpha_1}} \, (s)$ & MSE($\alpha_1$)  & $\overline{\widehat{\beta_1}} \, (s)$  & MSE($\beta_1$)& $\overline{\widehat{\alpha_2}} \, (s)$ & MSE($\alpha_2$)  & $\overline{\widehat{\beta_2}} \, (s)$  & MSE($\beta_2$)& $\overline{\widehat{\alpha_3}} \, (s)$ & MSE($\alpha_3$)  & $\overline{\widehat{\beta_3}} \, (s)$  & MSE($\beta_3$)\\ \toprule
\multirow{8} {1.2cm}{Low variability} & \multirow{4} {1cm}{High linearity} & 10 & 1.9247 (0.8816) & 0.7820 & 1.0367 (0.8390) & 0.7046 & 0.7799 (0.9285) & 0.9397 & 2.9414 (1.3009) & 1.6941 & 1.5213 (1.0289) & 1.0579 & 0.9711 (0.8998) & 0.8097 \\
                                   &                                 & 30 & 1.9912 (0.4426) & 0.1958 & 0.9954 (0.4330) & 0.1874 & 0.5167 (0.4119) & 0.1697  & 3.0225 (0.4770) & 0.2278 & 1.5212 (0.5678) & 0.3225 & 0.9815 (0.5585) & 0.3119\\
                                   &                                 & 100 & 2.0081 (0.2224) & 0.0495 & 1.0000 (0.2244) & 0.0503 & 0.4970 (0.2458) & 0.0603 & 2.9956 (0.2468) & 0.0609 & 1.4927 (0.2693) & 0.0725 & 1.0018 (0.2570) & 0.0660 \\
                                   &                                 & 250 &  2.0039 (0.1581) & 0.0250 & 1.0012 (0.1531) & 0.0234 & 0.4987 (0.1536) & 0.0236 & 2.9996 (0.1508) & 0.0227 & 1.4904 (0.1596) & 0.0255 & 1.0051 (0.1605) & 0.0258\\ \cline{2-15}
                                   & \multirow{4} {0.8cm}{Low linearity} & 10 & 1.2467 (1.8961) & 4.1591 & 1.1982 (1.8335) & 3.3976 & 2.7543 (4.4820) & 25.1500 & 3.4780 (4.7935) & 23.1829 & 1.5193 (2.3382) & 5.4621 & 1.2395 (2.1548) & 4.6958\\
\end{tabular}}
\caption{\footnotesize  Results of the \textit{DSD Model} $\Psi_{\widehat{Y}(j)}^{-1}(t)=-1+2\Psi_{X_{1}(j)}^{-1}(t)-1\Psi_{X_{1}(j)}^{-1}(1-t)+0.5\Psi_{X_{2}(j)}^{-1}(t)-3\Psi_{X_{2}(j)}^{-1}(1-t)+1.5\Psi_{X_{3}(j)}^{-1}(t)-1\Psi_{X_{3}(j)}^{-1}(1-t)$ in different conditions.}
\label{table4SA3}
\end{center}
\end{table}

\end{landscape}

\newpage

\begin{landscape}
\vspace{10cm}
\renewcommand{\arraystretch}{1.2}
\begin{table}
\begin{center}
{\scriptsize
\begin{tabular}{p{2cm}|p{2cm}|c|c|c|c|c|c|c}
 \toprule
   \multirow{2} {2cm} {\textit{Degree of variability}}  & \multirow{2} {2cm} {\textit{Degree of linearity }} & \multirow{2} {*} {$n$}& \multicolumn{2}{|c|}{\textit{Parameter estimated }} &\multicolumn{4}{|c|}{\textit{Goodness-of-fit measures}} \\ \cline{4-9}
      & & & $\overline{\widehat{\gamma}} \, (s)$ & MSE($\gamma$) & $\overline{\Omega} \, (s)$ & $\overline{RMSE}_{M} \, (s)$ & $\overline{RMSE}_{L} \, (s)$ & $\overline{RMSE}_{U} \, (s)$ \\ \toprule
\multirow{12} {*}{Low variability} & \multirow{4} {1cm}{High linearity} & 10 & -2.1536 (19.2674) & 372.1926 & 0.9851 (0.0051) & 1.3216  (0.2356) & 1.5383 (0.3481) & 1.5233 (0.3366)\\
                                   &                                 & 30 & -1.2743 (9.3471) & 87.3563 & 0.976 (0.0041) & 1.7268 (0.1514) & 1.903 (0.2096) & 1.912 (0.2111)\\
                                   &                                 & 100 & -0.9628 (3.9252) & 15.3933 & 0.9716 (0.0024) & 1.867 (0.0792) & 2.0315 (0.118) & 2.0399 (0.116)\\
                                   &                                 & 250 &  -0.8874 (2.4337) & 5.9297 & 0.9707 (0.0015) & 1.9269 (0.0508) & 2.0735 (0.069) & 2.0789 (0.0743)\\ \cline{2-9}
                                   & \multirow{4} {1cm}{Low linearity} & 10 & -2.6576 (35.6539) & 1272.6805  & 0.5126 (0.1076) & 11.539 (1.8376) & 13.3151 (2.8115) & 13.3202 (2.8765)\\
                                   &                                 & 30 & -0.5548 (28.5401) & 813.9222 & 0.4074 (0.0563) & 14.0863 (1.1283) & 15.5937 (1.6976) & 15.5885 (1.7508)\\
\end{tabular}}
\caption{\footnotesize  Results of the \textit{DSD Model} $\Psi_{\widehat{Y}(j)}^{-1}(t)=-1+2\Psi_{X_{1}(j)}^{-1}(t)-1\Psi_{X_{1}(j)}^{-1}(1-t)+0.5\Psi_{X_{2}(j)}^{-1}(t)-3\Psi_{X_{2}(j)}^{-1}(1-t)+1.5\Psi_{X_{3}(j)}^{-1}(t)-1\Psi_{X_{3}(j)}^{-1}(1-t)$ in different conditions (continuation of the \textit{Table \ref{table4SA3}}).}
\label{table5SA3}
\end{center}
\end{table}

\end{landscape}

\newpage

\begin{landscape}
\vspace{10cm}
\renewcommand{\arraystretch}{1.2}
\begin{table}
\begin{center}
{\scriptsize
\begin{tabular}{p{1.2cm}|p{1cm}|p{0.25cm}|c|@{}c@{}|c|@{}c@{}|c|@{}c@{}|c|@{}c@{}|c|@{}c@{}|c|@{}c@{}}
\toprule
   \multirow{2} {1.2cm} {\textit{Degree of variability}}  & \multirow{2} {1cm} {\textit{Degree of linearity }} & \multirow{2} {0.25cm} {$n$}&  \multicolumn{12}{|c|}{\textit{Parameters estimated}} \\ \cline{4-15}
      & & & $\overline{\widehat{\alpha_1}} \, (s)$ & MSE($\alpha_1$)  & $\overline{\widehat{\beta_1}} \, (s)$  & MSE($\beta_1$)& $\overline{\widehat{\alpha_2}} \, (s)$ & MSE($\alpha_2$)  & $\overline{\widehat{\beta_2}} \, (s)$  & MSE($\beta_2$)& $\overline{\widehat{\alpha_3}} \, (s)$ & MSE($\alpha_3$)  & $\overline{\widehat{\beta_3}} \, (s)$  & MSE($\beta_3$)\\ \toprule
\multirow{8} {1.2cm}{Low variability} & \multirow{4} {1cm}{High linearity} & 10 & 6.1694 (3.1361) & 9.8541 & 1.3428 (2.0324) & 5.9296 & 2.891 (3.3623) & 12.0875 & 7.9978 (4.5123) & 20.3402 & 8.7821 (3.0319) & 10.6667 & 4.1545(3.4200) & 12.3998 \\
                                   &                                 & 30 & 6.0988 (1.5960) & 2.5544 & 0.5709 (0.8460) & 1.0409  &1.9224 (1.4279) & 2.0428 & 7.8815 (1.6203) & 2.6369 & 9.7278 (2.1155) &4.5448 &4.5571 (2.0011) &4.1966\\
                                   &                                 & 100 &  6.0955 (0.8501) & 0.7311 & 0.2904 (0.4423) & 0.2798 & 1.9635 (0.8737) & 0.7639 & 7.8969 (0.8893) & 0.8007 & 9.7801 (0.9584) & 0.9660 & 4.8221 (0.9858) & 1.0025\\
                                   &                                 & 250 &  6.0455 (0.5802) & 0.3384 & 0.2176 (0.3295) & 0.1558 & 1.9635 (0.5389) & 0.2914 & 7.9428 (0.5316) & 0.2856 & 9.862 (0.5637) & 0.3365 & 4.8699 (0.5815) & 0.3547\\ \cline{2-15}
                                   & \multirow{4} {0.8cm}{Low linearity} & 10 & 4.5523 (6.6216) & 45.8982 & 3.8118 (6.4589) & 56.206 & 8.6424 (14.3291) & 249.2382 & 12.8654 (17.2332) & 320.3577 & 5.4207 (8.0094) & 85.0558 & 4.7002 (7.8248) & 61.2565 \\
\end{tabular}}
\caption{\footnotesize  Results of the \textit{DSD Model} $\Psi_{\widehat{Y}(j)}^{-1}(t)=3+6\Psi_{X_{1}(j)}^{-1}(t)-0\Psi_{X_{1}(j)}^{-1}(1-t)+2\Psi_{X_{2}(j)}^{-1}(t)-8\Psi_{X_{2}(j)}^{-1}(1-t)+10\Psi_{X_{3}(j)}^{-1}(t)-5\Psi_{X_{3}(j)}^{-1}(1-t)$ in different conditions.}
\label{table6SA3}
\end{center}
\end{table}

\end{landscape}

\newpage

\begin{landscape}
\vspace{10cm}
\renewcommand{\arraystretch}{1.2}
\begin{table}
\begin{center}
{\scriptsize
\begin{tabular}{p{2cm}|p{2cm}|c|c|c|c|c|c|c}
\toprule
   \multirow{2} {2cm} {\textit{Degree of variability}}  & \multirow{2} {2cm} {\textit{Degree of linearity }} & \multirow{2} {*} {$n$}& \multicolumn{2}{|c|}{\textit{Parameter estimated }} &\multicolumn{4}{|c|}{\textit{Goodness-of-fit measures}} \\ \cline{4-9}
      & & & $\overline{\widehat{\gamma}} \, (s)$ & MSE($\gamma$) & $\overline{\Omega} \, (s)$ & $\overline{RMSE}_{M} \, (s)$ & $\overline{RMSE}_{L} \, (s)$ & $\overline{RMSE}_{U} \, (s)$ \\\toprule
\multirow{12} {*}{Low variability} & \multirow{4} {1cm}{High linearity} & 10 & 5.5837 (62.6561) & 3928.5307 & 0.9847 (0.0052) & 4.7065 (0.8275) & 5.4394 (1.1555) & 5.5045 (1.2423)\\
                                   &                                 & 30 & 2.5514 (33.1331) & 1096.9076 & 0.9750 (0.0040) & 6.1476 (0.5048) & 6.7387 (0.7676) & 6.7623 (0.7482)\\
                                   &                                 & 100 & 3.5040 (13.995) & 195.9175 & 0.9699 (0.0025) & 6.7096 (0.2856) & 7.2891 (0.4075) & 7.3016 (0.3988)\\
                                   &                                 & 250 & 3.3424 (8.3900) & 70.4392 & 0.9713 (0.0015) & 6.6510 (0.1779) & 7.1877 (0.2533) & 7.1793 (0.2544)\\ \cline{2-9}
                                   & \multirow{4} {1cm}{Low linearity} & 10 & 40.0872 (123.3892) & 16585.1230 & 0.5104 (0.1054) & 40.7149 (6.4713) & 46.7226 (10.1200) & 47.1164 (10.1294)\\
                                   &                                 & 30 & 24.5761 (110.7202) & 12712.2417 & 0.3998 (0.0548) & 49.6452 (3.9681) & 54.3399 (5.9191) & 54.8069 (5.9734)\\
\end{tabular}}
\caption{\footnotesize  Results of the \textit{DSD Model} $\Psi_{\widehat{Y}(j)}^{-1}(t)=3+6\Psi_{X_{1}(j)}^{-1}(t)-0\Psi_{X_{1}(j)}^{-1}(1-t)+2\Psi_{X_{2}(j)}^{-1}(t)-8\Psi_{X_{2}(j)}^{-1}(1-t)+10\Psi_{X_{3}(j)}^{-1}(t)-5\Psi_{X_{3}(j)}^{-1}(1-t)$ in different conditions (continuation of the \textit{Table \ref{table4SA3}}).}
\label{table7SA3}
\end{center}
\end{table}

\end{landscape}

\end{document}